\documentclass[journal=jacsat,manuscript=preprint]{achemso}

\usepackage{chemformula}  
\usepackage{amsmath}
\DeclareMathOperator*{\atan2}{atan2}
\usepackage{braket}
\usepackage{setspace}
\usepackage{subfigure}
\usepackage{float}
\usepackage{soul}
\usepackage{comment}
\usepackage{tabularx}
\usepackage[export]{adjustbox}
\usepackage{xr}
\mciteErrorOnUnknownfalse

\usepackage{caption}
\author{Omar Ibrahim}
\affiliation{Department of Chemistry, University of Pennsylvania, Philadelphia, PA, 19104}
\email{omaribr@sas.upenn.edu}
\author{Sunghee Lee}
\affiliation{Department of Chemistry and Nanoscience, Ewha Womans University, 52 Ewhayeodae-gil, Seodaemun-gu, Seoul 03760, South Korea}
\author{Sung Wook Kim}
\affiliation{Department of Chemical Engineering, Hanyang University, Seoul 04763, Republic of Korea}
\author{Seung Beom Pyun}
\affiliation{Department of Chemical Engineering, Hanyang University, Seoul 04763, Republic of Korea}
\author{Connor Woods}
\affiliation{Department of Chemistry, University of Pennsylvania, Philadelphia, PA, 19104}
\author{Eun Chul Cho}
\affiliation{Department of Chemical Engineering, Hanyang University, Seoul 04763, Republic of Korea}
\author{So-Jung Park}
\affiliation{Department of Chemistry and Nanoscience, Ewha Womans University, 52 Ewhayeodae-gil, Seodaemun-gu, Seoul 03760, South Korea}
\author{Zahra Fakhraai}
\affiliation{Department of Chemistry, University of Pennsylvania, Philadelphia, PA, 19104}
\email{fakhraai@sas.upenn.edu} 

\title[An \textsf{achemso} demo]
  {Optical Magnetic Multipolar Resonances in Large Dynamic Metamolecules}


\begin{document}

\begin{abstract}
Dynamic metamolecules (DMMs) are composed of a dielectric core made of hydrogel surrounded by randomly-packed plasmonic beads that can display magnetic resonances when excited by light at optical frequencies. Their optical properties can be controlled by controlling their core diameter through temperature variations. We have recently shown that DMMs display strong optical magnetism, including magnetic dipole and magnetic quadrupole resonances, offering significant potential for novel applications. Here, we use a T-matrix approach to characterize the magnetic multipole resonance modes of model metamolecules and explore their presence in experimental data.
We show that high-order multipole resonances become prominent as the bead size and the overall structure sizes are increased, and when the the inter-bead gap is decreased. In this limit, mode mixing among high-order magnetic multipole modes also become significant, particularly in the directional scattering spectra. We discuss trends in magnetic scattering observed in both experiments and simulations, and provide suggestions for experimental design and verification of high-order optical magnetic resonances in the forward or backward scattering spectra. In addition, angular scattering of higher-order magnetic modes can display Fano-like interference patterns that should be experimentally detectable. 
\end{abstract}

\singlespacing
\footnotesize
\section{Introduction} 
Magnetic resonances can be achieved at optical frequencies through emergent optical response of specifically-designed arrangements of  metallic and dielectric nanostructures, that produce rotating displacement currents that are in phase with the magnetic field of light\cite{Nader2,QuestOptMagnetismReview,Parker}. Optical magnetism has been realized in two-dimensional (2D) arrangements and clusters of plasmonic metal nanoparticles ring-resonators\cite{meng2020plasmon,liu2012magnetic,stevenson2020active,cherqui2016stem,Ogut2012toroidal}, arrangements with broken symmetry\cite{pakizeh2008structural,cherqui2016stem}, 2D arrays of nanoparticles\cite{wang2019magnetic,MasielloNanoclusters,ballantine2020optical,kang2019interference,linden2006photonic,li2016optical,liu2007magnetic}, as well as hybrid metal-dielectric nanostructures and arrays\cite{ginn2012realizing,verre2015optical,tserkezis2008understanding}.
There significant recent interest in optical magnetism stems\cite{MagneticHotspotReview,QuestOptMagnetismReview,AnotherReviewPaper,SelfAssembledNanoshells,chen2017engineering,qin2013optical,urzhumov2008optical} from their potential to produce materials with unique optical properties in applications, such as near-zero and negative indices of refraction\cite{urzhumov2008optical,NearZeroRF,NegativeRIPhysics,NegativeRIMetamaterials}, optical cloaking \cite{Cloaking, RockstuhlCloaking}, sensing of chiral properties\cite{ChiralEnhancement}, and surface enhanced spectroscopies\cite{OpticalSensing,SubDiffractionMicroscopy}.

Most 2D structures and assemblies used to achieve optical magnetism, require top-down fabrication and need the incident light to have a specific polarization to produce magnetic moments. However, 3D self-assembled structures can also produce magnetic responses in optical frequencies\cite{ACSNano,ponsinet2015resonant,bourgeois2017self,MagneticMetafluid}, and in contrast to 2D assemblies, these clusters produce magnetic properties regardless of the orientation and polarization of the incident field. 
A specific subclass of these materials, named magnetic metamolecules, consist of tightly packed clusters of plasmonic nanoparticles (beads). These metamolecules have seen great interest in recent years due to their ease of synthesis, tunability, and potent magnetic resonances \cite{ACSNano,lim2014gold,JPCC,ChenJPCC,ParkerFirst,RockstuhlPlasmonic,RockstuhlReview,bourgeois2017self, CoreShellModel,ResonantIsotropicMagnetism,3DNanoclusters,MagneticMetafluid}.   
We have recently demonstrated that the magnetic resonance modes of these metamolecules can be dynamically controlled using hydrogel cores that can change their core size upon heating, which we call dynamic metamolecules (DMMs).\cite{JPCC} 
We have shown that colloidal solutions of DMMs can produce experimentally realizable magnetic dipole and magnetic quadrupole resonances in the far-field, in both gold and silver, and their properties can be controlled by adjusting the temperature of their aqueous solution \cite{JPCC}. 
This is significant, as tunable higher order magnetic resonances can have unique scattering behaviors \cite{ParkerBackscattering,GoldArrayFano} and non-linear optical effects\cite{kruk2017nonlinear} that could serve to further expand the range of applications of these systems. 

While previous studies have explored the optical behavior of magnetic metamolecules, the majority of these studies have focused on magnetic dipolar resonances\cite{Nader2,ChenJPCC,ponsinet2015resonant,bourgeois2017self}. Recent studies have also investigated higher order modes \cite{Parker,ParkerArXiv,ParkerFirst} but many details remain poorly understood. In particular, a direct connection between theoretical predictions and experimental observations of these higher order modes have not been made and a clear recipe to experimentally realize and verify the existence of high-order magnetic multipole modes have not been provided. Here, we fully characterize the scattering behavior of model MMs using a T-matrix approach\cite{barber1990light,stout2008recursive,duan2015experimental,khlebtsov2007multipole,khlebtsov2013t,arya2006scattering,RockstuhlTMatrix} based on finite-difference time domain (FDTD) simulations. The T-matrix approach is especially powerful compared to standard basis decomposition techniques as it allows us to visualize not only the contribution of each scattering mode on the total scattering\cite{RockstuhlTMatrix,johnson1988invariant,Simon} and extinction cross-sections\cite{zhao2003extinction}, but also provides information on directional and anisotropic scattering\cite{kiselev2002light} as well as the influence of each multipole mode and mode-mixing on these quantities\cite{Simon}
Through a single T-matrix calculation, we can obtain this information at any orientation of incident and scattered light \cite{Simon, RockstuhlTMatrix}. 

Using the T-matrix approach to solve the full scattering matrix of model magnetic metamolecules (MMs), We demonstrate that  high-order magnetic resonances can emerge in the optical frequencies as the bead size or the overall structure size grows, and when the DMMs cores are shrunk, such that the interbead gap is reduced. We also demonstrate the potential significance of mode mixing in the differential scattering intensities and the total scattering cross-sections, which has not been previously explored. We characterize the angular scattering of model DMMs and show how the high-order magnetic resonances can interfere with electric modes to produce Fano-like resonances in certain directions, which can be used as a strategy to experimentally observe high-order magnetic modes. Using this concept we use backscattering analysis to propose a method to detect the magnetic octupole resonance in the far-field, which has not been experimentally observed. We compare these results with experimental data that show magnetic quadrupole resonances in the far-field.  Our results provide a recipe for accurately identifying the magnetic quadruple and octupole resonances using comparisons between forward and backward scattering experiments.

\section{Methods}

\subsection{Experimental Methods}

\paragraph{Fabrication of Gold Dynamic Metamolecules.} Gold DMMs were fabricated by following a previous literature procedure.\cite{JPCC} A solution of freshly prepared gold nanobeads dispersed in 1 mM sodium citrate (4 mL) and a 10 $\mu$L solution of 1 wt\% poly(N-isopropylacrylamide-co-allyamine)  hydrogel (PNIPAM) were mixed in a 10 mL glass vial. The concentration of the gold nanobead solution and other structural parameters are summarized in Table S1. This solution was kept undisturbed at room temperature for about 12 h. Since DMMs are heavy, once formed, they settle to the bottom of the vial over time, while unattached nanobeads remain in the supernatant. The supernatant was removed, the precipitate was redispersed in 1 mM sodium citrate solution (4 mL), followed by sonication for 10 s. gold DMMs were characterized using scanning electron microscope (SEM, JEOL JSM-7610F at an accelerating voltage of 15 kV).
 
\paragraph{Temperature-Dependent Extinction Measurement of Gold DMMs.} Temperature-dependent extinction spectra of gold DMMs were recorded with an Agilent 8453 UV-Vis spectrometer. For this, gold DMMs solutions (3 mL) were taken in a cuvette (dimensions (H×W×D): 45×12.5×12.5 mm$^3$, optical path length: 10 mm) and the extinction spectra were measured by increasing the temperature at the rate of 1°~C/min from 20~°C to 55~°C under magnetic stirring. Samples were equilibrated for 1~min at each temperature.

\paragraph{Temperature-Dependent Hydrodynamic Diameter Measurement of PNIPAM Hydrogel.}
Temperature-dependent hydrodynamic diameter measurements of PNIPAM hydrogel cores were
conducted using dynamic light scattering (DLS, Malvern Zetasizer Nano ZS) with a He-Ne laser (633 nm). The PNIPAM hydrogel solution (0.01 wt\%, 1 mL) were measured by increasing the temperature from 25 °C to 50 °C. The solution was equilibrated for 10 min at each temperature. This data can be seen in Figure S2

\subsection{Simulation Methods}

To generate model metamolecules (MMs), points were randomly distributed on a unit sphere. A Monte Carlo method (with $T=0$~K) was then used to uniformly separate the points by treating them as point charges. These structures were then scaled up and imported into Lumerical FDTD Solutions where optical properties of the structure could be determined (see our previous publication for details \cite{JPCC}). The T-matrix method involving multiple plane wave illumination was adopted \cite{Simon, RockstuhlTMatrix}, where both the incident field and the scattered field were expanded into vector spherical harmonic basis functions at many random orientations. This provides an over-determined set of equations, which were then numerically solved to obtain the T-Matrix. This T-Matrix can then be used to characterize the complete scattering properties of the MMs for any orientations of incident and scattered light. 

\paragraph{FDTD Simulations.}
The structures generated by the Monte Carlo method were imported into Lumerical FDTD Solutions Software. Each point charge was replaced by a gold spherical bead with optical properties based on the CRC Handbook of Chemistry and Physics. \cite{CRCHandbook} 
The core index was set to 1.56 to mimic the hydrogel index of refraction for DMMs and a background index of 1.333 was used to model the fact that the experimental DMMs were measured in aqueous solutions.\cite{JPCC} See Table S2 for all structural parameters for MMs and DMMs used in this study.  

\begin{figure}[H]
\includegraphics[width=0.8\linewidth]{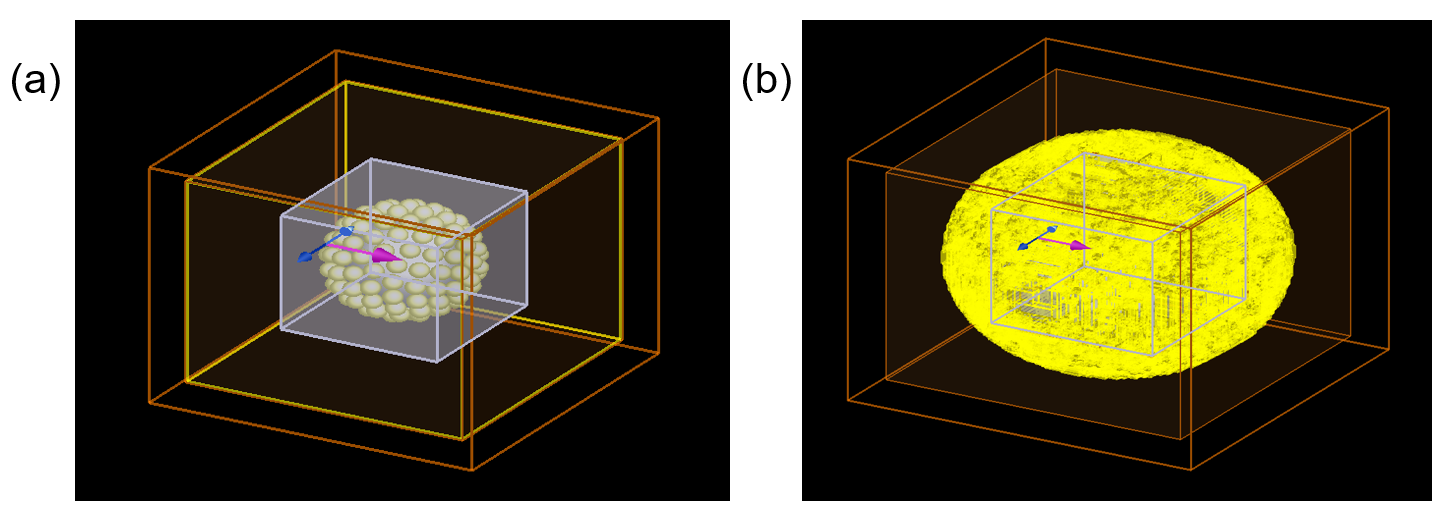} 
\caption{a) Simulation region set up in Lumerical FDTD package with a cubic field monitor. b) The cubic shell field monitors set up in the Lumerical FDTD package. The grey box in (a) represents Lumerical's built-in TFSF source and the yellow boxes in both figures represent field monitors.}
\label{fig:Setup}
\end{figure}

To distinguish between the incident light and scattered light in our simulations we used Lumerical's built-in total-field scattered field (TFSF) source. TFSF source specifies a box whereby a plane wave is introduced on one side and subtracted on the other. This has the effect of separating the region into a total field region and a scattered field region \cite{Simon}.
It is in this scattered field region where the $\vec{E}$ field and $\vec{H}$ fields can be measured in a cubic box (Figure \ref{fig:Setup}a) and these fields can then be interpolated to a spherical shell of interest to calculate the T-matrix. To ensure that none of the incident field is contained in this sphere the TFSF source must be completely inscribed in the sphere of interest which requires that the field monitors be at least a factor of $\sqrt{3}$ away from the edge of the TFSF source. This can be computationally expensive. In this work, instead of saving the entire inscribing sphere as has been done in our previous works \cite{Simon}, we use a large number of cubic monitors to construct a thin shell about the radius of interest as shown in Figure \ref{fig:Setup}b (details of this approach are in Section SII of SI). This approach reduces the requirements for disk space, allowing simulations on larger systems.

Throughout this paper T-matrix results are also compared to the scattering cross-section and the differential scattering cross-section outputted directly by Lumerical FDTD as a benchmark. To obtain the scattering cross-sections, the MM is surrounded by a box composed of six plane power monitors that when factored in together provide the total power scattered by the model MM. To obtain the differential scattering cross-sections, the far field projection feature offered by Lumerical FDTD was used.

\paragraph{T-Matrix Calculations.}
Here we use a T-matrix calculation technique involving multiple plane wave illuminations \cite{Simon,RockstuhlTMatrix}. Both the incident field ($\vec{E}_i(\vec{r})$), which is a plane polarized wave here, within a circumscribing sphere, and the scattered field ($\vec{E}_s(\vec{r})$), outside a circumscribing sphere, can be written as infinite linear combinations of spherical waves as:

\begin{eqnarray}
\vec{E}_i(\vec{r})&=&\sum \limits_{n=1}^{\infty} \sum \limits_{m=-n}^{n} [a^M_{mn} Rg \vec{M}_{mn}(kr, \theta, \phi) + a^N_{mn} Rg \vec{N}_{mn}(kr, \theta, \phi)]\label{eqn1}\\  
 \vec{E}_s(\vec{r})&=&\sum \limits_{n=1}^{\infty} \sum \limits_{m=-n}^{n} [b^M_{mn} \vec{M}_{mn}(kr, \theta, \phi) + b^N_{mn} \vec{N}_{mn}(kr, \theta, \phi)]\label{eqn2}
\end{eqnarray}
where $\vec{M}$ and $\vec{N}$  in equations \ref{eqn1} \& \ref{eqn2} constitute a complete basis and $Rg \vec{M}$ and $Rg \vec{N}$ are regularized to contain no singularity within the sphere. $\vec{M}$ and $Rg \vec{M}$ represent the electric field components generated due to magnetic modes and $\vec{N}$ and $Rg \vec{N}$ represents the electric field components generated due to electric modes. $n$ represents the mode order (monopole, dipole, etc.) while $m$ represents a given sub-mode of $n$, where $-n\leq m\leq n$. $\{a\}_{mn}$ and $\{b\}_{mn}$ are the basis coefficients of the incident and scattered waves respectively while the superscript $N$ and $M$ denote the electric and magnetic components respectively.  Due to the linearity of the Maxwell's equations and their boundary conditions \cite{Bohren},
the incident and scattered fields are linearly related through the T-matrix ($\overline{\overline{T}}$):

\begin{eqnarray}
	\begin{bmatrix}
		\vec{b}_{mn}^{N} \\
		\vec{b}_{mn}^M
	\end{bmatrix}
	&=&
	\overline{\overline{T}}
	\begin{bmatrix}
		\vec{a}_{mn}^{N} \\
		\vec{a}_{mn}^{M}
		\end{bmatrix}
\\
	\overline{\overline{T}}
	&=& 
	\begin{bmatrix}
		\overline{\overline{T}}^{EE} & 	\overline{\overline{T}}^{EM} \\
		\overline{\overline{T}}^{ME} &
		\overline{\overline{T}}^{MM}
	\end{bmatrix}
\end{eqnarray}

We note that all T-matrix calculations here are performed only on the electric component of light as indicated in equations 1 \& 2. $\overline{\overline{T}}^{EE}$ is the sub-matrix that gives the electric influence on the scattered electric field while $\overline{\overline{T}}^{MM}$ is the sub-matrix that gives the magnetic influence on the scattered electric field. $\overline{\overline{T}}^{EM}$ and $\overline{\overline{T}}^{ME}$ are the sub-matrices that provide electric-magnetic mode mixing. It is important to note that $\overline{\overline{T}}$ and its sub-matrices are infinite matrices and thus they must be truncated at a given mode. Here, we use up to $n=4$, which corresponds to hexadecapole modes. $\vec{a}^M_{mn}$ and $\vec{a}^N_{mn}$ can be directly calculated for a plane wave propagating in any direction. 
To determine $\vec{b}^M_{mn}$ and $\vec{b}^N_{mn}$ we obtain $\vec{E}$ and $\vec{H}$ at every point within a shell surrounding the particle and outside the TFSF source. We then define a sphere within this shell to which $\vec{E}_s$ and $\vec{H}_s$ can be interpolated and write: 

\begin{eqnarray}
b_{mn}^{M} &=&\frac{i}{c \epsilon_0 N_i} \frac{[\oint (\hat{r} \cdot \vec{H}_s) Y^{-m}_{n}(\theta, \phi)dA] k (2n + 1)}{4 \pi r_0^2 \gamma_{mn} (-1)^m n(n+1) h_n(kr_0)}\\ b_{mn}^{N} &=&\frac{[\oint (\hat{r} \cdot \vec{E}_s) Y^{-m}_{n}(\theta, \phi)dA] k (2n + 1)}{4 \pi r_0^2 \gamma_{mn} (-1)^m n(n+1) h_n(kr_0)}
\end{eqnarray}
\noindent
where $r_0$ is the radius of the integration sphere, $N_i$ is the refractive index,
and $\gamma_{mn} \equiv \sqrt{\frac{(2n + 1) (n-m)!}{4 \pi n (n+1) (n+m)!}}$ \cite{SimonThesis,Simon,RockstuhlTMatrix}. In these simulations, rather than rotating the incident light, we choose to rotate the structure, which is more convenient for FDTD simulations. $\vec{a}_{mn}^M$, $\vec{a}_{mn}^N$, $\vec{b}_{mn}^M$, and $\vec{b}_{mn}^N$ are determined for $K$ 
different random orientations of incident light. As such, we can write:

\begin{equation}
	\begin{bmatrix}
		\vec{b}_{mn}^{N, 1}  & \vec{b}_{mn}^{N, 2} & \dots & \vec{b}_{mn}^{N, K} \\
		\vec{b}_{mn}^{M, 1} & \vec{b}_{mn}^{M, 2} & \dots & \vec{b}_{mn}^{M, K}
	\end{bmatrix}
	= \overline{\overline{T}}
	\begin{bmatrix}
		\vec{a}_{mn}^{N, 1}  & \vec{a}_{mn}^{N, 2} & \dots & \vec{a}_{mn}^{N, K} \\
		\vec{a}_{mn}^{M, 1} & \vec{a}_{mn}^{M, 2} & \dots & \vec{a}_{mn}^{M, K}
		\end{bmatrix}
\end{equation}

To compute the T-matrix from the above expression we do a minimization over the T-matrix elements
using the objective function $||\overline{\overline{T}}  \overline{\overline{A}} - \overline{\overline{B}} ||^2_F$ (where the subscript $F$ denotes the Frobenius norm). In this optimization we take advantage of the reciprocity relation $T^{ij}_{-m'n'(-m)n} = (-1)^{m' + m} T^{ji}_{mnm'n'}$ for $i, j = \{E, M\}$, cutting the number of free parameters by a factor of almost 2 \cite{Tsang}. A possible set of free parameters implied by this reciprocity relation can be visualized in Figure S4.

This approach is advantageous as it allows for the T-Matrix of a large number of frequencies to be determined simultaneously whereas traditionally the T-matrix is calculated separately for each desired frequency. Since we only require the fields scattered in any sphere about the scattering body, this approach is also highly robust to sharp features \cite{Simon}. Once we calculate the T-matrix for each desired frequency all desired optical properties can be determined from it. In particular, from the T-matrix we obtain the scattering amplitude dyad $\overline{\overline{F}}$ from which numerous parameters of interest are obtained. Relevant parameters derived from $\overline{\overline{F}}$ include $\vec{E}_s$ and $\vec{J}_D$, the far field scattered field, and the far field displacement current at all points. We also obtain $\vec{X}(\theta, \phi)$ ($\theta \in [0, \pi]$, $\phi \in [0, 2 \pi])$, the vector scattering amplitude, which we take to be 
\begin{equation}
\vec{X} = \overline{\overline{F}} \cdot \hat{\beta}
\label{eqn:amplitude}
\end{equation}
(where $\hat{\beta}$ is the polarization vector) such that in the far field, $\vec{X} \cdot \hat{r} = 0$ and $\vec{E}_s = \frac{e^{ikr}}{r} \vec{X}$. Finally we also derive the scattering cross-section $C_{scat}$ and the differential scattering cross-section $\frac{d C_{scat}}{d \Omega}$ for any given direction. To see the details of how these functions are obtained from the T-matrix see Section SVI of SI. \cite{Tsang,Bohren}

\paragraph{Basis Decomposition Calculations.} 
To make a more direct comparison and assignments of the peaks observed in modeled MMs with those observed in experiments, we simulated series of structures where specific structural motifs were varied, and calculated their magnetic resonance modes. To reduce the computational costs of these calculations, we performed a basis decomposition of the FDTD data as opposed to running the full T-matrix, which requires multiple FDTD simulations at various angles of the incident light. This method similarly involves computing the scattering coefficients, however we only perform a single simulation to perform the calculations. We then use these coefficients to compute the far-field scattered electric and magnetic fields for each mode from which the scattering cross-section can be obtained through numerical integration of the Poynting vector via $C_{scat} = \frac{1}{2I} \int_{\Omega} Re(\vec{E} \times \vec{H}^*) \cdot d\vec{A}$ where $I$ is the intensity of the incident plane wave.

This approach stops short of calculating the full T-matrix, but still allows us to see the modal scattering strength for the simulated orientation (Figure \ref{fig:Simulation}). Given that these structures are highly symmetric, the error due to this technique is negligible. For example, there is near perfect agreement between the $C_{scat}$ calculated using this simplified version and the full T-matrix calculated for the largest structure studied here (shown in Figure \ref{Fig:Csct}) For all other structures, the sum of the modal cross-sections also agrees well with $C_{scat}$ directly calculated from FDTD (see Figures S13-S15).

\section{Results and Discussions}

\begin{figure}[H]

\includegraphics[width=0.8\linewidth]{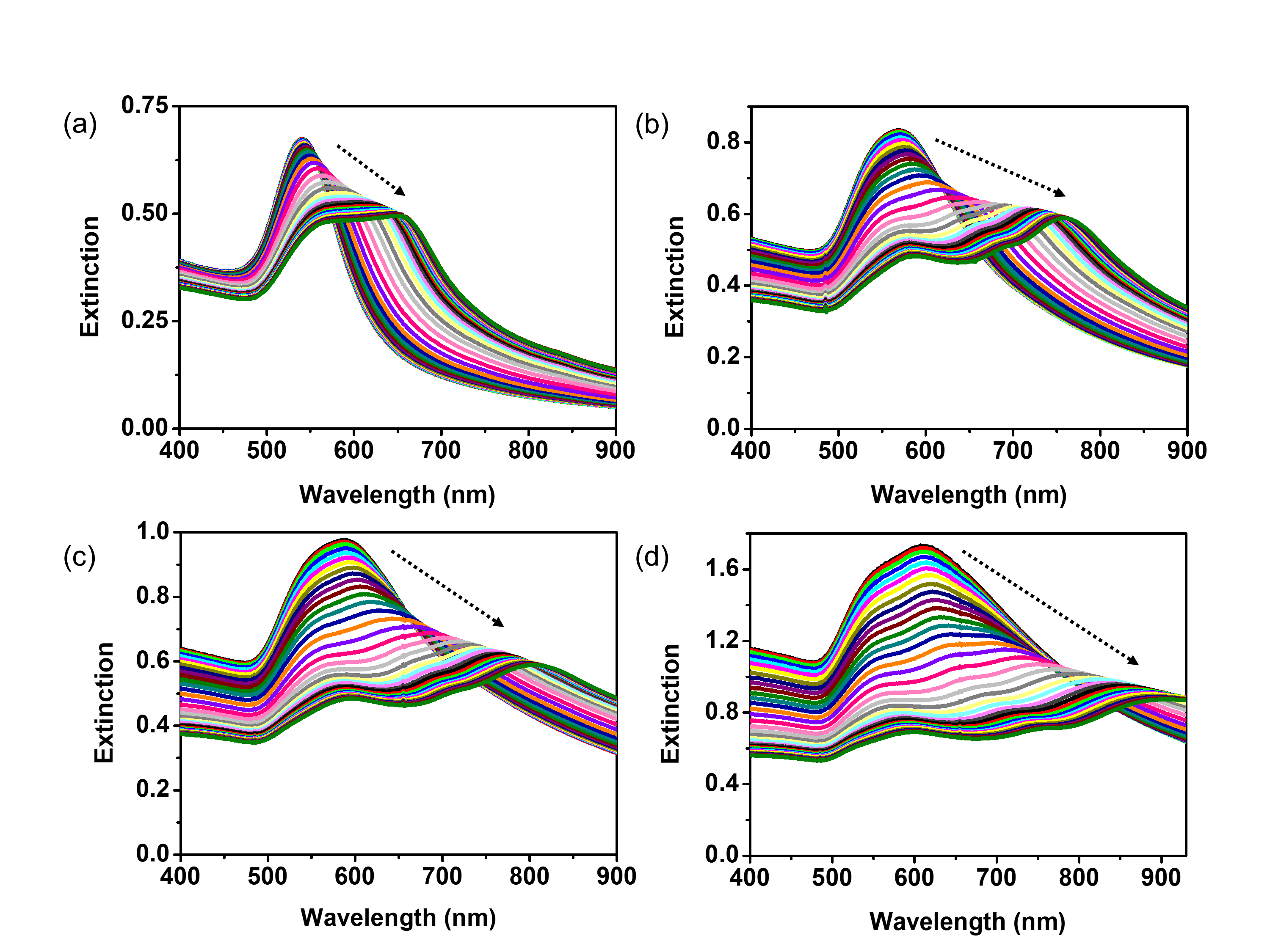}
\caption{(a-d) Temperature-dependent experimental extinction spectra of gold DMMs with nanobeads of average number $N$ and size $D$; (a) $N=57\pm3$, $D=35\pm3$~nm, (b) $N=61 \pm 4$, $D=45\pm2$~nm (c) $N=59 \pm 4$, $D=49\pm2$~nm, and (d) $N=41 \pm 4$, $D=60\pm3$~nm. Other structural parameters are summarized in Table S1 and Figures S1 and S2.
The dashed arrow in each figure shows the direction of increasing temperature from 20~°C to 55~°C, resulting in shrinking values of the core diameter, $Z$.}
\label{fig:experiment}
\end{figure}

\paragraph{Emergent High-Order Magnetic Resonances.} We have recently reported experimental and FDTD simulation results for optical properties of dynamic metamolecules (DMMs), demonstrating the emergence of strong magnetic dipole resonance modes when DMMs are heated above the hyrogel core's lower critical solution temperature(LCST), which for this PNIPAM hydrogel was determined to be LSCT=33~°C\cite{JPCC}. 
As can be seen from Figure S2a
as the temperature is increased above this 
temperature, the hydrodnamics diameter ($D_h$) of the hydrogel core size shrinks from $\sim$360~nm to $\sim$160~nm. 
Above LCST, the core diameter shrinks\cite{JPCC}, resulting in closer packing of the nanobeads. The strong coupling between the local electric dipole modes of these nanobeads arranged in a spherical geometry results in emergent magnetic dipole resonance, consistent with previous experimental and theoretical predictions\cite{ChenJPCC,ACSNano,MagneticMetafluid,Nader1,Nader2,MasielloNanoclusters,3DNanoclusters,bourgeois2017self}. 
However, it is important to note that as we have previously reported\cite{JPCC}, in DMMs, the gold beads provide geometric frustration that prevents the core size from fully shrinking. As such, above LSCT, the effective core diameter, $Z$ can be larger than the $D_h$ measured here. The electrostatic interaction between the nanobeads as well as potential coating of the gold beads by a monolayer of the hydrogel, keep the nanobeads from touching each other. Indeed, we have not observed any evidence that the nanobeads are in physical contact.  This is strengthen by the observation that a direct contact between gold beads would result in the loss of magnetic resonances \cite{JPCC,ACSNano}, which is not observed here. 

We also observed the emergence of an additional resonance mode as the structures size was increased, which we hypothesized was due to magnetic quadrupole modes, similar to previous observations\cite{ACSNano,Parker}. 
In this work, we have performed a set of experiments using gold nanobeads with larger diameters ($D$), to further investigate the emergence and the details of higher-order magnetic multipole resonances. Increasing $D$ is expected to increase the local electric dipole strength of the beads\cite{ChenJPCC,Nader2}, which can facilitate the observation of these high-order resonances within the range of optical frequencies.
Figures \ref{fig:experiment}a-d show the extinction spectra of colloidal DMMs when heated from 20~°C to 55~°C. The diameter of the gold nanobeads is increased from $D=35\pm3$~nm to $D=60\pm3$~nm from Figure \ref{fig:experiment}a to Figure \ref{fig:experiment}d, respectively. As we increase the temperature, the hydrogel core diameter ($Z$) shrinks, making tighter packings of the beads, and thus increasing the strengths of the emergent modes and shifting the modes to longer wavelengths (lower energy). For each $D$, as we increase the temperature above LCST, decreasing $Z$ and therefore the average inter-bead gap distance ($d_g$), a prominent middle peak emerges. This effect is more prominent for larger diameter nanobeads (Figure \ref{fig:experiment}c \& d). In order to better understand the nature and the rise of these plasmon resonances and gain insight in how to experimentally tune them, we simulated various structures and calculated their modal scattering, to visualize the modal contributions to each of the observed features. 

\begin{figure}[H]  
\includegraphics[width=0.6\linewidth]{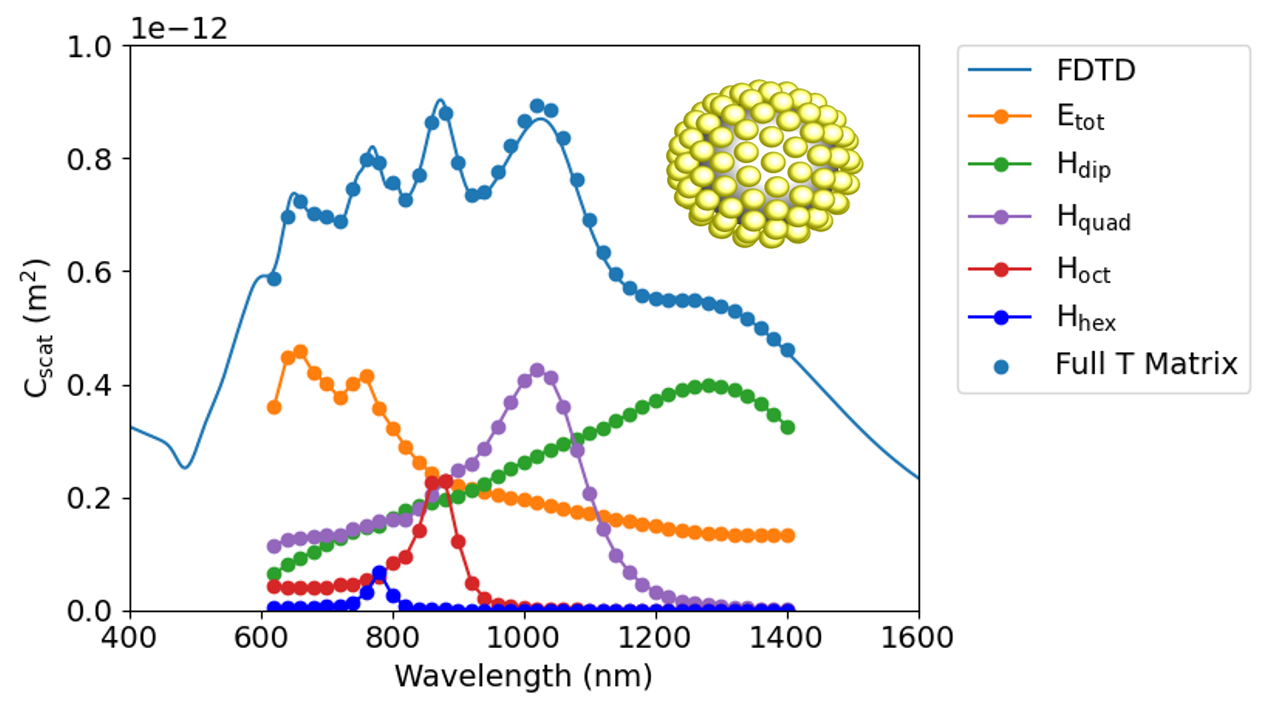}
\caption{Scattering cross-section ($C_{scat}$), calculated based on the full T-matrix with the order $n=4$ (modal sum, light blue circles), along with its relevant modal contributions; the sum of all electric modes; ($E_{tot}$, orange), magnetic dipole ($H_{dip}$, green), magnetic quadrupole ($H_{quad}$, purple), magnetic octopole ($H_{oct}$, red), and magnetic hexadecapole ($H_{hex}$, blue) modes, for a $Z=415$~nm diameter MM with $N=126$ gold nanobeads of $D=70$~nm diameter. The solid line is the total scattering cross-section directly obtained from FDTD simulations. The simulated structure is shown in the inset.} 
\label{Fig:Csct}
\end{figure}

\paragraph{Model Metamolecule with High-order Magnetic Multipolar Resonances.}We first focus our attention to one specific model MM structure, which is much larger than the experimentally available sizes; a simulated MM with $N=126$ number of nanobeads with $D=70$~nm, placed on a dielectric core with diameter, $Z=415$~nm (shown in the inset of Figure \ref{Fig:Csct}). A histogram of inter-bead gap distances ($d_g$) is shown in Figure S14. We chose this structure because it has stronger magnetic multipole resonances than the other structures studied here as well as structures that we have synthesized (Figure \ref{fig:experiment}). As such, this structure has the largest number of magnetic multipole modes active in the optical domain, while still being computationally tractable. This bounds the maximum number of dimensions that we need to consider in our T-matrix calculations.  We have computed the T-matrix for this structure at 40 different wavelengths and have decomposed the scattering cross-section into its individual modes, as detailed in the T-matrix calculation section as well as Sections SIII and SVI of SI.

To calculate the T-matrix, we chose  the T-matrix order $n=4$, which includes modes up to the electric and magnetic hexadecapole modes, assuming that the contribution from $n \geq 5$ modes are negligible. We also choose to fit every element of the T-matrix, as not doing so resulted significant error in the differential scattering cross sections (see Section SIII of SI and Figure S5 for more details). 
A representative image of the T-matrix results of the intensity of the scattered field calculated at a wavelength of $\lambda=780~nm$ is shown in Figure S7. This figure also demonstrates the relative strength of off-diagonal terms, particularly within the magnetic block of the T-matrix. We observe that off-diagonal terms corresponding to electric-magnetic mode mixing is mostly prominent for the magnetic hexadecapole ($H_{hex}$) modes, which couple with the electric dipole ($E_{dip}$) modes at their resonant frequency.

Figure \ref{Fig:Csct} shows the calculated scattering cross-section ($C_{scat}$) of the full T-matrix with $n=4$ for this structure, as well as the contribution of various multipole modes to the spectrum. As we can see in this Figure, the calculated scattering cross-section based on $n=4$ agrees well with the cross-section  obtained directly from the FDTD calculations, demonstrating the validity of our approach. We see that using $n=4$ captures all of the observed resonances in the structure, thus there is no need to include more modes with $n\geq5$, which would increase the dimension of the T-matrix from 48 to 70, leading to potentially dramatic over-fitting. Given that a magnetic hexadecapole ($n=4$) resonance, at the wavelength of $\lambda=780$~nm, is clearly observed in the far-field scattering cross-section, 
it is necessary to include $n=4$ modes in our T-matrix calculations for this structure, in order to capture this resonance. Further justification for using the full T-matrix as opposed to the simplified diagonal or intra-mode mixing (IMM, where modes with the same $n$ are allowed to mix) forms is provided in section SIII of SI (Figures S5 and S6). These discussion also show that for modes with $n\leq3$ (when $H_{hex}$ is not of interest), a block diagonal T-matrix can be used as a reasonable approximation. 

Figure \ref{Fig:Csct} shows that magnetic resonances dominate the scattering spectrum, particularly at long wavelengths ($\lambda\geq780$~nm) where the contribution from the electric modes slowly decays. Surprisingly, as $n$ increases from $n=1$ the magnetic multipole resonances remain strong, and all three magnetic resonances;  quadrupole resonance ($n=2$,  $\lambda_{Hquad}=1020$~nm), octupole resonance ($n=3$, $\lambda_{Hoct}=860$~nm), and hexadecapole resonance ($n=4$, $\lambda_{Hhex}=780$~nm) have apparent far-field scattering intensities that are stronger than that of the magnetic dipole resonance ($n=1$,  $\lambda_{Hdip}=1280$~nm).   However, when one focuses on the contribution from each multipole mode, the actual intensity of each resonance decreases for $n\geq2$. 
This is due to the added background scattering contribution from the combined electric multipole modes. The electric modes only have resonance features at low wavelengths ($\lambda~800$~nm), but decay slowly with increasing wavelength and as such, remain significant in the total scattering cross-section, despite being broad and featureless at $\lambda>800$~nm.
As such, each peak in the total scattering at $\lambda\geq780$~nm corresponds to a magnetic, rather than an electric multipole resonance.

\begin{figure}[H]
\includegraphics[width=1.2\linewidth,center]{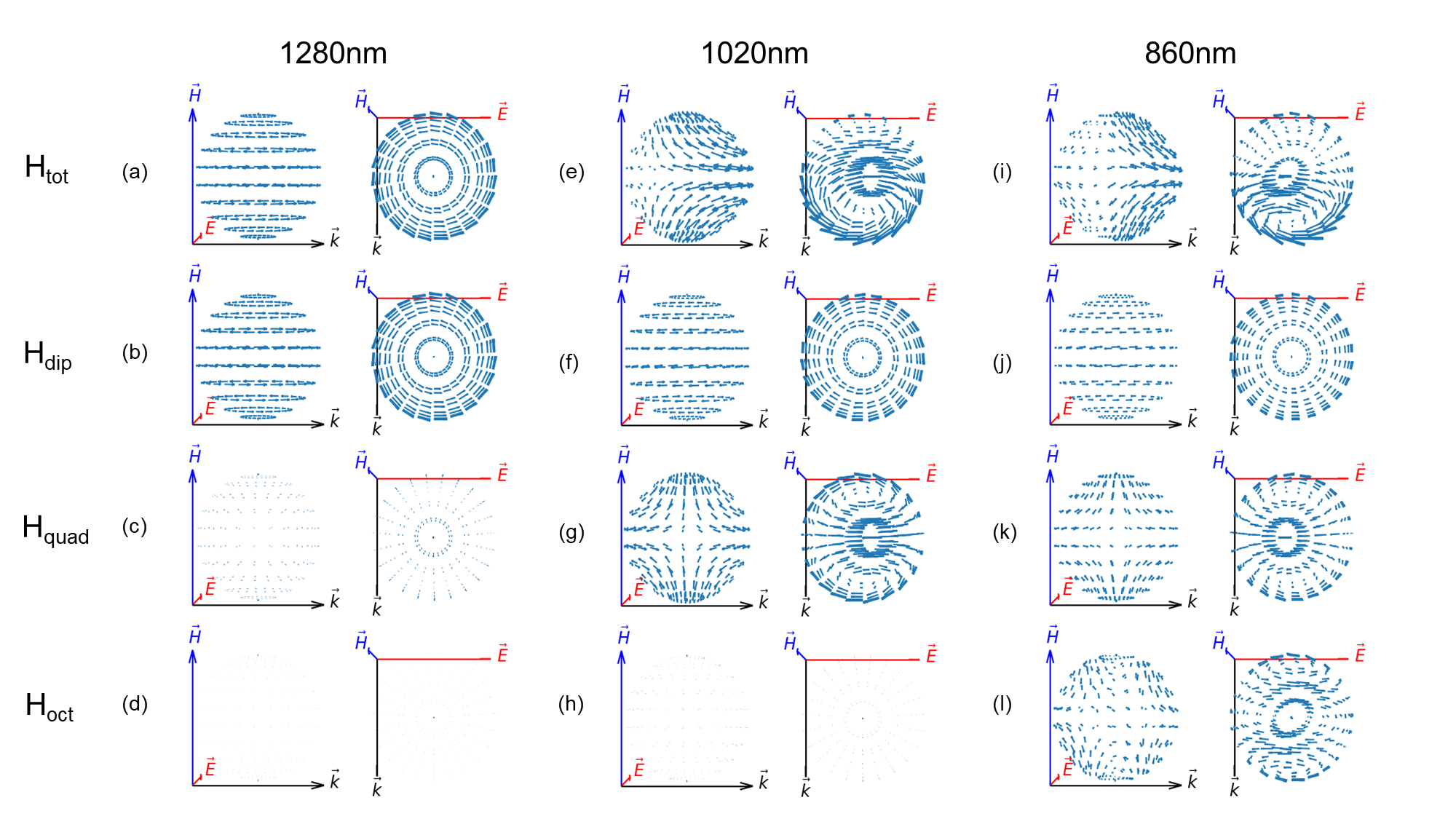}
\caption{Vector plots of far field displacement currents contributing to the total magnetic scattering modes ($H_{tot}$, top row) as well as the contribution the magnetic dipole ($H_{dip}$, row 2), quadrupole ($H_{quad}$, row 3), and octupole mode ($H_{oct}$, row 4), respectively, at the a-d) magnetic dipole resonance ($\lambda_{Hdip}=1280$~nm), e-h) magnetic quadrupole resonance ($\lambda_{Hquad}=1020$~nm), and i-l) magnetic octupole resonance ($\lambda_{Hoct}=860$~nm) wavelengths, for the structure shown in Figure \ref{Fig:Csct}.  For each condition, the figure on the left shows the plane of $\vec{H}$ and $\vec{k}$ (side view) and the figure on the right shows the plane of $\vec{E}$ and $\vec{k}$ (top view). The length of the arrows in all figures are plotted on the same scale and are proportional to the value of the displacement current.}
\label{fig:current}
\end{figure} 

\paragraph{Spatial Distribution of Magnetic Multipole Resonances.} Optical magnetism is an emergent phenomenon, formed due to rotating displacement currents that are in-phase with the incident light\cite{Nader1,Nader2,ChenJPCC}. These current loops are formed due to strong displacement fields localized in dielectric medium of the small interbead gaps (hot-spots). Due to the boundary conditions, the displacement currents in these gaps are forced to be normal to the surface of the nanobeads forcing the phase to slightly rotate.  The orientation and location of these current loops can help elucidate how high-order magnetic multipole resonances emerge in MMs. Using the full T-matrix, we can calculate the far-field displacement currents as well as the contribution from each mode (see SI section SVI for details on the calculation). Figure \ref{fig:current} shows vector plots of normalized displacement currents and their breakdown for each mode at the resonance wavelength of magnetic modes. 
As shown in Figure \ref{fig:current}a-d, at the magnetic dipole resonance ($\lambda_{Hdip}=1280$~nm), we observe the formation of a global loop of induced displacement currents circulating in-phase throughout the entire structure. This global displacement current loop, results in an induced magnetic dipole, which is in phase with the magnetic field of the incident light. This observation is consistent with previous predictions\cite{Parker,Nader1,Nader2} and experimental works \cite{ACSNano,ChenJPCC}. 
At this wavelength, the contribution from the magnetic quadrupole and octupole modes are negligible, as observed in nearly-zero values of displacement current contributions of $H_{quad}$ and $H_{oct}$ modes.

As the wavelength is decreased, currents become phase lagged across the structure, resulting in the emergence of a second current loop at the quadrupole resonance ($\lambda_{Hquad}=1020$~nm, Figures \ref{fig:current}e-h). The two distinct current loops are split across the two hemispheres, and rotate perfectly out of phase (top view of Figure \ref{fig:current}g). This manifests itself through the strengthening of the quadrupole mode, which is superimposed on a weakened, but non-zero dipole  mode (in-phase rotation seen in Figure \ref{fig:current}f). At the quadrupole resonance ($\lambda_{Hquad}=1020$~nm), the magnetic dipole is in phase with the magnetic quadrupole, resulting in enhanced forward-scattering and weakened back-scattering (visible in the side view of \ref{fig:current}e). The magnetic octupole mode remains week at these wavelengths. As the wavelength is further decreased, three loops of displacement currents emerge. The loops at the top and bottom of the structure are out of phase with the central loop, generating a magnetic octupole resonance at $\lambda_{Hoct}=860$~nm (Figures \ref{fig:current}i-f). These currents cannot be trivially described through a linear superposition of magnetic dipole modes and magnetic quadrupole modes. We also see as the wavelength is further decreased the emergence of hexadecapolar currents (see SI Figure S9). Overall, the larger the structure, the more we observe a phase lag between currents at various positions across the structure, which results in the vertical stacking of various magnetic dipole modes that are spatially distinct. The formation of three out of phase loops of current across the structure, requires well-separated displacement currents, which can only be sustained in large structures, with strong individual dipole strengths of nanobeads. Given that the dipole strength of spherical nanoparticles is proportional to their volume\cite{Nader2,ChenJPCC}, the large diameter of this model structure, as well as the large size of the MM ($Z=415$~nm) are key factors in the observation of the magnetic octupole resonance in this system.

\begin{figure}[H]
\includegraphics[width=\linewidth]{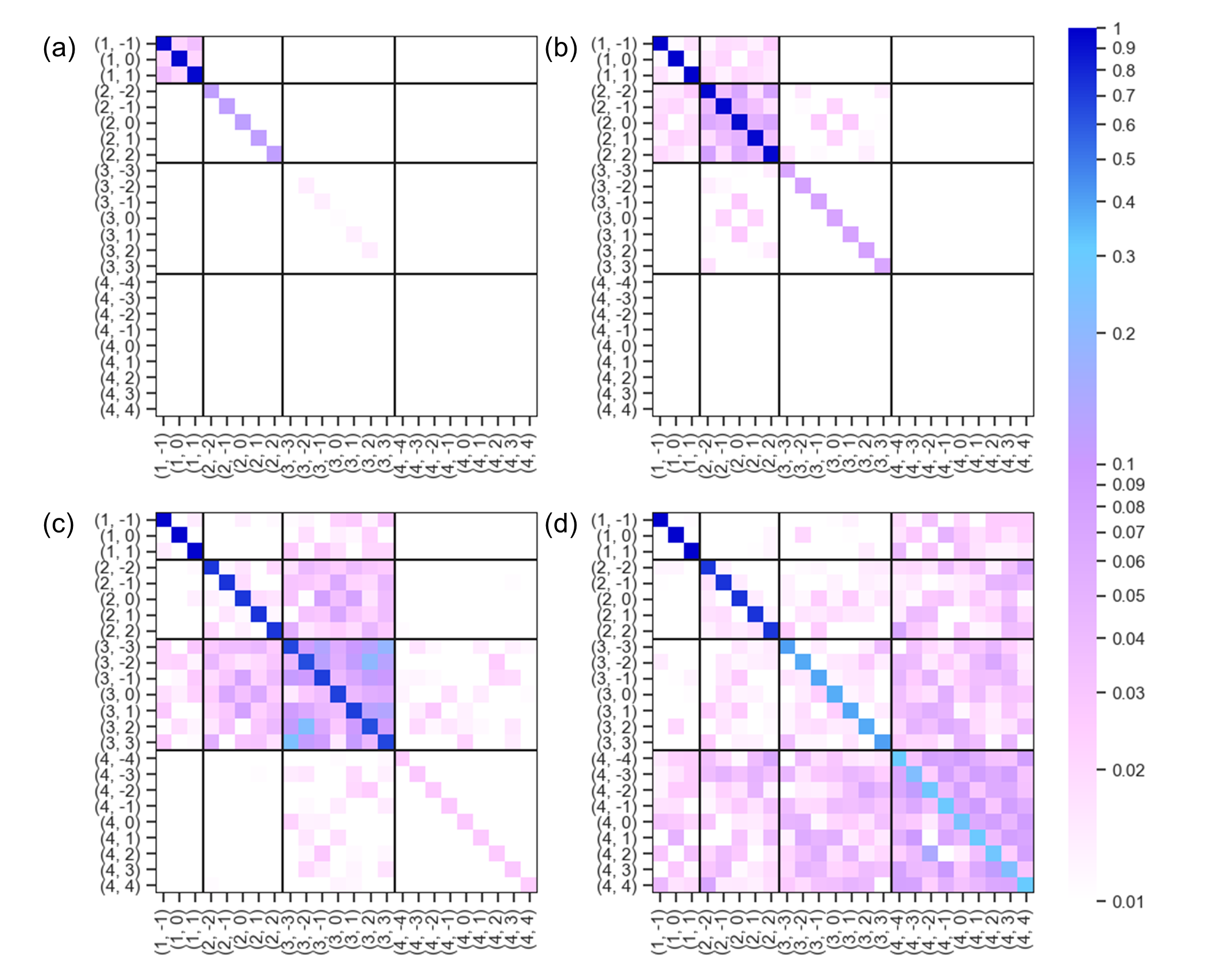}
\caption{The amplitude of the magnetic elements of the full T-matrix for the structure shown in Figure \ref{Fig:Csct} at the (a) magnetic dipole ($\lambda_{Hdip}=1280$~nm), (b) magnetic quadrupole ($\lambda_{Hquad}=1020$~nm), (c) magnetic octupole ($\lambda_{Hoct}=860$~nm), and (d) the magnetic hexadecapole  ($\lambda_{Hhex}=780$~nm) resonance wavelengths. The labels indicate $(n, m)$ where $n$ is the multipole mode order and $m$ is the submode ($-n \leq m \leq n$). All elements with magnitude $\leq 0.01$ are shown as white squares.}
\label{fig:modemixing}
\end{figure}

\paragraph{Mode Mixing in Magnetic Multipole Modes.} As discussed earlier (Figures S5 and S6),
the scattering cross-section of the diagonal T-matrix or a T-matrix that only allows intra-mode mixing (IMM, where modes with the same $n$ are allowed to mix) deviate significantly from the FDTD results at the resonance wavelengths of  high order magnetic modes ($H_{oct}$ and $H_{quad}$). Both the diagonal and IMM T-matrices underestimate the contributions of both modes to the peak intensity (see Figure S5).

To properly predict the high-order magnetic multipole resonances, at least a block-diagonal T-matrix is necessary, where all the magnetic (or electric) modes are allowed to intermix. To visualize the interference terms of the T-matrix that are critical for the accurate description of the structure, we visualize the amplitude of the magnetic elements of the full T-matrix elements at the magnetic dipole ($\lambda_{Hdip}=1280$~nm), magnetic quadrupole ($\lambda_{Hquad}=1020$~nm), magnetic octupole ($\lambda_{Hoct}=860$~nm), and magnetic hexadecapole ($\lambda_{Hhex}=780$~nm) resonance wavelengths, as shown  in Figure \ref{fig:modemixing}.
As seen in this figure, the strength of the mode mixing (off diagonal terms of the magnetic block of the T-matrix) relative to the main excited mode(diagonal terms) increases at wavelengths corresponding to higher order resonances (larger $n$). These terms are especially strong at the octupole resonance ($n=3$, $\lambda_{Hoct}=860$~nm) and hexadecapole ($n=4$,$\lambda_{Hhex}=780$~nm), with the off diagonal terms of the magnetic octupole and magnetic hexadecapole modes reaching 1/3 and 1/2 of the strength of their respective diagonal terms.
Both the increase in relative strength of off diagonal terms, and the relative number of terms, compared to the diagonal terms ensures that the mixing becomes more apparent as the mode order ($n$) grows. This can be seen as a manifestation of the fact that higher order modes are more localized and thus have more complex currents. As such, it is possible that the polarization currents are more susceptible to distortion by the variations in the local geometry. Interestingly, we see that higher order modes tend to also mix with lower order modes, as seen especially in the case of the octupole and hexadecapole modes.  The broad distribution of the dipole mode towards lower wavelengths as seen in Figure \ref{fig:Simulation} as well as the global nature of this mode, that distributes currents across the structure both contribute to this strong mode-mixing. This higher order inter-mode mixing is also significant in the directional scattering as seen from the high error of the IMM T-matrix compared to the full T-matrix error in Figure S5. 

Beyond a spectral effect on the intensity of the high-order multiple modes, the magnetic mode-mixing generates significant effects on the directional scattering patterns of MMs. We also note that the plane wave used to excite the structure, when decomposed into a sum of spherical waves, only contains a subset of nonzero amplitudes. For example, for the octupole mode ($n=2$) only the $m=-2$, $m=0$, and $m=2$ sub-modes have nonzero excitations. However, significant mode mixing produces scattering for all submodes, which can also distort the expected directional scattering patterns. These effects can be amplified for more selective polarizations as seen in single mode excitation simulations (Section SIV of SI).

\begin{figure}[H]
\includegraphics[width=\linewidth]{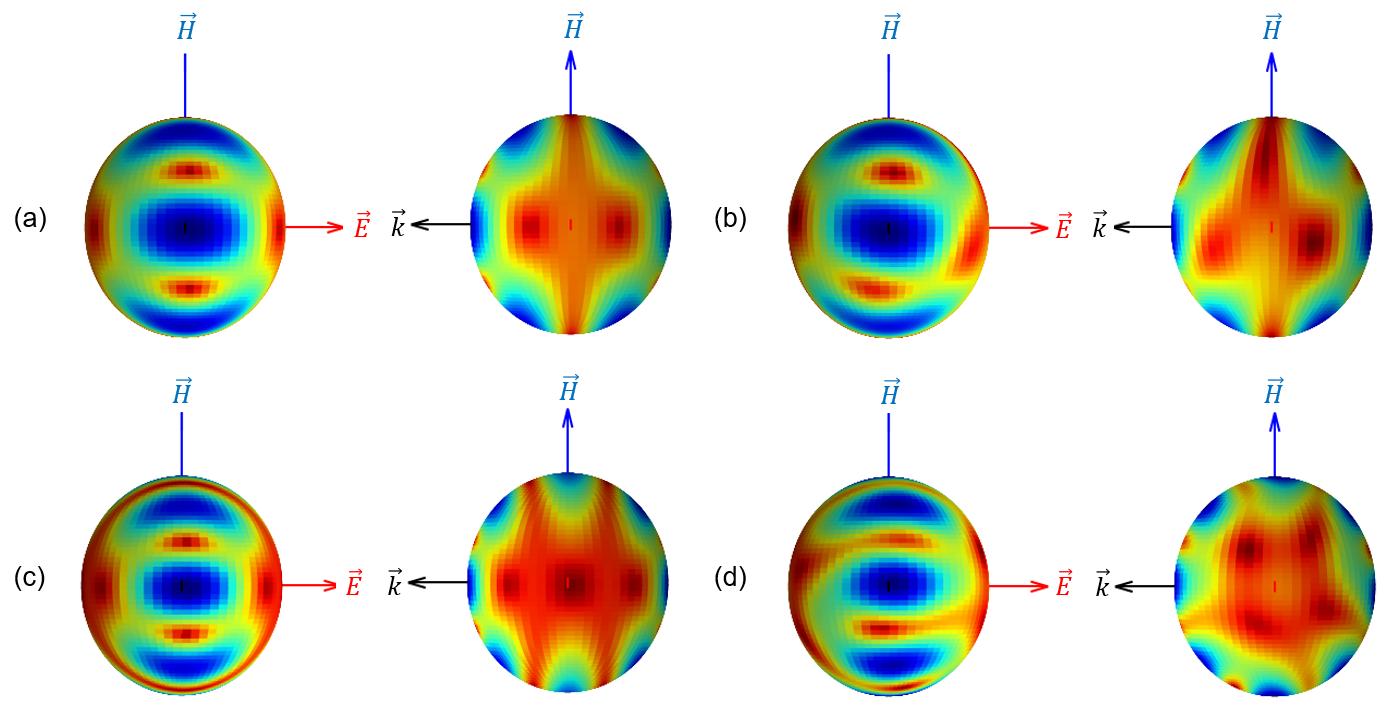}
\caption{The amplitude ($|\vec{X}|$) of the directional scattering of the magnetic octupole mode at ($\lambda_{Hoct}=860$~nm, octupole resonance) for (a) a pure magnetic octupole mode and (b) the magnetic octupole mode generated by a mixture of all magnetic modes. $|\vec{X}|$ of the directional scattering of the magnetic hexadecapole mode at $\lambda_{Hhex}=780$~nm, hexadecapole resonance) for (c) a pure magnetic hexadecapole mode and (d) the mixed magnetic hexadecapole mode generated by a mixture of all magnetic modes. The blue regions represent areas of high scattering amplitude while the red regions represent areas of low scattering amplitude. The MM structure is shown in Figure \ref{Fig:Csct}}.
\label{fig:spatialshape}
\end{figure}

\begin{figure}[H]
\includegraphics[width=1.2\linewidth,center]{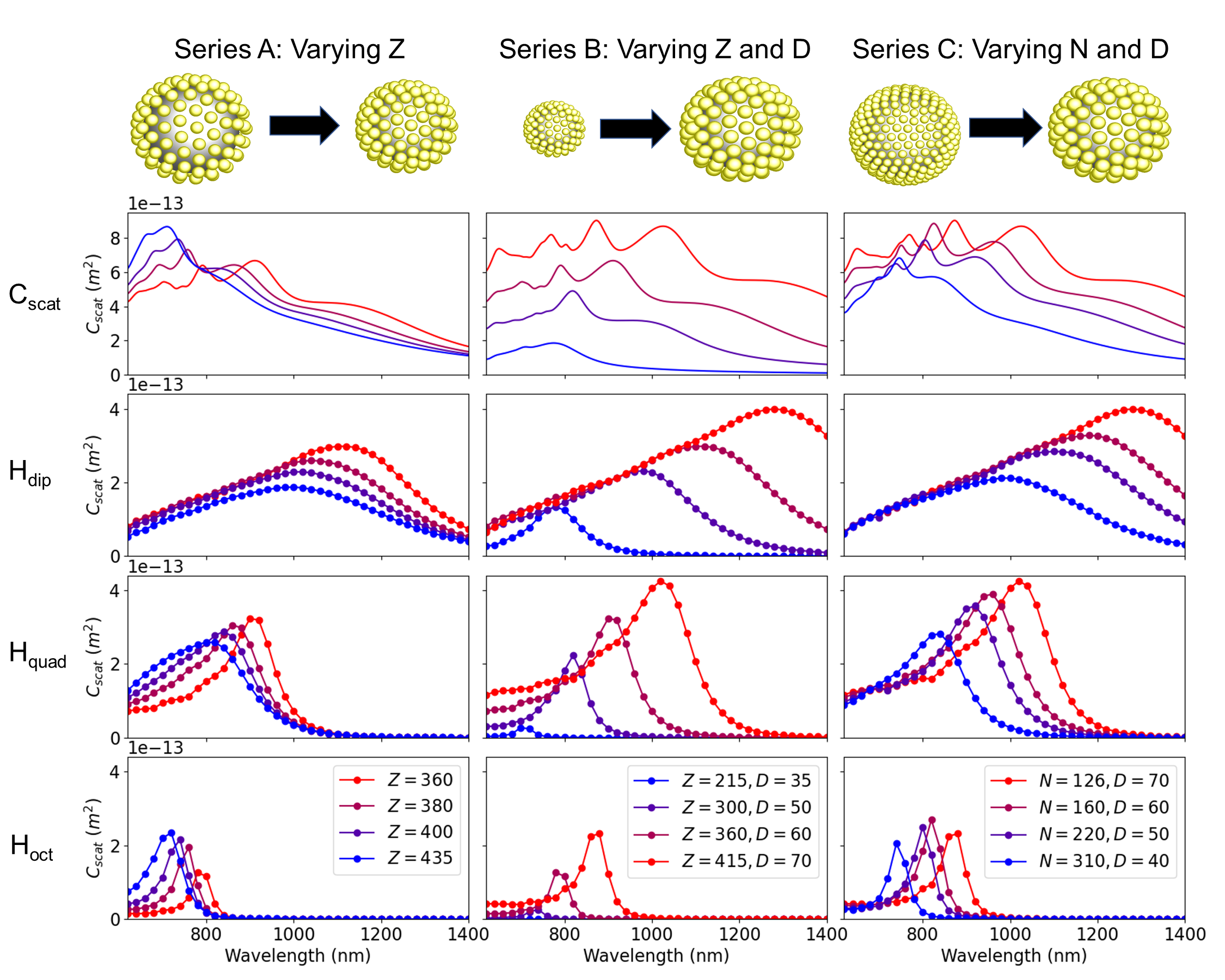}
\caption{Simulations exploring resonance trends for simulated MMs. From top to bottom, the simulated structure at the two extremes for each series, the total scattering cross-section (FDTD), the calculated magnetic dipole, magnetic quadrupole, and magnetic octupole scattering cross-sections are shown for series {\bf A} (left column), MMs with $N=126$, and $D=60$~nm, with varying $Z$, analogous to temperature change in experiments; series {\bf B} (middle column), MMs with $N=126$, varying both $Z$ and $D$, representing scale up of a constant structure; and series {\bf C} (right column), MMs with $Z=415$~nm, varying $N$ and $D$, representing constant nanobead density.The values of  variables $Z$ and $D$ are reported in units of nanometer. }
\label{fig:Simulation}
\end{figure}

We can visualize the spatial distortion effect of mode mixing on the scattering behavior and displacement currents of MMs by plotting the dirctional amplitude of the scattering of $H_{oct}$ and $H_{hex}$ modes at  their resonance wavelengths.
By zeroing out all rows of the T-matrix except for the mode of interest, we can calculate the scattering amplitude $|\vec{X}(\theta, \phi)|$ of an individual multipole modes in the far-field. Figures \ref{fig:spatialshape}a and b show the scattering generated by $H_{oct}$ at its resonance wavelength ($\lambda_{Hoct}=860$~nm).  Figure \ref{fig:spatialshape}a shows the magnitude of  $|\vec{X}|$ for a pure octupole mode (diagonal term) while  Figure \ref{fig:spatialshape}b shows the amplitude of $H_{oct}$ calculated based on the full T-matrix, which represents the magnetic octupole generated by a mixture of all magnetic modes (including weak hexadecapole contribution as seen in Figure \ref{fig:modemixing}).

In both Figures \ref{fig:spatialshape}a and b we observe the expected characteristic 6-lobe octupolar scattering pattern. However, in the mixed case (Figure \ref{fig:spatialshape}b) the spatial patterns of the octupole mode are clearly distorted. Thus, though for a plane wave excitation we only expect amplitude from the $m=-2$, $m=0$, and $m=2$ submodes, we can still visually observe active albeit weak currents from all other octupolar modes. These results are even more pronounced in the case of the magnetic hexadecapole mode, $n=4$, at its respective resonance wavelength of $\lambda_{Hhex}=780$~nm (Figure \ref{fig:spatialshape}c and d for diagonal and mixed modes, respectively). It is likely that the randomness of the self-assembled structure contributes to this distortion effect. We have previously shown that mode mixing can be significant in disordered, asymmetric optical systems, such as spiky nanoshells \cite{Simon}. We also note that while the pure modes are relatively symmetric and achiral, the distortion from the mixed modes breaks this symmetry causing the scattering pattern to gain a slight amount of chirality, especially in the magnetic hexadecapole mode, where mode mixing is also observed with the electric dipole modes (Figure S7). This can be potentially useful for various applications, where the electric dipole can act as an antenna to report the magnetic properties of these systems .

\paragraph{Dependence of Magnetic Multipole Resonances on Structural Variables.} Given the strong role of spatially-separated current loops in the emergence of high-order magnetic modes, it is important to investigate the role of structural parameters such as the core size $Z$, nanobead diameter $D$, and the interbead gap, $d_g$, on the strength and the energy of high-order magnetic resonances. Previous studies have shown that both $d_g$ and $D$ play a strong role in the magnetic dipole resonance of raspberry metamolecules\cite{ChenJPCC}, which are structurally similar to the experimental DMMs studied here. These dependencies can be then compared with our experimental results.

Figure \ref{fig:Simulation} shows three series of structural variations explored in this study. Series {\bf A} (left column in Figure \ref{fig:Simulation}) shows the modal break down when the structure's core size is decreased from $Z=435$~nm to $Z=360$~nm. See Table S2 and Figure S13 in SI for the full modal breakdown calculations. This is analogous to performing experiments on DMMs, where the temperature is increased above LCST to shrink the core size (Figure \ref{fig:experiment}). 
As the core size is reduced while the bead diameter is kept constant, the average inter-bead gap distance ($d_g$) decreases dramatically, increasing the coupling between the local dipole moments of the gold nanobeads. In agreement with previous studies \cite{ChenJPCC} as well as experiments shown in Figure \ref{fig:experiment}, the magnetic dipole resonance ($H_{dip}$, second row) red-shifts and strengthens as a result. The quality factor of this resonance also increases with decreasing $Z$. Data in Figure \ref{fig:Simulation} also shows a similar trend for the red-shifting magnetic quadrupole resonance ($H_{quad}$, third row). However, while the magnetic octupole resonance ($H_{quad}$, last row) red-shifts and becomes sharper, its strength is reduced, resulting in less prominent features in the far-field total scattering cross-section ($C_{scat}$, top row). This is because, as the $Z$ is reduced, it becomes more difficult to spatially separate loops of current across the structure, and the distinction between $H_{quad}$ and randomly coupled electric dipole modes become less prominent both spatially and spectrally. To design a structure with a prominent $H_{oct}$, one needs to build structures with a large overall size, along with large-size nanobeads. 

Indeed, an overall scale-up of the structure, as explored in series {\bf B} (middle column in Figure \ref{fig:Simulation}) can achieve this goal. In series {\bf B}, both $Z$ and $D$ are increased such that the overall structure remains relatively similar, with an approximately constant inter-bead gap distance ($d_g$).  See Table S2 and Figures S11 and S14 in SI for the full modal breakdown calculations and the distribution of $d_g$ for each structure. As both $Z$ and $D$ are increased, increasing the strength of the local electric dipole of the nanobeads, as well as providing a larger overall structure, all three magnetic resonances red shift and their strength increases. The quality factor is also significantly increased for the $H_{quad}$ and $H_{oct}$ resonances, making these peaks prominently observable in the far-field scattering spectrum. However, the magnetic dipole resonance $H_{dip}$ remains broad and its contribution to the total scattering cross-section becomes weaker relative to $H_{quad}$, despite their similar intensities in the modal scattering cross-sections. As explained in the previous sections, one possible explanation is the fact that as $H_{dip}$ shits to longer wavelengths, the broad background scattering from the electric multiple modes decays to zero. As shown in Figure \ref{Fig:Csct}, the background scattering from the electric multipole modes may be a factor in increasing the apparent strength of the high order ($n\geq2$) resonances in the far-field total scattering cross-section.
The contributions from the electric multipole modes of these series are shown in SI Figure S14. 

As seen in this figure (last column), even in the sum total of the magnetic modes alone, show stronger cross-sections at $H_{quad}$ resonance compared to the $H_{dip}$ resonance. This is because the very broad shoulder of $H_{dip}$ resonance towards the lower wavelength region itself also contributes to the background intensity at $H_{quad}$ resonance. Not surprisingly, this observation is also consistent with the strong forward scattering seen at $\lambda_{Hquad}=1020$~nm in Figure \ref{fig:current}, where the magnetic dipole moment is still strong, despite being off-resonance.  

The significant broadening of the $H_{dip}$ resonance is due to the large overall size of the structure. As the size of the structure is increased, there is a larger heterogeneity in the diameter of the loops of currents that are resonating in phase from the top of the structure to the center (look for example at the loops shown in the side views of Figure \ref{fig:current}a and b). This results in the broadening of the magnetic dipole resonance as the overall structure increases, increasing the energy required for the displacement current loops to remain in-phase, explaining the broad shoulder of the peak towards lower wavelengths (higher energy). A similar broadening towards lower wavelengths is also observed for $H_{quad}$  in the largest structures in both series {\bf B} and {\bf C}. 

Given the strong role of spatial separation in the emergent $H_{quad}$ and $H_{oct}$ resonances, it is difficult to immediately observe the role of the bead size $D$ in these resonances, as has been previously explored for the magnetic dipole resonance\cite{ChenJPCC}. To this end, we have also simulated a series of MMs, shown in the right column of Figure \ref{fig:Simulation}, where we maintain a constant $Z$ while increasing $D$ from $D=40$~nm to $D=70$~nm. To keep the inter-bead gap distance ($d_g$) relatively constant, such that the coupling strength between the nanobeads can also be kept invariant, we increased $N$. See Table S2 and Figures S12 and S15 of SI 
for full modal breakdown and structural variables for this series. As $D$ is increased, all three magnetic resonances red shift. However, while the trends observed in $H_{dip}$ and $H_{quad}$ are similar to the observations in series {\bf B}, $H_{oct}$ shits more modestly and its strength remains relatively constant across the structures. 

Combined, the data in Figure \ref{fig:Simulation} show that the bead diameter $D$ plays a strong role in the location of the magnetic multipole resonances, in particular for the magnetic dipole and magnetic quadrupole resonances, resulting in strong red-shifting when $D$ is increased. The overall size of the structure $Z$ contributes to the broadening of $H_{dip}$ and $H_{quad}$, resulting in their eventual splitting into multiple out of phase currents, and the emergence of $H_{oct}$ (and potentially higher order resonances). However, given its high-energy and overlap with the electric dipole modes, it is more difficult to tune and strengthen the magnetic octupole mode. While we expect these modes to be present in most structures, they are only prominently visible when both $D$ and $Z$ are significantly large. We thus predict that it is easier to experimentally observe magnetic dipole and magnetic quadrupole resonances in smaller structures, when strong inter-bead coupling exists, as they have been previously reported in DMMs\cite{ACSNano,JPCC} and are also prominently observed in our experimental data (Figure \ref{fig:experiment}.

A simplified explanation can also be provided through the Mie theory. In Mie theory, the magnetic component of the incident electric field for a given mode and wavelength peaks at a certain distance from the center of the structure  (More details in SI section SIX and Figure S16). 
All resonances becomes narrower and blue shift as the mode order increases. The resonance for all modes also broaden and red shift as we move further from the center of the structure. When the location of the nanobeads fall within this resonant zone for a certain mode, the dipoles within the beads can rotate the phase of light sufficiently for the displacement currents to resonate with the incident magnetic field, leading to strong magnetic moments and scattering. This explains why higher order modes have sharper resonances, and require larger overall structures or core sizes with high refractive indices\cite{Parker} to be excited (see Section SIX of SI for more details). 

This simple scaling argument can for example explain why peaks tend to broaden and red shift when nanobead are located farther from the center (series {\bf B}). However, it is important to note that this argument does not take into account the role of inter-bead coupling, and thus the inter-bead gap distance, $d_g$. In Series {\bf A}, we see an opposite trend than predicted by this argument, as when the core (and therefore the structure) size is increased, all modes blue shift and weaken. This latter effect is mostly due to decreasing $d_g$, which generates strong inter-bead coupling when $D$ (corresponding to the dipole moment of each bead) is constant.

\begin{figure}[H]
\includegraphics[width=\linewidth]{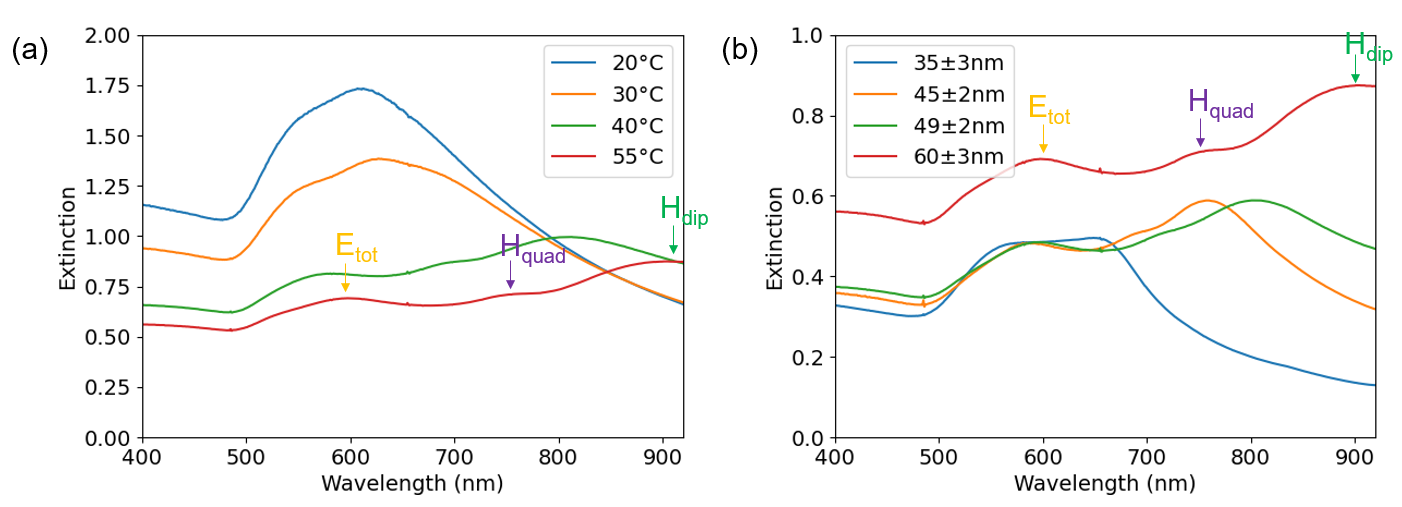}
\caption{a) Experimental extinction spectra of DMMs with gold nanobeads of average size $D=60.0\pm3$~nm at various temperatures. b) Experimental extinction spectra of DMMs with various average nanobead sizes measured at 55°C, where the core is compact. We identify the resonances based on comparison with simulation data shown in Figure \ref{fig:Simulation}. The predicted magnetic dipole resonance is marked with $\mathrm{H_{dip}}$, the predicted magnetic quadrupole resonance is marked with $\mathrm{H_{quad}}$, and the predicted electric resonance is marked with $\mathrm{E_{tot}}$.}
\label{fig:exp2}
\end{figure} 

Our simulated results are consistent with the trends observed in our experimental data for large DMMs (Figure \ref{fig:experiment}). For example, for clarity, Figure \ref{fig:exp2}a shows a subset of data in Figure \ref{fig:experiment}d, for the largest experimental structures with $N=41\pm4$ gold nanobeads with diameter $D=60\pm3$~nm. At $T\geq$40~°C, a prominent middle peak emerges between the red shifting magnetic dipole
resonance ($H_{dip}$ located at $\lambda>800$~nm) and the electric dipole resonance ($E_{tot}$, wit resonance at $\lambda_{Etot}\sim550-600$~nm). The hydrogel core experiences a sharp transition from swollen state into collapsed state LCST ($T=33$~°C), and the core sizes $Z$ decrease sharply above this temperature (Figure S2 of SI)\cite{lim2014gold,JPCC} 
Given the dramatic decrease in $Z$ above LSCT, the effective interparticle gap distance ($d_g$) also decreases dramatically. For example, for $N=41$ number of $D=60$~nm gold nanobeads, $d_g$ is estimated to be $d_g\sim110$~nm at 25~°C, which provides sufficient distance between beads to prevent strong coupling. At 50~°C however, the gold nanobeads are closely associated with a gap distance distribution of 2~nm$<d_g<$16~nm, which provides strong coupling between the nanobeads. 

As such, the trend at these temperature ranges is similar to the Series {\bf A} simulations (left column of Figure \ref{fig:Simulation}). We assign this peak as the magnetic quadrupole mode ($H_quad$). When looking at all collaped structures at $T=55$°C (Figure \ref{fig:exp2}), analogous to Series {\bf B} (middle column in Figure \ref{fig:Simulation}), where the structure is scaled with relatively similar inter-bead gap distance $d_g$, we also observe significant red-shifting in the assigned $H_{dip}$ and $H_{quad}$ as $D$ is increased. However, in experiments, $H_{quad}$ remains weaker than $H_{dip}$, presumably due to the reduced quality factor due to the heterogeneity of the structures both at the single particle level and at ensemble level in colloidal experiments. This needs to be further explored using single particle measurements detailed in the next section. Similarly, it is not immediately apparent that the magnetic quadrupole resonances exist in our experimental data. Compared to the simulations, as the experiments have fewer number of nanobeads, smaller core sizes, and increased heterogeneity. All of these factors can contribute to the blue shifting and reduced quality factor of the quadrupole resonance.   

\paragraph{Detecting Magnetic Multipole Resonances through Angular Scattering.} A major challenge with identifying magnetic multipole resonances in experiments is the broad nature of the observed resonance features. Compared to the modeled MMs, the experimental DMMs tend to show significantly broadened peak features, likely due to heterogeneity of DMMs in ensemble solution, as well as the heterogeneity of the nanobead size and number on individual DMMS. As such, it becomes quite difficult to assign a resonance feature to a specific magnetic multipole mode, both in the data shown in Figure \ref{fig:exp2} as well as in our previous experimental studies \cite{lim2014gold,JPCC}. In particular, while it is possible to identify magnetic quadrupole resonances in large structures with small interparticle gaps, as shown here, the spectral overlap between higher order ($n\geq3$) and the electric dipole modes, the identification becomes subjective, if one only uses total scattering or extinction cross-sections. 

However, given the differences observed in the directional scattering shown in Figures \ref{fig:current} and \ref{fig:spatialshape}, it may be significantly easier to assign these resonances using directional scattering experiments. Each magnetic multipole mode has a characteristic angular scattering pattern (Figure \ref{fig:spatialshape}), which can aid in the mode's identification, through comparisons between forward and backward scattering, directional scattering experiments, as well as experiments using polarized light. Similar strategies have been previously used to reliably identify magnetic dipole resonances\cite{MagneticMetafluid,ACSNano,schmidt2012dielectric}, and have been demonstrated for magnetic quadrupole resonance detection\cite{Parker,ParkerBackscattering}. Nonlinear optical experiments can also elucidate the existence of high-order magnetic multipolar resonances\cite{kruk2017nonlinear} due to their polarization selectivity rules.

\begin{figure}[H]
\includegraphics[width=0.6\linewidth]{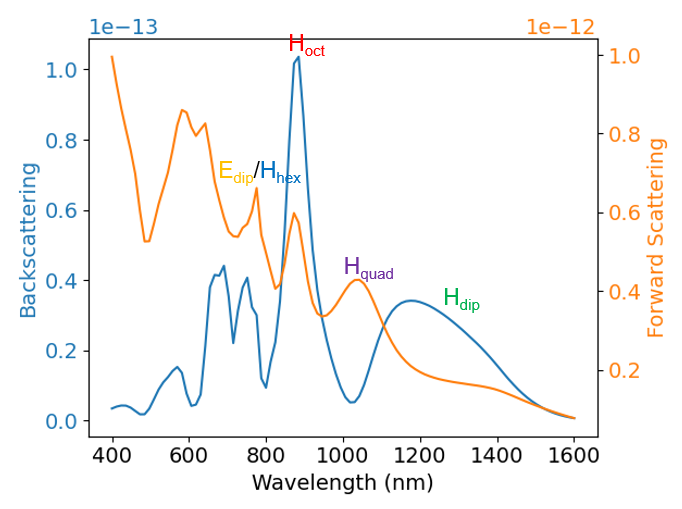}
\caption{forward (orange) and backward (blue) scattering cross-sections for the model MM with $N=126$, $Z=415$~nm, $D=70$~nm (structure shown in Figure \ref{Fig:Csct}). The relevant resonances are labeled above the peaks. Note that the electric dipole and magnetic hexadecapole resonances overlap.}
\label{forwardAndBack}
\end{figure}

Figure \ref{forwardAndBack} shows a comparison between the forward and backward scattering cross-sections for our model MM structure (Figure \ref{Fig:Csct}) simply obtained from the FDTD calculations. 
We can see in this figure that even a simple comparison between cross-sections for these two directions can provide large contrast to identify the $H_{quad}$ resonance and possibly $H_{hex}$, both of which have negligible backscattering cross-sections, but show reasonably strong forward scattering cross-sections. Interestingly, in the forward direction, the higher order modes show progressively stronger cross-sections, which as discussed earlier can be due to the overlap with the broad electric dipole cross-section (which has uniform intensity in all $\phi$ directions. 
In contrast, in the backscattering direction, a constructive/destructive interference occurs for modes with modes with odd/even numbers, respectively. Figure S18 shows that a similar, but weaker effect can occure in MMs with smaller bead sizes $D$ (series {\bf C}) in Figure \ref{fig:Simulation}.

To better understand the origin of this behavior, we can focus on the behavior of the components of the full T-matrix in a given direction. We will proceed here to focus on the backscattering direction, given the apparent destructive interferences for the odd modes in this direction. However, we note that this generalized approach can be readily applied to any direction of interest once the T-matrix is calculated. In addition, the exact experimental conditions can also be simulated by considering the cone of angles and polarization conditions for the incident and scattered light probed in a certain apparatus, as well as the addition of any types of  substrates or change in the index of refraction of the ambient medium (for example from water here to air). Such in-depth analyses are beyond the scope of the current study and will be explored in the future. 

\begin{figure}[H]
\includegraphics[width=\linewidth]{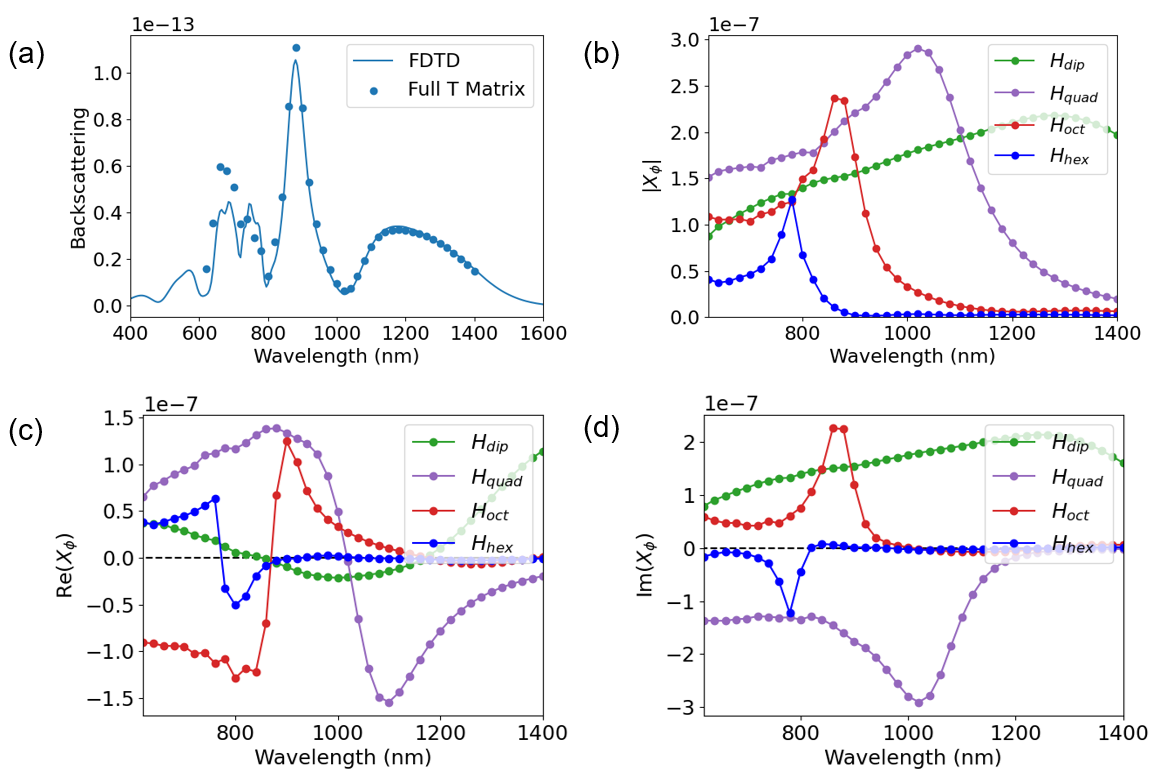}
\caption{a) Differential backscattering cross-section obtained from FDTD (solid lines) and T-matrix (blue symbols) calculations for the structure shown in Figure \ref{Fig:Csct}. b) The magnetic components of the scattering amplitude ($|{X}_{\phi}|$) in the backscattering direction, obtained from T-matrix. c) The corresponding Re(${X}_{\phi}$) and d) Im(${X}_{\phi}$) values of $|{X}_{\phi}|$ for the magnetic multipole modes in the backscattering direction.  For the scattering amplitude of the electric modes as well as the phase of both electric and magnetic modes see Figure S17.}
\label{backscattering}
\end{figure}

Figure \ref{backscattering} shows the differential backscattering cross-section of the default MM structure (Structure shown in Figure \ref{Fig:Csct}) calculated using the full T-matrix (Figure \ref{backscattering}a) as compared to the FDTD calculations in Figure \ref{forwardAndBack}, as well as the T-matrix values of the amplitude ($|X_{\phi}|$, Figure \ref{backscattering}b ), its real ($Re X_{\phi}$, Figure \ref{backscattering}c ) and the imaginary ($Im X_{\phi}$, Figure \ref{backscattering}d) components in the backscattering direction, corresponding to each magnetic multipole mode. Note that $|\vec{X}(\theta, \phi)|$ is defined in  Equation \ref{eqn:amplitude} as $\vec{E}_s(r, \theta, \phi) = \frac{e^{ikr}}{r} \vec{X}(\theta, \phi)$ and the scattering cross-section is given by $dC_{scat}/d\Omega = |\vec{X}(\theta, \phi)|^2$. As such,  $\vec{X}$ has both a $\theta$ and $\phi$ ($\theta \in [0, \pi]$, $\phi \in [0, 2 \pi])$ components, but here we refer only to the $\phi$ component as the $\theta$ component is small in the backscattering direction, thus a different notation is used. The data in Figures \ref{backscattering}b-d show the corresponding magnetic components of the T-matrix in this specific direction, while the corresponding electric components, as well as the phases for all components are shown in Figure S17. 
These patterns allow us to visualize the modal contributions to the differential backscattering cross-section as well as the role of the constructive and destructive interferences in the observed signal attenuations.

As we can see in Figure \ref{backscattering}a, the full T-matrix successfully predicts the major features of the differential backscattering cross-section for  $\lambda\geq700$~nm,  including the sharp feature observed at $\lambda=800$~nm 
and $\lambda=1110$~nm, and the peaks at $\lambda=880$ nm and $\lambda=1200$ nm.
The modal decomposition of the backscattering cross-section is quite rich in the information content that can explain these rather sharp features in this direction. As seen in Figure \ref{backscattering}b the magnetic dipole is spectrally broad, and extends well into the high-frequency region. This is consistent but slightly less with the observations of the total scattering cross-section shown in Figures \ref{Fig:Csct} and \ref{fig:Simulation}. However, the spectral decay is even weaker in this direction. As explained above, this is because of the large heterogeneity in the size of the in-phase current loops across the structure. In contrast, we see that the higher order magnetic multipole modes ($n\geq2$) are sharper (Figure \ref{backscattering}b) which give resonances with higher quality factors with a shape similar to that of a Lorentz oscillator\cite{Bohren}.  This spectral overlap between the broad magnetic dipole mode and the sharp multipole mode resonances with $n\geq2$, generate Fano-like interference patterns in the backscattering cross-section similar to interferences observed between the magnetic and electric dipole resonances
\cite{QuestOptMagnetismReview,GoldArrayFano,cherqui2016stem,verre2015optical,mirin2009fano}. However, the origin of the resonance here is mostly due to interferences between the magnetic multipole modes themselves, as is also highlighted the fano-like shape that occurs at the boundary between the magnetic octupole resonance and the magnetic quadrupole resonance (at around $800$~nm-$1100$~nm). This fano-like effect is more apparent when one investigates Re(${X}_{\phi}$) and Im(${X}_{\phi}$) shown in Figures \ref{backscattering}c and d, respectively. There is a sharp change in the sign of Re(${X}_{\phi}$) at each magnetic multipole resonance with $n\geq2$, which corresponds to the large slopes observed in the backscattering cross section in  Figures \ref{backscattering}a. As such, the minima observed at $\lambda=800$~nm and $\lambda=1100$~nm correspond to the minimum value of Re(${X}_{\phi}$) as opposed to the $H_{hex}$ and $H_{quad}$ resonances, while the inflection point is the actual resonance frequency. For the $H_{oct}$ resonance, there Im(${X}_{\phi}$) is positive, implying a constructive interference, as opposed to a destructive one, causing a sharp apparent maximum in the backscattering at the maximum value of Re(${X}_{\phi}$), which occurs at 
$\lambda=900$~nm, again red-shifted compared to the $\lambda_{Hoct}$. The inflection point in the backscattering cross-section is the actual resonance. 

We note that $H_{hex}$ is rather unique as it directly overlaps with the electric dipole resonance. As discussed earlier, in the T-matrix, the $H_{hex}$ resonance spectrally overlaps with the $E_{dip}$ resonance which also shows fano-like effects (Figure S17c) and the two modes have strong off-diagonal mixing terms (Figure S7), making the spectral region with $\lambda\leq800$~nm more congested and likely impossible to distinguish in the backscattering directions. For this mode to be identified, scattering in other directions relevant to this mode are necessary, which is beyond the focus of this study. 

We note that the phase inversion that can extenuate the $H_{oct}$ and suppress  $H_{hex}$ and  $H_{quad}$ in the backscattering directions is not unique to this model structure. As we can see in Figure S18 as an example for series {\bf C}, where we reduce the bead size ($D$), while keeping the interparticle gap $d_g$ constant by increasing $N$, which the strength of $H_{oct}$ and $H_{quad}$ resonances decrease significantly, the Fano-like patterns should still be observable in comparing the forward and backward scattering cross-sections. The important factor here is that the structure needs to be large enough to support magnetic octupole resonances, as discussed earlier. Thus we can take advantage of modal interferences in the backscattering to observe far field magnetic octupoles for the first time. More extended directional scattering calculations/measurements using selective polarization directions can indeed be more conclusive, but is beyond the scope of this work.

\section{Summary and Conclusions}

In this study we have used a T-matrix approach to characterize high-order magnetic multipole modes and their resonances in the visible and near-IR region of the spectrum for a model magnetic metamolecule with large size nanobeads, randomly packed on a dielectric shell. We observed that as the interparticle gap discance is decreased, for systems with large-size nanobeads and large overall size, multipolar magnetic resonances up to  magnetic hexadecapoles can be identified with a combination of directional and total scattering cross-section measurements.  We show that there is significant mode-mixing in the magnetic block of the T-matrix, particularly for large structures, resulting in unique observable features in the far-field. We explored the experimental parameters and conditions that can sustain high-order magnetic multipole modes, and provide strong evidence that large constructs, with tight interparticle gaps are critical. Some of the results are compared with experiments on dynamic metamolecules in colloidal suspensions, that can reduce their gap by shrinking their hydrogel core size upon heating. While the experimental systems have significant size heterogeneity, they are still able to show magnetic quadrupole resonances in the far-field in their compact form.

By exploring the size parameters important for high-order magnetic multipole resonances, we also provide a recipe for future experiments that can more conclusively identify these resonances in the optical domain. To do so, one can focus on directional scattering experiments as well as polarization specific ones (not explored here). These high-order multipolar resonances can have chrial effects which are of interest for various experiments and applications. The T-matrix approach provides a wealth of data on the directional scattering features that can be further explored in the future to better experimentally identify these resonances.

\begin{acknowledgement}
S.-J.P. acknowledges the financial support from the National Research Foundation (NRF) of Korea ( NRF-2018R1A2B3001049) and the Science Research Center (SRC) funded by NRF (NRF-2017R1A5A1015365). Z.F., C.N.W., and O.I. acknowledge funding from the Penn Laboratory for Research in Structure of Matter (LRSM) funded by NSF MRSEC grant (DMR-1720530), the Department of Chemistry (summer fellowship for O.I.), and the School of Arts and Sciences Dean’s Global Inquiries Fund at the University of Pennsylvania (Funded visit to South Korea for C.N.W.). C.N.W. was supported by a fellowship from the REACT for the Human Habitats Program at Penn (through NSF PIRE Grant \#OISE-1545884). E.C.C. acknowledges financial support from National Research Foundation (NRF) of Korea (NRF-2019R1A2C1004306) 
 
\end{acknowledgement}

\bibliography{references}

\providecommand{\latin}[1]{#1}
\makeatletter
\providecommand{\doi}
  {\begingroup\let\do\@makeother\dospecials
  \catcode`\{=1 \catcode`\}=2 \doi@aux}
\providecommand{\doi@aux}[1]{\endgroup\texttt{#1}}
\makeatother
\providecommand*\mcitethebibliography{\thebibliography}
\csname @ifundefined\endcsname{endmcitethebibliography}
  {\let\endmcitethebibliography\endthebibliography}{}
\begin{mcitethebibliography}{70}
\providecommand*\natexlab[1]{#1}
\providecommand*\mciteSetBstSublistMode[1]{}
\providecommand*\mciteSetBstMaxWidthForm[2]{}
\providecommand*\mciteBstWouldAddEndPuncttrue
  {\def\EndOfBibitem{\unskip.}}
\providecommand*\mciteBstWouldAddEndPunctfalse
  {\let\EndOfBibitem\relax}
\providecommand*\mciteSetBstMidEndSepPunct[3]{}
\providecommand*\mciteSetBstSublistLabelBeginEnd[3]{}
\providecommand*\EndOfBibitem{}
\mciteSetBstSublistMode{f}
\mciteSetBstMaxWidthForm{subitem}{(\alph{mcitesubitemcount})}
\mciteSetBstSublistLabelBeginEnd
  {\mcitemaxwidthsubitemform\space}
  {\relax}
  {\relax}

\bibitem[Al\`u and Engheta(2008)Al\`u, and Engheta]{Nader2}
Al\`u,~A.; Engheta,~N. Dynamical theory of artificial optical magnetism
  produced by rings of plasmonic nanoparticles. \emph{Phys. Rev. B}
  \textbf{2008}, \emph{78}, 085112\relax
\mciteBstWouldAddEndPuncttrue
\mciteSetBstMidEndSepPunct{\mcitedefaultmidpunct}
{\mcitedefaultendpunct}{\mcitedefaultseppunct}\relax
\EndOfBibitem
\bibitem[Monticone and Alù(2014)Monticone, and Alù]{QuestOptMagnetismReview}
Monticone,~F.; Alù,~A. The quest for optical magnetism: from split-ring
  resonators to plasmonic nanoparticles and nanoclusters. \emph{J. Mater. Chem.
  C} \textbf{2014}, \emph{2}, 9059--9072\relax
\mciteBstWouldAddEndPuncttrue
\mciteSetBstMidEndSepPunct{\mcitedefaultmidpunct}
{\mcitedefaultendpunct}{\mcitedefaultseppunct}\relax
\EndOfBibitem
\bibitem[Parker \latin{et~al.}(2018)Parker, Gray, and Scherer]{Parker}
Parker,~J.; Gray,~S.; Scherer,~N.~F. {Optical magnetism in core-satellite
  nanostructures excited by vector beams}. Photonic and Phononic Properties of
  Engineered Nanostructures VIII. 2018; pp 60 -- 68\relax
\mciteBstWouldAddEndPuncttrue
\mciteSetBstMidEndSepPunct{\mcitedefaultmidpunct}
{\mcitedefaultendpunct}{\mcitedefaultseppunct}\relax
\EndOfBibitem
\bibitem[Meng \latin{et~al.}(2020)Meng, Zhang, Lei, Li, Li, Liu, Xie, and
  Leung]{meng2020plasmon}
Meng,~Y.; Zhang,~Q.; Lei,~D.; Li,~Y.; Li,~S.; Liu,~Z.; Xie,~W.; Leung,~C.~W.
  Plasmon-Induced Optical Magnetism in an Ultrathin Metal Nanosphere-Based
  Dimer-on-Film Nanocavity. \emph{Laser \& Photonics Reviews} \textbf{2020},
  \emph{14}, 2000068\relax
\mciteBstWouldAddEndPuncttrue
\mciteSetBstMidEndSepPunct{\mcitedefaultmidpunct}
{\mcitedefaultendpunct}{\mcitedefaultseppunct}\relax
\EndOfBibitem
\bibitem[Liu \latin{et~al.}(2012)Liu, Mukherjee, Bao, Brown, Dorfm\"{u}ller,
  Nordlander, and Halas]{liu2012magnetic}
Liu,~N.; Mukherjee,~S.; Bao,~K.; Brown,~L.~V.; Dorfm\"{u}ller,~J.;
  Nordlander,~P.; Halas,~N.~J. Magnetic plasmon formation and propagation in
  artificial aromatic molecules. \emph{Nano letters} \textbf{2012}, \emph{12},
  364--369\relax
\mciteBstWouldAddEndPuncttrue
\mciteSetBstMidEndSepPunct{\mcitedefaultmidpunct}
{\mcitedefaultendpunct}{\mcitedefaultseppunct}\relax
\EndOfBibitem
\bibitem[Stevenson \latin{et~al.}(2020)Stevenson, Du, Cherqui, Bourgeois,
  Rodriguez, Neff, Abreu, Meiler, Tamma, Apkarian, \latin{et~al.}
  others]{stevenson2020active}
Stevenson,~P.~R.; Du,~M.; Cherqui,~C.; Bourgeois,~M.~R.; Rodriguez,~K.;
  Neff,~J.~R.; Abreu,~E.; Meiler,~I.~M.; Tamma,~V.~A.; Apkarian,~V.~A.,
  \latin{et~al.}  Active Plasmonics and Active Chiral Plasmonics through
  Orientation-Dependent Multipolar Interactions. \emph{ACS nano} \textbf{2020},
  \emph{14}, 11518--11532\relax
\mciteBstWouldAddEndPuncttrue
\mciteSetBstMidEndSepPunct{\mcitedefaultmidpunct}
{\mcitedefaultendpunct}{\mcitedefaultseppunct}\relax
\EndOfBibitem
\bibitem[Cherqui \latin{et~al.}(2016)Cherqui, Wu, Li, Quillin, Busche, Thakkar,
  West, Montoni, Rack, Camden, \latin{et~al.} others]{cherqui2016stem}
Cherqui,~C.; Wu,~Y.; Li,~G.; Quillin,~S.~C.; Busche,~J.~A.; Thakkar,~N.;
  West,~C.~A.; Montoni,~N.~P.; Rack,~P.~D.; Camden,~J.~P., \latin{et~al.}
  STEM/EELS imaging of magnetic hybridization in symmetric and symmetry-broken
  plasmon oligomer dimers and all-magnetic Fano interference. \emph{Nano
  letters} \textbf{2016}, \emph{16}, 6668--6676\relax
\mciteBstWouldAddEndPuncttrue
\mciteSetBstMidEndSepPunct{\mcitedefaultmidpunct}
{\mcitedefaultendpunct}{\mcitedefaultseppunct}\relax
\EndOfBibitem
\bibitem[\"{O}g\"{u}t \latin{et~al.}(2012)\"{O}g\"{u}t, Talebi, Vogelgesang,
  Sigle, and Van~Aken]{Ogut2012toroidal}
\"{O}g\"{u}t,~B.; Talebi,~N.; Vogelgesang,~R.; Sigle,~W.; Van~Aken,~P.~A.
  Toroidal plasmonic eigenmodes in oligomer nanocavities for the visible.
  \emph{Nano letters} \textbf{2012}, \emph{12}, 5239--5244\relax
\mciteBstWouldAddEndPuncttrue
\mciteSetBstMidEndSepPunct{\mcitedefaultmidpunct}
{\mcitedefaultendpunct}{\mcitedefaultseppunct}\relax
\EndOfBibitem
\bibitem[Pakizeh \latin{et~al.}(2008)Pakizeh, Dmitriev, Abrishamian, Granpayeh,
  and K{\"a}ll]{pakizeh2008structural}
Pakizeh,~T.; Dmitriev,~A.; Abrishamian,~M.; Granpayeh,~N.; K{\"a}ll,~M.
  Structural asymmetry and induced optical magnetism in plasmonic
  nanosandwiches. \emph{JOSA B} \textbf{2008}, \emph{25}, 659--667\relax
\mciteBstWouldAddEndPuncttrue
\mciteSetBstMidEndSepPunct{\mcitedefaultmidpunct}
{\mcitedefaultendpunct}{\mcitedefaultseppunct}\relax
\EndOfBibitem
\bibitem[Wang \latin{et~al.}(2019)Wang, Huh, Lee, Kim, Park, Lee, and
  Ke]{wang2019magnetic}
Wang,~P.; Huh,~J.-H.; Lee,~J.; Kim,~K.; Park,~K.~J.; Lee,~S.; Ke,~Y. Magnetic
  Plasmon Networks Programmed by Molecular Self-Assembly. \emph{Advanced
  Materials} \textbf{2019}, \emph{31}, 1901364\relax
\mciteBstWouldAddEndPuncttrue
\mciteSetBstMidEndSepPunct{\mcitedefaultmidpunct}
{\mcitedefaultendpunct}{\mcitedefaultseppunct}\relax
\EndOfBibitem
\bibitem[Montoni \latin{et~al.}(2018)Montoni, Quillin, Cherqui, and
  Masiello]{MasielloNanoclusters}
Montoni,~N.~P.; Quillin,~S.~C.; Cherqui,~C.; Masiello,~D.~J. Tunable Spectral
  Ordering of Magnetic Plasmon Resonances in Noble Metal Nanoclusters.
  \emph{ACS Photonics} \textbf{2018}, \emph{5}, 3272--3281\relax
\mciteBstWouldAddEndPuncttrue
\mciteSetBstMidEndSepPunct{\mcitedefaultmidpunct}
{\mcitedefaultendpunct}{\mcitedefaultseppunct}\relax
\EndOfBibitem
\bibitem[Ballantine and Ruostekoski(2020)Ballantine, and
  Ruostekoski]{ballantine2020optical}
Ballantine,~K.; Ruostekoski,~J. Optical magnetism and huygens’ surfaces in
  arrays of atoms induced by cooperative responses. \emph{Physical review
  letters} \textbf{2020}, \emph{125}, 143604\relax
\mciteBstWouldAddEndPuncttrue
\mciteSetBstMidEndSepPunct{\mcitedefaultmidpunct}
{\mcitedefaultendpunct}{\mcitedefaultseppunct}\relax
\EndOfBibitem
\bibitem[Kang \latin{et~al.}(2019)Kang, Bao, and Werner]{kang2019interference}
Kang,~L.; Bao,~H.; Werner,~D.~H. Interference-enhanced optical magnetism in
  surface high-index resonators: a pathway toward high-performance ultracompact
  linear and nonlinear meta-optics. \emph{Photonics Research} \textbf{2019},
  \emph{7}, 1296--1305\relax
\mciteBstWouldAddEndPuncttrue
\mciteSetBstMidEndSepPunct{\mcitedefaultmidpunct}
{\mcitedefaultendpunct}{\mcitedefaultseppunct}\relax
\EndOfBibitem
\bibitem[Linden \latin{et~al.}(2006)Linden, Enkrich, Dolling, Klein, Zhou,
  Koschny, Soukoulis, Burger, Schmidt, and Wegener]{linden2006photonic}
Linden,~S.; Enkrich,~C.; Dolling,~G.; Klein,~M.~W.; Zhou,~J.; Koschny,~T.;
  Soukoulis,~C.~M.; Burger,~S.; Schmidt,~F.; Wegener,~M. Photonic
  metamaterials: magnetism at optical frequencies. \emph{IEEE Journal of
  Selected Topics in Quantum Electronics} \textbf{2006}, \emph{12},
  1097--1105\relax
\mciteBstWouldAddEndPuncttrue
\mciteSetBstMidEndSepPunct{\mcitedefaultmidpunct}
{\mcitedefaultendpunct}{\mcitedefaultseppunct}\relax
\EndOfBibitem
\bibitem[Li \latin{et~al.}(2016)Li, Li, Yang, and Wu]{li2016optical}
Li,~G.; Li,~Q.; Yang,~L.; Wu,~L. Optical magnetism and optical activity in
  nonchiral planar plasmonic metamaterials. \emph{Optics letters}
  \textbf{2016}, \emph{41}, 2911--2914\relax
\mciteBstWouldAddEndPuncttrue
\mciteSetBstMidEndSepPunct{\mcitedefaultmidpunct}
{\mcitedefaultendpunct}{\mcitedefaultseppunct}\relax
\EndOfBibitem
\bibitem[Liu \latin{et~al.}(2007)Liu, Genov, Wu, Liu, Liu, Sun, Zhu, and
  Zhang]{liu2007magnetic}
Liu,~H.; Genov,~D.; Wu,~D.; Liu,~Y.; Liu,~Z.; Sun,~C.; Zhu,~S.; Zhang,~X.
  Magnetic plasmon hybridization and optical activity at optical frequencies in
  metallic nanostructures. \emph{Physical review B} \textbf{2007}, \emph{76},
  073101\relax
\mciteBstWouldAddEndPuncttrue
\mciteSetBstMidEndSepPunct{\mcitedefaultmidpunct}
{\mcitedefaultendpunct}{\mcitedefaultseppunct}\relax
\EndOfBibitem
\bibitem[Ginn \latin{et~al.}(2012)Ginn, Brener, Peters, Wendt, Stevens, Hines,
  Basilio, Warne, Ihlefeld, Clem, \latin{et~al.} others]{ginn2012realizing}
Ginn,~J.~C.; Brener,~I.; Peters,~D.~W.; Wendt,~J.~R.; Stevens,~J.~O.;
  Hines,~P.~F.; Basilio,~L.~I.; Warne,~L.~K.; Ihlefeld,~J.~F.; Clem,~P.~G.,
  \latin{et~al.}  Realizing optical magnetism from dielectric metamaterials.
  \emph{Physical review letters} \textbf{2012}, \emph{108}, 097402\relax
\mciteBstWouldAddEndPuncttrue
\mciteSetBstMidEndSepPunct{\mcitedefaultmidpunct}
{\mcitedefaultendpunct}{\mcitedefaultseppunct}\relax
\EndOfBibitem
\bibitem[Verre \latin{et~al.}(2015)Verre, Yang, Shegai, and
  Kall]{verre2015optical}
Verre,~R.; Yang,~Z.-J.; Shegai,~T.; Kall,~M. Optical magnetism and plasmonic
  Fano resonances in metal--insulator--metal oligomers. \emph{Nano letters}
  \textbf{2015}, \emph{15}, 1952--1958\relax
\mciteBstWouldAddEndPuncttrue
\mciteSetBstMidEndSepPunct{\mcitedefaultmidpunct}
{\mcitedefaultendpunct}{\mcitedefaultseppunct}\relax
\EndOfBibitem
\bibitem[Tserkezis \latin{et~al.}(2008)Tserkezis, Papanikolaou, Gantzounis, and
  Stefanou]{tserkezis2008understanding}
Tserkezis,~C.; Papanikolaou,~N.; Gantzounis,~G.; Stefanou,~N. Understanding
  artificial optical magnetism of periodic metal-dielectric-metal layered
  structures. \emph{Physical Review B} \textbf{2008}, \emph{78}, 165114\relax
\mciteBstWouldAddEndPuncttrue
\mciteSetBstMidEndSepPunct{\mcitedefaultmidpunct}
{\mcitedefaultendpunct}{\mcitedefaultseppunct}\relax
\EndOfBibitem
\bibitem[Calandrini \latin{et~al.}(01 Jan. 2019)Calandrini, Cerea, Angelis,
  Zaccaria, and Toma]{MagneticHotspotReview}
Calandrini,~E.; Cerea,~A.; Angelis,~F.~D.; Zaccaria,~R.~P.; Toma,~A. Magnetic
  hot-spot generation at optical frequencies: from plasmonic metamolecules to
  all-dielectric nanoclusters. \emph{Nanophotonics} \textbf{01 Jan. 2019},
  \emph{8}, 45 -- 62\relax
\mciteBstWouldAddEndPuncttrue
\mciteSetBstMidEndSepPunct{\mcitedefaultmidpunct}
{\mcitedefaultendpunct}{\mcitedefaultseppunct}\relax
\EndOfBibitem
\bibitem[Ha \latin{et~al.}(2019)Ha, Kim, You, Li, Fan, and
  Nam]{AnotherReviewPaper}
Ha,~M.; Kim,~J.-H.; You,~M.; Li,~Q.; Fan,~C.; Nam,~J.-M. Multicomponent
  Plasmonic Nanoparticles: From Heterostructured Nanoparticles to Colloidal
  Composite Nanostructures. \emph{Chemical Reviews} \textbf{2019}, \emph{119},
  12208--12278, PMID: 31794202\relax
\mciteBstWouldAddEndPuncttrue
\mciteSetBstMidEndSepPunct{\mcitedefaultmidpunct}
{\mcitedefaultendpunct}{\mcitedefaultseppunct}\relax
\EndOfBibitem
\bibitem[Fan \latin{et~al.}(2010)Fan, Wu, Bao, Bao, Bardhan, Halas, Manoharan,
  Nordlander, Shvets, and Capasso]{SelfAssembledNanoshells}
Fan,~J.~A.; Wu,~C.; Bao,~K.; Bao,~J.; Bardhan,~R.; Halas,~N.~J.;
  Manoharan,~V.~N.; Nordlander,~P.; Shvets,~G.; Capasso,~F. Self-Assembled
  Plasmonic Nanoparticle Clusters. \emph{Science} \textbf{2010}, \emph{328},
  1135--1138\relax
\mciteBstWouldAddEndPuncttrue
\mciteSetBstMidEndSepPunct{\mcitedefaultmidpunct}
{\mcitedefaultendpunct}{\mcitedefaultseppunct}\relax
\EndOfBibitem
\bibitem[Chen \latin{et~al.}(2017)Chen, Fan, Zhang, Tang, Chen, Wu, Li, and
  Yu]{chen2017engineering}
Chen,~J.; Fan,~W.; Zhang,~T.; Tang,~C.; Chen,~X.; Wu,~J.; Li,~D.; Yu,~Y.
  Engineering the magnetic plasmon resonances of metamaterials for high-quality
  sensing. \emph{Optics express} \textbf{2017}, \emph{25}, 3675--3681\relax
\mciteBstWouldAddEndPuncttrue
\mciteSetBstMidEndSepPunct{\mcitedefaultmidpunct}
{\mcitedefaultendpunct}{\mcitedefaultseppunct}\relax
\EndOfBibitem
\bibitem[Qin \latin{et~al.}(2013)Qin, Zhang, Peng, Xiong, Zhang, Huang, and
  Wang]{qin2013optical}
Qin,~L.; Zhang,~K.; Peng,~R.-W.; Xiong,~X.; Zhang,~W.; Huang,~X.-R.; Wang,~M.
  Optical-magnetism-induced transparency in a metamaterial. \emph{Physical
  Review B} \textbf{2013}, \emph{87}, 125136\relax
\mciteBstWouldAddEndPuncttrue
\mciteSetBstMidEndSepPunct{\mcitedefaultmidpunct}
{\mcitedefaultendpunct}{\mcitedefaultseppunct}\relax
\EndOfBibitem
\bibitem[Urzhumov and Shvets(2008)Urzhumov, and Shvets]{urzhumov2008optical}
Urzhumov,~Y.~A.; Shvets,~G. Optical magnetism and negative refraction in
  plasmonic metamaterials. \emph{Solid State Communications} \textbf{2008},
  \emph{146}, 208--220\relax
\mciteBstWouldAddEndPuncttrue
\mciteSetBstMidEndSepPunct{\mcitedefaultmidpunct}
{\mcitedefaultendpunct}{\mcitedefaultseppunct}\relax
\EndOfBibitem
\bibitem[{n}igo Liberal and Engheta(2017){n}igo Liberal, and
  Engheta]{NearZeroRF}
{n}igo Liberal,~I.; Engheta,~N. Near-zero refractive index photonics.
  \emph{Nature Photonics} \textbf{2017}, \emph{11}, 149--158\relax
\mciteBstWouldAddEndPuncttrue
\mciteSetBstMidEndSepPunct{\mcitedefaultmidpunct}
{\mcitedefaultendpunct}{\mcitedefaultseppunct}\relax
\EndOfBibitem
\bibitem[Ramakrishna(2005)]{NegativeRIPhysics}
Ramakrishna,~S. Physics of negative refractive index materials. \emph{Reports
  on Progress in Physics} \textbf{2005}, \emph{68}, 449--521\relax
\mciteBstWouldAddEndPuncttrue
\mciteSetBstMidEndSepPunct{\mcitedefaultmidpunct}
{\mcitedefaultendpunct}{\mcitedefaultseppunct}\relax
\EndOfBibitem
\bibitem[Padilla \latin{et~al.}(2006)Padilla, Basov, and
  Smith]{NegativeRIMetamaterials}
Padilla,~W.~J.; Basov,~D.~N.; Smith,~D.~R. Negative refractive index
  metamaterials. \emph{Materials Today} \textbf{2006}, \emph{9}, 28 -- 35\relax
\mciteBstWouldAddEndPuncttrue
\mciteSetBstMidEndSepPunct{\mcitedefaultmidpunct}
{\mcitedefaultendpunct}{\mcitedefaultseppunct}\relax
\EndOfBibitem
\bibitem[Cai \latin{et~al.}(2006)Cai, Chettiar, Kildishev, and
  Shalaev]{Cloaking}
Cai,~W.; Chettiar,~U.; Kildishev,~A.; Shalaev,~V. Optical Cloaking with
  Non-Magnetic Metamaterials. \textbf{2006}, \emph{1}, 224--227\relax
\mciteBstWouldAddEndPuncttrue
\mciteSetBstMidEndSepPunct{\mcitedefaultmidpunct}
{\mcitedefaultendpunct}{\mcitedefaultseppunct}\relax
\EndOfBibitem
\bibitem[Mühlig \latin{et~al.}(2013)Mühlig, Cunningham, Dintinger, Farhat,
  Hasan, Scharf, Bürgi, Lederer, and Rockstuhl]{RockstuhlCloaking}
Mühlig,~S.; Cunningham,~A.; Dintinger,~J.; Farhat,~M.; Hasan,~S.~B.;
  Scharf,~T.; Bürgi,~T.; Lederer,~F.; Rockstuhl,~C. A self-assembled
  three-dimensional cloak in the visible. \emph{Scientific Reports}
  \textbf{2013}, \emph{3}, 2328--\relax
\mciteBstWouldAddEndPuncttrue
\mciteSetBstMidEndSepPunct{\mcitedefaultmidpunct}
{\mcitedefaultendpunct}{\mcitedefaultseppunct}\relax
\EndOfBibitem
\bibitem[Zambrana-Puyalto and Bonod(2016)Zambrana-Puyalto, and
  Bonod]{ChiralEnhancement}
Zambrana-Puyalto,~X.; Bonod,~N. Tailoring the chirality of light emission with
  spherical Si-based antennas. \emph{Nanoscale} \textbf{2016}, \emph{8},
  10441--10452\relax
\mciteBstWouldAddEndPuncttrue
\mciteSetBstMidEndSepPunct{\mcitedefaultmidpunct}
{\mcitedefaultendpunct}{\mcitedefaultseppunct}\relax
\EndOfBibitem
\bibitem[Chen \latin{et~al.}(2017)Chen, Fan, Zhang, Tang, Chen, Wu, Li, and
  Yu]{OpticalSensing}
Chen,~J.; Fan,~W.; Zhang,~T.; Tang,~C.; Chen,~X.; Wu,~J.; Li,~D.; Yu,~Y.
  Engineering the magnetic plasmon resonances of metamaterials for high-quality
  sensing. \emph{Optics Express} \textbf{2017}, \emph{25}, 3675--3681\relax
\mciteBstWouldAddEndPuncttrue
\mciteSetBstMidEndSepPunct{\mcitedefaultmidpunct}
{\mcitedefaultendpunct}{\mcitedefaultseppunct}\relax
\EndOfBibitem
\bibitem[Fang \latin{et~al.}(2005)Fang, Lee, Sun, and
  Zhang]{SubDiffractionMicroscopy}
Fang,~N.; Lee,~H.; Sun,~C.; Zhang,~X. Sub-Diffraction-Limited Optical Imaging
  with a Silver Superlens. \emph{Science} \textbf{2005}, \emph{308},
  534--537\relax
\mciteBstWouldAddEndPuncttrue
\mciteSetBstMidEndSepPunct{\mcitedefaultmidpunct}
{\mcitedefaultendpunct}{\mcitedefaultseppunct}\relax
\EndOfBibitem
\bibitem[Qian \latin{et~al.}(2015)Qian, Hastings, Li, Edward, McGinn, Engheta,
  Fakhraai, and Park]{ACSNano}
Qian,~Z.; Hastings,~S.~P.; Li,~C.; Edward,~B.; McGinn,~C.~K.; Engheta,~N.;
  Fakhraai,~Z.; Park,~S.-J. Raspberry-like Metamolecules Exhibiting Strong
  Magnetic Resonances. \emph{ACS Nano} \textbf{2015}, \emph{9}, 1263--1270,
  PMID: 25621502\relax
\mciteBstWouldAddEndPuncttrue
\mciteSetBstMidEndSepPunct{\mcitedefaultmidpunct}
{\mcitedefaultendpunct}{\mcitedefaultseppunct}\relax
\EndOfBibitem
\bibitem[Ponsinet \latin{et~al.}(2015)Ponsinet, Barois, Gali, Richetti, Salmon,
  Vallecchi, Albani, Le~Beulze, Gomez-Grana, Duguet, \latin{et~al.}
  others]{ponsinet2015resonant}
Ponsinet,~V.; Barois,~P.; Gali,~S.~M.; Richetti,~P.; Salmon,~J.-B.;
  Vallecchi,~A.; Albani,~M.; Le~Beulze,~A.; Gomez-Grana,~S.; Duguet,~E.,
  \latin{et~al.}  Resonant isotropic optical magnetism of plasmonic
  nanoclusters in visible light. \emph{Physical Review B} \textbf{2015},
  \emph{92}, 220414\relax
\mciteBstWouldAddEndPuncttrue
\mciteSetBstMidEndSepPunct{\mcitedefaultmidpunct}
{\mcitedefaultendpunct}{\mcitedefaultseppunct}\relax
\EndOfBibitem
\bibitem[Bourgeois \latin{et~al.}(2017)Bourgeois, Liu, Ross, Berlin, and
  Schatz]{bourgeois2017self}
Bourgeois,~M.~R.; Liu,~A.~T.; Ross,~M.~B.; Berlin,~J.~M.; Schatz,~G.~C.
  Self-Assembled plasmonic metamolecules exhibiting tunable magnetic response
  at optical frequencies. \emph{The Journal of Physical Chemistry C}
  \textbf{2017}, \emph{121}, 15915--15921\relax
\mciteBstWouldAddEndPuncttrue
\mciteSetBstMidEndSepPunct{\mcitedefaultmidpunct}
{\mcitedefaultendpunct}{\mcitedefaultseppunct}\relax
\EndOfBibitem
\bibitem[Sheikholeslami \latin{et~al.}(2013)Sheikholeslami, Alaeian, Koh, and
  Dionne]{MagneticMetafluid}
Sheikholeslami,~S.~N.; Alaeian,~H.; Koh,~A.~L.; Dionne,~J.~A. A Metafluid
  Exhibiting Strong Optical Magnetism. \emph{Nano Letters} \textbf{2013},
  \emph{13}, 4137--4141, PMID: 23919764\relax
\mciteBstWouldAddEndPuncttrue
\mciteSetBstMidEndSepPunct{\mcitedefaultmidpunct}
{\mcitedefaultendpunct}{\mcitedefaultseppunct}\relax
\EndOfBibitem
\bibitem[Lim \latin{et~al.}(2014)Lim, Song, La, and Cho]{lim2014gold}
Lim,~S.; Song,~J.~E.; La,~J.~A.; Cho,~E.~C. Gold nanospheres assembled on
  hydrogel colloids display a wide range of thermoreversible changes in optical
  bandwidth for various plasmonic-based color switches. \emph{Chemistry of
  Materials} \textbf{2014}, \emph{26}, 3272--3279\relax
\mciteBstWouldAddEndPuncttrue
\mciteSetBstMidEndSepPunct{\mcitedefaultmidpunct}
{\mcitedefaultendpunct}{\mcitedefaultseppunct}\relax
\EndOfBibitem
\bibitem[Lee \latin{et~al.}(2020)Lee, Woods, Ibrahim, Kim, Pyun, Cho, Fakhraai,
  and Park]{JPCC}
Lee,~S.; Woods,~C.~N.; Ibrahim,~O.; Kim,~S.~W.; Pyun,~S.~B.; Cho,~E.~C.;
  Fakhraai,~Z.; Park,~S.-J. Distinct Optical Magnetism in Gold and Silver
  Probed by Dynamic Metamolecules. \emph{The Journal of Physical Chemistry C}
  \textbf{2020}, \emph{124}, 20436--20444\relax
\mciteBstWouldAddEndPuncttrue
\mciteSetBstMidEndSepPunct{\mcitedefaultmidpunct}
{\mcitedefaultendpunct}{\mcitedefaultseppunct}\relax
\EndOfBibitem
\bibitem[Li \latin{et~al.}(2018)Li, Lee, Qian, Woods, Park, and
  Fakhraai]{ChenJPCC}
Li,~C.; Lee,~S.; Qian,~Z.; Woods,~C.; Park,~S.-J.; Fakhraai,~Z. Controlling
  Magnetic Dipole Resonance in Raspberry-like Metamolecules. \emph{The Journal
  of Physical Chemistry C} \textbf{2018}, \emph{122}, 6808--6817\relax
\mciteBstWouldAddEndPuncttrue
\mciteSetBstMidEndSepPunct{\mcitedefaultmidpunct}
{\mcitedefaultendpunct}{\mcitedefaultseppunct}\relax
\EndOfBibitem
\bibitem[Manna \latin{et~al.}(2017)Manna, Lee, Deng, Parker, Shepherd,
  Weizmann, and Scherer]{ParkerFirst}
Manna,~U.; Lee,~J.-H.; Deng,~T.-S.; Parker,~J.; Shepherd,~N.; Weizmann,~Y.;
  Scherer,~N.~F. Selective Induction of Optical Magnetism. \emph{Nano Letters}
  \textbf{2017}, \emph{17}, 7196--7206, PMID: 29111760\relax
\mciteBstWouldAddEndPuncttrue
\mciteSetBstMidEndSepPunct{\mcitedefaultmidpunct}
{\mcitedefaultendpunct}{\mcitedefaultseppunct}\relax
\EndOfBibitem
\bibitem[Mühlig \latin{et~al.}(2011)Mühlig, Cunningham, Scheeler, Pacholski,
  Bürgi, Rockstuhl, and Lederer]{RockstuhlPlasmonic}
Mühlig,~S.; Cunningham,~A.; Scheeler,~S.; Pacholski,~C.; Bürgi,~T.;
  Rockstuhl,~C.; Lederer,~F. Self-Assembled Plasmonic Core–Shell Clusters
  with an Isotropic Magnetic Dipole Response in the Visible Range. \emph{ACS
  Nano} \textbf{2011}, \emph{5}, 6586--6592, PMID: 21714523\relax
\mciteBstWouldAddEndPuncttrue
\mciteSetBstMidEndSepPunct{\mcitedefaultmidpunct}
{\mcitedefaultendpunct}{\mcitedefaultseppunct}\relax
\EndOfBibitem
\bibitem[Mühlig \latin{et~al.}(2013)Mühlig, Cunningham, Dintinger, Scharf,
  Bürgi, Lederer, and Rockstuhl]{RockstuhlReview}
Mühlig,~S.; Cunningham,~A.; Dintinger,~J.; Scharf,~T.; Bürgi,~T.;
  Lederer,~F.; Rockstuhl,~C. Self-assembled plasmonic metamaterials.
  \emph{Nanophotonics} \textbf{2013}, \emph{2}, 211--240\relax
\mciteBstWouldAddEndPuncttrue
\mciteSetBstMidEndSepPunct{\mcitedefaultmidpunct}
{\mcitedefaultendpunct}{\mcitedefaultseppunct}\relax
\EndOfBibitem
\bibitem[Simovski and Tretyakov(2009)Simovski, and Tretyakov]{CoreShellModel}
Simovski,~C.~R.; Tretyakov,~S.~A. Model of isotropic resonant magnetism in the
  visible range based on core-shell clusters. \emph{Phys. Rev. B}
  \textbf{2009}, \emph{79}, 045111\relax
\mciteBstWouldAddEndPuncttrue
\mciteSetBstMidEndSepPunct{\mcitedefaultmidpunct}
{\mcitedefaultendpunct}{\mcitedefaultseppunct}\relax
\EndOfBibitem
\bibitem[Ponsinet \latin{et~al.}(2015)Ponsinet, Barois, Gali, Richetti, Salmon,
  Vallecchi, Albani, Le~Beulze, Gomez-Grana, Duguet, Mornet, and
  Treguer-Delapierre]{ResonantIsotropicMagnetism}
Ponsinet,~V.; Barois,~P.; Gali,~S.~M.; Richetti,~P.; Salmon,~J.~B.;
  Vallecchi,~A.; Albani,~M.; Le~Beulze,~A.; Gomez-Grana,~S.; Duguet,~E.;
  Mornet,~S.; Treguer-Delapierre,~M. Resonant isotropic optical magnetism of
  plasmonic nanoclusters in visible light. \emph{Phys. Rev. B} \textbf{2015},
  \emph{92}, 220414\relax
\mciteBstWouldAddEndPuncttrue
\mciteSetBstMidEndSepPunct{\mcitedefaultmidpunct}
{\mcitedefaultendpunct}{\mcitedefaultseppunct}\relax
\EndOfBibitem
\bibitem[Urban \latin{et~al.}(2013)Urban, Shen, Wang, Large, Wang, Knight,
  Nordlander, Chen, and Halas]{3DNanoclusters}
Urban,~A.~S.; Shen,~X.; Wang,~Y.; Large,~N.; Wang,~H.; Knight,~M.~W.;
  Nordlander,~P.; Chen,~H.; Halas,~N.~J. Three-Dimensional Plasmonic
  Nanoclusters. \emph{Nano Letters} \textbf{2013}, \emph{13}, 4399--4403, PMID:
  23977943\relax
\mciteBstWouldAddEndPuncttrue
\mciteSetBstMidEndSepPunct{\mcitedefaultmidpunct}
{\mcitedefaultendpunct}{\mcitedefaultseppunct}\relax
\EndOfBibitem
\bibitem[Deng \latin{et~al.}(2020)Deng, Parker, Hirai, Shepherd, Yabu, and
  Scherer]{ParkerBackscattering}
Deng,~T.-S.; Parker,~J.; Hirai,~Y.; Shepherd,~N.; Yabu,~H.; Scherer,~N.~F.
  Designing "Metamolecules" for Photonic Function: Reduced Backscattering.
  \emph{physica status solidi (b)} \textbf{2020}, \emph{257}, 2000169\relax
\mciteBstWouldAddEndPuncttrue
\mciteSetBstMidEndSepPunct{\mcitedefaultmidpunct}
{\mcitedefaultendpunct}{\mcitedefaultseppunct}\relax
\EndOfBibitem
\bibitem[Bakhti \latin{et~al.}(2016)Bakhti, Tishchenko, Zambrana-Puyalto,
  Bonod, Dhuey, Schuck, Cabrini, Alayoglu, and Destouches]{GoldArrayFano}
Bakhti,~S.; Tishchenko,~A.~V.; Zambrana-Puyalto,~X.; Bonod,~N.; Dhuey,~S.~D.;
  Schuck,~P.~J.; Cabrini,~S.; Alayoglu,~S.; Destouches,~N. Fano-like resonance
  emerging from magnetic and electric plasmon mode coupling in small arrays of
  gold particles. \emph{Scientific Reports} \textbf{2016}, \emph{6}\relax
\mciteBstWouldAddEndPuncttrue
\mciteSetBstMidEndSepPunct{\mcitedefaultmidpunct}
{\mcitedefaultendpunct}{\mcitedefaultseppunct}\relax
\EndOfBibitem
\bibitem[Kruk \latin{et~al.}(2017)Kruk, Camacho-Morales, Xu, Rahmani, Smirnova,
  Wang, Tan, Jagadish, Neshev, and Kivshar]{kruk2017nonlinear}
Kruk,~S.~S.; Camacho-Morales,~R.; Xu,~L.; Rahmani,~M.; Smirnova,~D.~A.;
  Wang,~L.; Tan,~H.~H.; Jagadish,~C.; Neshev,~D.~N.; Kivshar,~Y.~S. Nonlinear
  optical magnetism revealed by second-harmonic generation in nanoantennas.
  \emph{Nano letters} \textbf{2017}, \emph{17}, 3914--3918\relax
\mciteBstWouldAddEndPuncttrue
\mciteSetBstMidEndSepPunct{\mcitedefaultmidpunct}
{\mcitedefaultendpunct}{\mcitedefaultseppunct}\relax
\EndOfBibitem
\bibitem[Parker \latin{et~al.}(2017)Parker, Gray, and Scherer]{ParkerArXiv}
Parker,~J.; Gray,~S.; Scherer,~N. Multipolar analysis of electric and magnetic
  modes excited by vector beams in core-satellite nano-structures. \emph{arXiv
  preprint} \textbf{2017}, arXiv:1711.06833\relax
\mciteBstWouldAddEndPuncttrue
\mciteSetBstMidEndSepPunct{\mcitedefaultmidpunct}
{\mcitedefaultendpunct}{\mcitedefaultseppunct}\relax
\EndOfBibitem
\bibitem[Barber and Hill(1990)Barber, and Hill]{barber1990light}
Barber,~P.~W.; Hill,~S.~C. \emph{Light scattering by particles: computational
  methods}; World scientific, 1990; Vol.~2\relax
\mciteBstWouldAddEndPuncttrue
\mciteSetBstMidEndSepPunct{\mcitedefaultmidpunct}
{\mcitedefaultendpunct}{\mcitedefaultseppunct}\relax
\EndOfBibitem
\bibitem[Stout \latin{et~al.}(2008)Stout, Auger, and
  Devilez]{stout2008recursive}
Stout,~B.; Auger,~J.-C.; Devilez,~A. Recursive T matrix algorithm for resonant
  multiple scattering: applications to localized plasmon excitations.
  \emph{JOSA A} \textbf{2008}, \emph{25}, 2549--2557\relax
\mciteBstWouldAddEndPuncttrue
\mciteSetBstMidEndSepPunct{\mcitedefaultmidpunct}
{\mcitedefaultendpunct}{\mcitedefaultseppunct}\relax
\EndOfBibitem
\bibitem[Duan \latin{et~al.}(2015)Duan, Haynes, and
  Moghaddam]{duan2015experimental}
Duan,~X.; Haynes,~M.; Moghaddam,~M. Experimental Verification of the Recursive
  T-Matrix Method Solutions at Microwave Frequencies. \emph{IEEE Transactions
  on Antennas and Propagation} \textbf{2015}, \emph{63}, 5727--5740\relax
\mciteBstWouldAddEndPuncttrue
\mciteSetBstMidEndSepPunct{\mcitedefaultmidpunct}
{\mcitedefaultendpunct}{\mcitedefaultseppunct}\relax
\EndOfBibitem
\bibitem[Khlebtsov and Khlebtsov(2007)Khlebtsov, and
  Khlebtsov]{khlebtsov2007multipole}
Khlebtsov,~B.~N.; Khlebtsov,~N.~G. Multipole plasmons in metal nanorods:
  scaling properties and dependence on particle size, shape, orientation, and
  dielectric environment. \emph{The Journal of Physical Chemistry C}
  \textbf{2007}, \emph{111}, 11516--11527\relax
\mciteBstWouldAddEndPuncttrue
\mciteSetBstMidEndSepPunct{\mcitedefaultmidpunct}
{\mcitedefaultendpunct}{\mcitedefaultseppunct}\relax
\EndOfBibitem
\bibitem[Khlebtsov(2013)]{khlebtsov2013t}
Khlebtsov,~N.~G. T-matrix method in plasmonics: An overview. \emph{Journal of
  Quantitative Spectroscopy and Radiative Transfer} \textbf{2013}, \emph{123},
  184--217\relax
\mciteBstWouldAddEndPuncttrue
\mciteSetBstMidEndSepPunct{\mcitedefaultmidpunct}
{\mcitedefaultendpunct}{\mcitedefaultseppunct}\relax
\EndOfBibitem
\bibitem[Arya(2006)]{arya2006scattering}
Arya,~K. Scattering T-matrix theory in wave-vector space for surface-enhanced
  Raman scattering in clusters of nanoscale spherical metal particles.
  \emph{Physical Review B} \textbf{2006}, \emph{74}, 195438\relax
\mciteBstWouldAddEndPuncttrue
\mciteSetBstMidEndSepPunct{\mcitedefaultmidpunct}
{\mcitedefaultendpunct}{\mcitedefaultseppunct}\relax
\EndOfBibitem
\bibitem[Fruhnert \latin{et~al.}(2017)Fruhnert, Fernandez-Corbaton, Yannopapas,
  and Rockstuhl]{RockstuhlTMatrix}
Fruhnert,~M.; Fernandez-Corbaton,~I.; Yannopapas,~V.; Rockstuhl,~C. Computing
  the T-matrix of a scattering object with multiple plane wave illuminations.
  \emph{Beilstein Journal of Nanotechnology} \textbf{2017}, \emph{8},
  614--626\relax
\mciteBstWouldAddEndPuncttrue
\mciteSetBstMidEndSepPunct{\mcitedefaultmidpunct}
{\mcitedefaultendpunct}{\mcitedefaultseppunct}\relax
\EndOfBibitem
\bibitem[Johnson(1988)]{johnson1988invariant}
Johnson,~B.~R. Invariant imbedding T matrix approach to electromagnetic
  scattering. \emph{Applied optics} \textbf{1988}, \emph{27}, 4861--4873\relax
\mciteBstWouldAddEndPuncttrue
\mciteSetBstMidEndSepPunct{\mcitedefaultmidpunct}
{\mcitedefaultendpunct}{\mcitedefaultseppunct}\relax
\EndOfBibitem
\bibitem[Hastings \latin{et~al.}(2015)Hastings, Qian, Swanglap, Fang, Engheta,
  Park, Link, and Fakhraai]{Simon}
Hastings,~S.~P.; Qian,~Z.; Swanglap,~P.; Fang,~Y.; Engheta,~N.; Park,~S.-J.;
  Link,~S.; Fakhraai,~Z. Modal interference in spiky nanoshells. \emph{Opt.
  Express} \textbf{2015}, \emph{23}, 11290--11311\relax
\mciteBstWouldAddEndPuncttrue
\mciteSetBstMidEndSepPunct{\mcitedefaultmidpunct}
{\mcitedefaultendpunct}{\mcitedefaultseppunct}\relax
\EndOfBibitem
\bibitem[Zhao \latin{et~al.}(2003)Zhao, Kelly, and Schatz]{zhao2003extinction}
Zhao,~L.; Kelly,~K.~L.; Schatz,~G.~C. The extinction spectra of silver
  nanoparticle arrays: influence of array structure on plasmon resonance
  wavelength and width. \emph{The Journal of Physical Chemistry B}
  \textbf{2003}, \emph{107}, 7343--7350\relax
\mciteBstWouldAddEndPuncttrue
\mciteSetBstMidEndSepPunct{\mcitedefaultmidpunct}
{\mcitedefaultendpunct}{\mcitedefaultseppunct}\relax
\EndOfBibitem
\bibitem[Kiselev \latin{et~al.}(2002)Kiselev, Reshetnyak, and
  Sluckin]{kiselev2002light}
Kiselev,~A.; Reshetnyak,~V.~Y.; Sluckin,~T. Light scattering by optically
  anisotropic scatterers: T-matrix theory for radial and uniform anisotropies.
  \emph{Physical Review E} \textbf{2002}, \emph{65}, 056609\relax
\mciteBstWouldAddEndPuncttrue
\mciteSetBstMidEndSepPunct{\mcitedefaultmidpunct}
{\mcitedefaultendpunct}{\mcitedefaultseppunct}\relax
\EndOfBibitem
\bibitem[Haynes(2014)]{CRCHandbook}
Haynes,~W.~M. \emph{CRC Handbook of Chemistry and Physics}, 95th ed.; CRC
  Press: USA, 2014\relax
\mciteBstWouldAddEndPuncttrue
\mciteSetBstMidEndSepPunct{\mcitedefaultmidpunct}
{\mcitedefaultendpunct}{\mcitedefaultseppunct}\relax
\EndOfBibitem
\bibitem[Boh(1998)]{Bohren}
\emph{Absorption and Scattering of Light by Small Particles}; John Wiley \&
  Sons, Ltd, 1998\relax
\mciteBstWouldAddEndPuncttrue
\mciteSetBstMidEndSepPunct{\mcitedefaultmidpunct}
{\mcitedefaultendpunct}{\mcitedefaultseppunct}\relax
\EndOfBibitem
\bibitem[Hastings(2014)]{SimonThesis}
Hastings,~S.~P. The Optical Properties of Spiky Gold Nanoshells. Ph.D.\ thesis,
  University of Pennsylvania, 2014\relax
\mciteBstWouldAddEndPuncttrue
\mciteSetBstMidEndSepPunct{\mcitedefaultmidpunct}
{\mcitedefaultendpunct}{\mcitedefaultseppunct}\relax
\EndOfBibitem
\bibitem[Tsa(2000)]{Tsang}
\emph{Scattering of Electromagnetic Waves: Theories and Applications}; John
  Wiley \& Sons, Ltd, 2000\relax
\mciteBstWouldAddEndPuncttrue
\mciteSetBstMidEndSepPunct{\mcitedefaultmidpunct}
{\mcitedefaultendpunct}{\mcitedefaultseppunct}\relax
\EndOfBibitem
\bibitem[Al\`{u} \latin{et~al.}(2006)Al\`{u}, Salandrino, and Engheta]{Nader1}
Al\`{u},~A.; Salandrino,~A.; Engheta,~N. Negative effective permeability and
  left-handed materials at optical frequencies. \emph{Opt. Express}
  \textbf{2006}, \emph{14}, 1557--1567\relax
\mciteBstWouldAddEndPuncttrue
\mciteSetBstMidEndSepPunct{\mcitedefaultmidpunct}
{\mcitedefaultendpunct}{\mcitedefaultseppunct}\relax
\EndOfBibitem
\bibitem[Schmidt \latin{et~al.}(2012)Schmidt, Esteban, S{\'a}enz,
  Su{\'a}rez-Lacalle, Mackowski, and Aizpurua]{schmidt2012dielectric}
Schmidt,~M.~K.; Esteban,~R.; S{\'a}enz,~J.; Su{\'a}rez-Lacalle,~I.;
  Mackowski,~S.; Aizpurua,~J. Dielectric antennas-a suitable platform for
  controlling magnetic dipolar emission. \emph{Optics express} \textbf{2012},
  \emph{20}, 13636--13650\relax
\mciteBstWouldAddEndPuncttrue
\mciteSetBstMidEndSepPunct{\mcitedefaultmidpunct}
{\mcitedefaultendpunct}{\mcitedefaultseppunct}\relax
\EndOfBibitem
\bibitem[Mirin \latin{et~al.}(2009)Mirin, Bao, and Nordlander]{mirin2009fano}
Mirin,~N.~A.; Bao,~K.; Nordlander,~P. Fano resonances in plasmonic nanoparticle
  aggregates. \emph{The Journal of Physical Chemistry A} \textbf{2009},
  \emph{113}, 4028--4034\relax
\mciteBstWouldAddEndPuncttrue
\mciteSetBstMidEndSepPunct{\mcitedefaultmidpunct}
{\mcitedefaultendpunct}{\mcitedefaultseppunct}\relax
\EndOfBibitem
\bibitem[Virtanen \latin{et~al.}(2020)Virtanen, Gommers, Oliphant, Haberland,
  Reddy, Cournapeau, Burovski, Peterson, Weckesser, Bright, van~der Walt,
  Brett, Wilson, Millman, Mayorov, Nelson, Jones, Kern, Larson, Carey, Polat,
  Feng, Moore, VanderPlas, Laxalde, Perktold, Cimrman, Henriksen, Quintero,
  Harris, Archibald, Ribeiro, Pedregosa, van Mulbregt, and Contributors]{scipy}
Virtanen,~P. \latin{et~al.}  SciPy 1.0: Fundamental Algorithms for Scientific
  Computing in Python. \emph{Nature Methods} \textbf{2020}, \emph{17},
  261--272\relax
\mciteBstWouldAddEndPuncttrue
\mciteSetBstMidEndSepPunct{\mcitedefaultmidpunct}
{\mcitedefaultendpunct}{\mcitedefaultseppunct}\relax
\EndOfBibitem
\end{mcitethebibliography}

\begin{figure}[H]
\includegraphics[width=\linewidth]{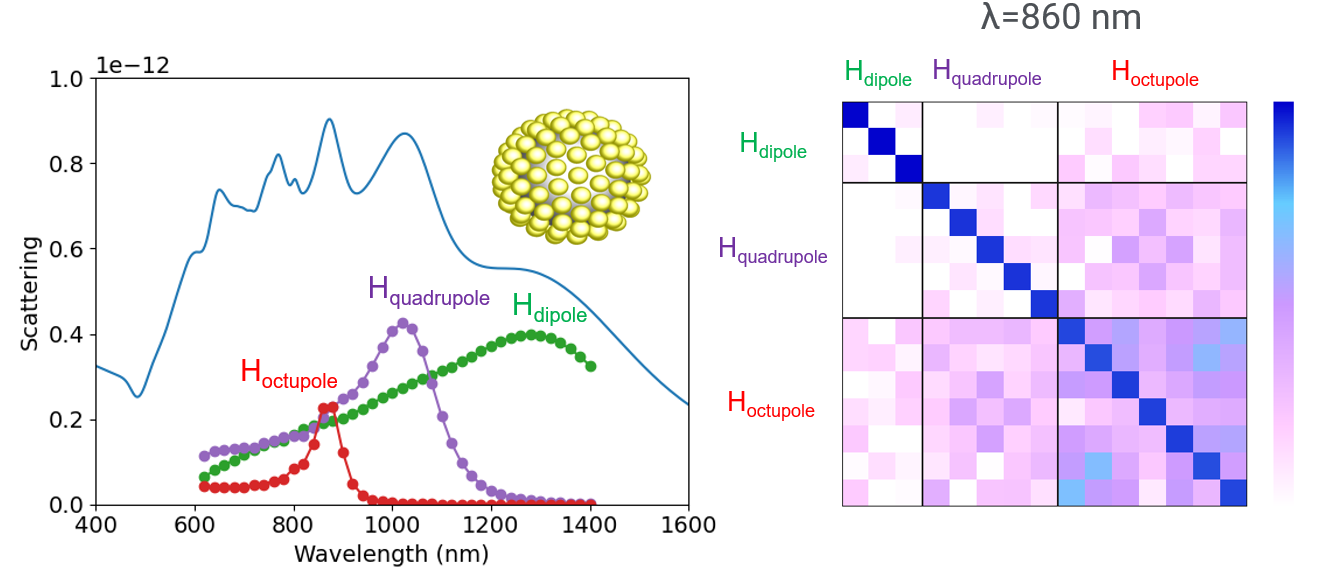}
\caption{For table of contents only}
\end{figure}

\pagebreak

\begin{suppinfo}

\renewcommand{\thefigure}{S\arabic{figure}}
\setcounter{figure}{0}

\singlespacing
\subsection{I. Experimental Details}

\begin{figure}[H]
\includegraphics[width=0.8\linewidth]{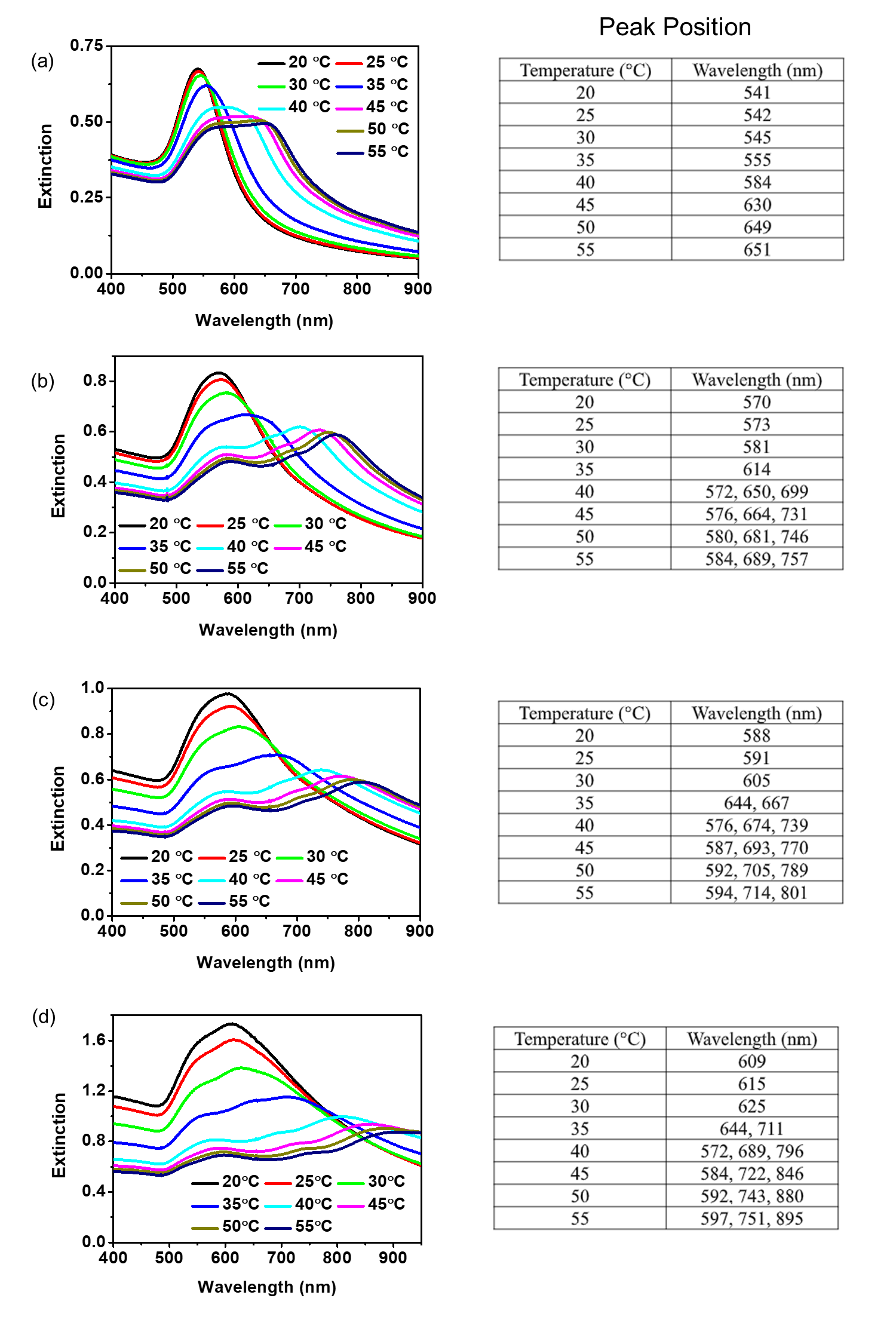}
\caption{Peak positions for experimental gold DMMs of average number and size (a) $N=57 \pm 3, D=34.8\pm2.7$~nm , (b) $N=61 \pm 4, D=44.6\pm1.9$~nm (c) $N=59 \pm 4, D=49.2\pm2.4$~nm, and (d) $N=41 \pm 4, D=60.0\pm2.5$~nm.}
\label{fig:SIExperimental}
\end{figure}

\begin{center}
Table S1: Experimental conditions and structural parameters for gold DMMs
\end{center}

\begin{tabularx}{1\textwidth} { 
  | >{\centering\arraybackslash}X 
  | >{\centering\arraybackslash}X 
  | >{\centering\arraybackslash}X 
  | >{\centering\arraybackslash}X
  | >{\centering\arraybackslash}X| }
\hline
 & Bead Size (nm) & Concentration of Nanobead Solution (nM) & Number of Beads & Spectral Data \\ 
\hline
1 & $34.8 \pm 2.7$ & $0.2$ & $57 \pm 3$ & Figure 2a \\ 
\hline
2 & $44.6 \pm 1.9$ & $0.1$ & $61 \pm 4$ & Figure 2b \\ 
\hline
3 & $49.2 \pm 2.4$ & $0.1$ & $59 \pm 4$ & Figure 2c \\ 
\hline
4 & $60.0 \pm 2.5$ & $0.03$ & $41 \pm 4$ & Figure 2d \\ 
\hline
\end{tabularx}

\begin{figure}[H]
\includegraphics[width=\linewidth]{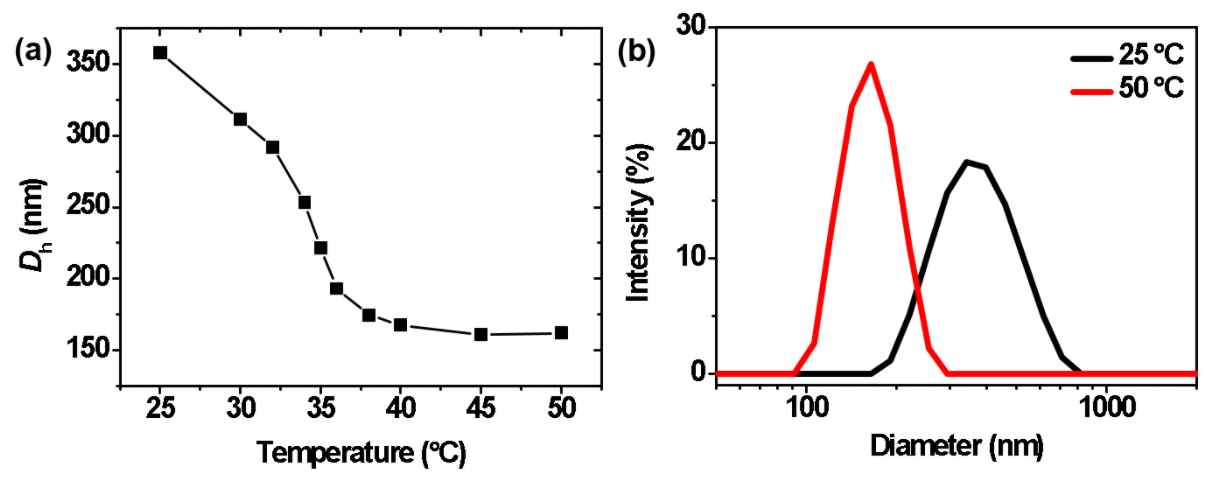}
\caption{(a) Temperature-dependent hydrodynamic diameter ($D_h$ of PNIPAM hydrogels. The transition temperature is measured to be 34°C. (b) DLS data of PNIPAM hydrogels obtained at 25 and 50°C. The $D_h$ of PNIPAM hydrogels was measured to be 355 $\pm$ 69 nm (polydispersity index (PDI): 0.08) at 25°C and 165 $\pm$ 33 nm (PDI: 0.01) at 50°C.}
\label{hydrodynamicR}
\end{figure}

\subsection{II. Additional Simulation Details}

In theory the $\vec{E}$ and $\vec{H}$ fields of the entire volume can be saved but this would be highly space inefficient, making T-matrix analysis slow or expensive. To address this, we recursively broke the mesh region of the FDTD simulations into octants with a cube monitor placed at each octant location that contains a mesh point within $r = [R_{int} - \frac{\sqrt{3}}{2} dx, R_{int} + \frac{\sqrt{3}}{2} dx]$ (where $R_{int}$ is the radius of integration and $dx$ is the maximum mesh step). This ensures that there is always a cube of data points around each point on the integration sphere for grid interpolation. This cuts the volume of data saved by a factor of up to 10 depending on the recursion depth used. These cubic monitors were then iteratively merged to form rectangular prisms in order to reduce the number of data files saved. 

\hspace*{2cm}\begin{figure}[H]
\includegraphics[width=\linewidth]{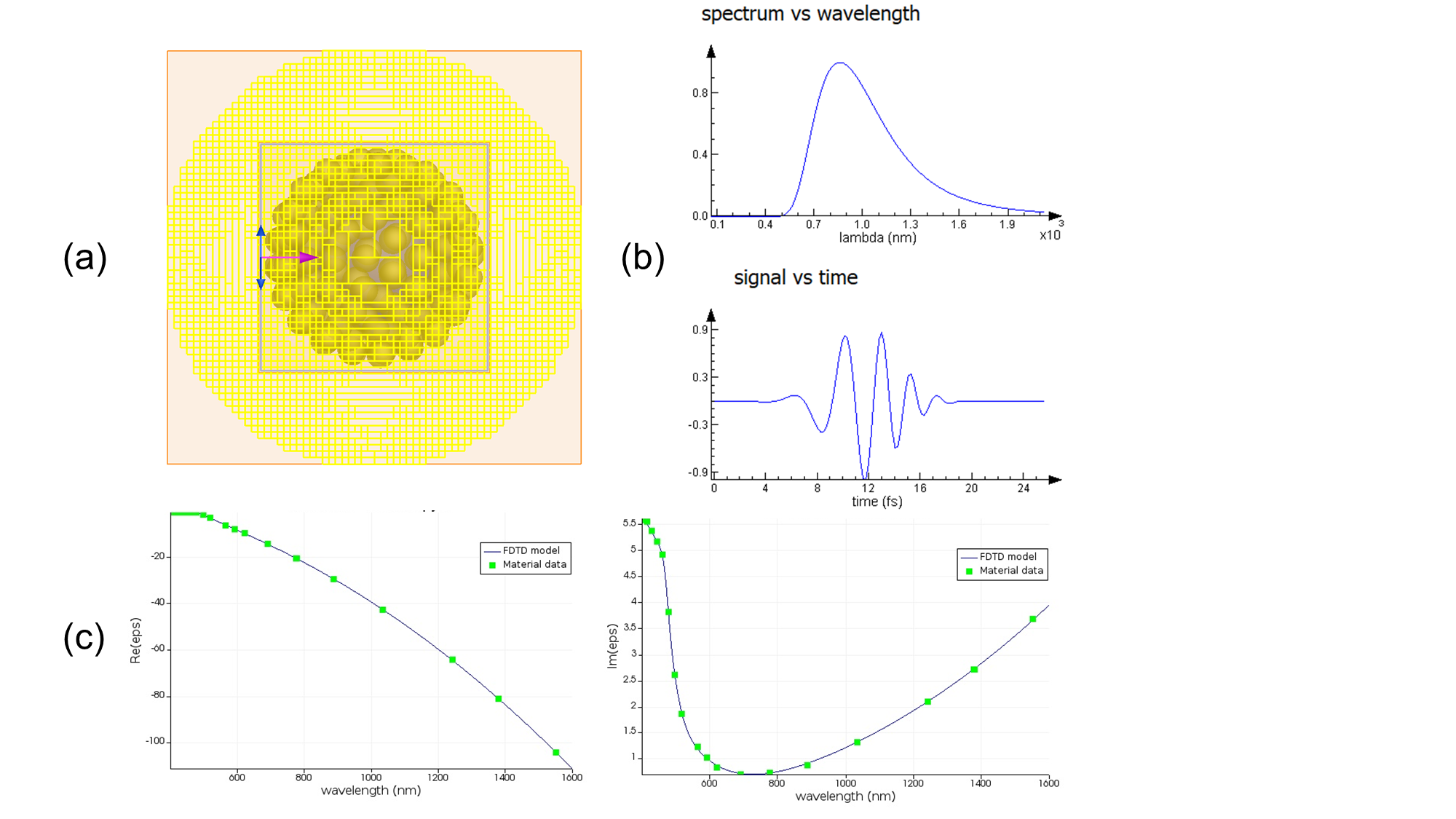} 
\caption{Details of (a) the T-Matrix simulation including simulation setup, (b) the plane wave pulse used to excite the particle, and (c) the material fit used for the gold NPs based on the CRC Handbook optical data \cite{CRCHandbook} }
\label{fig:SimDetails}
\end{figure}

The setup for the T matrix is shown in Figure \ref{fig:SimDetails}a. Two meshes were used: a fine mesh (with a 2.5 nm mesh size) encapsulating both the MM and the plane wave source (both inside the grey box) and a coarse mesh (with a 5 nm mesh size) on the outside where the scattered field was collected (outer orange box). The yellow shell made of rectangular monitors can be seen above in the region within the outer mesh but outside the inner mesh. 

\subsection{III. Calculation of the T-Matrix}

\begin{figure}[H]
\includegraphics[width=0.7\linewidth]{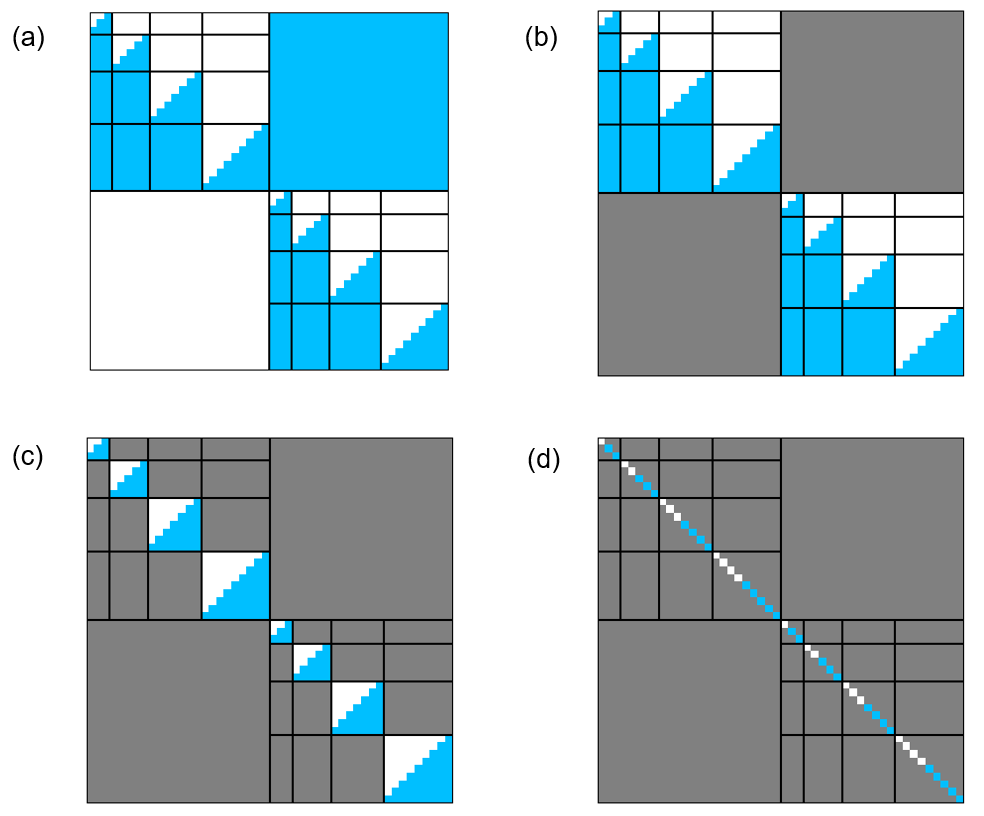} 
\caption{A possible set of free fit parameters for (a) the full T-matrix, (b) the block diagonal T matrix, (c) the IMM T matrix, and (d) the diagonal T matrix with a maximum $n$ of 4. Note that a blue square is an independent element, a white square is a dependent element, and a grey square is fixed at 0}
\label{fig:FreeFitParameters}
\end{figure} 

\paragraph{Reciprocity relations.} As stated in the main text, we take advantage of the T matrix reciprocity relation $T^{ij}_{-m'n'(-m)n} = (-1)^{m' + m} T^{ji}_{mnm'n'}$ for $i, j = \{E, M\}$ in order to reduce the number of free fit parameters for the T matrix \cite{Tsang}. Figure \ref{fig:FreeFitParameters} shows a set of free fit parameters for each type of T matrix that we use to minimize error. To compute the T-matrix from the above expression we do a minimization over the T-matrix elements using the scipy library for Python. In particular, we use the function scipy.optimize.minimize (using the SLSQP solver) \cite{scipy}.

\begin{figure}[H]
\includegraphics[width=\linewidth]{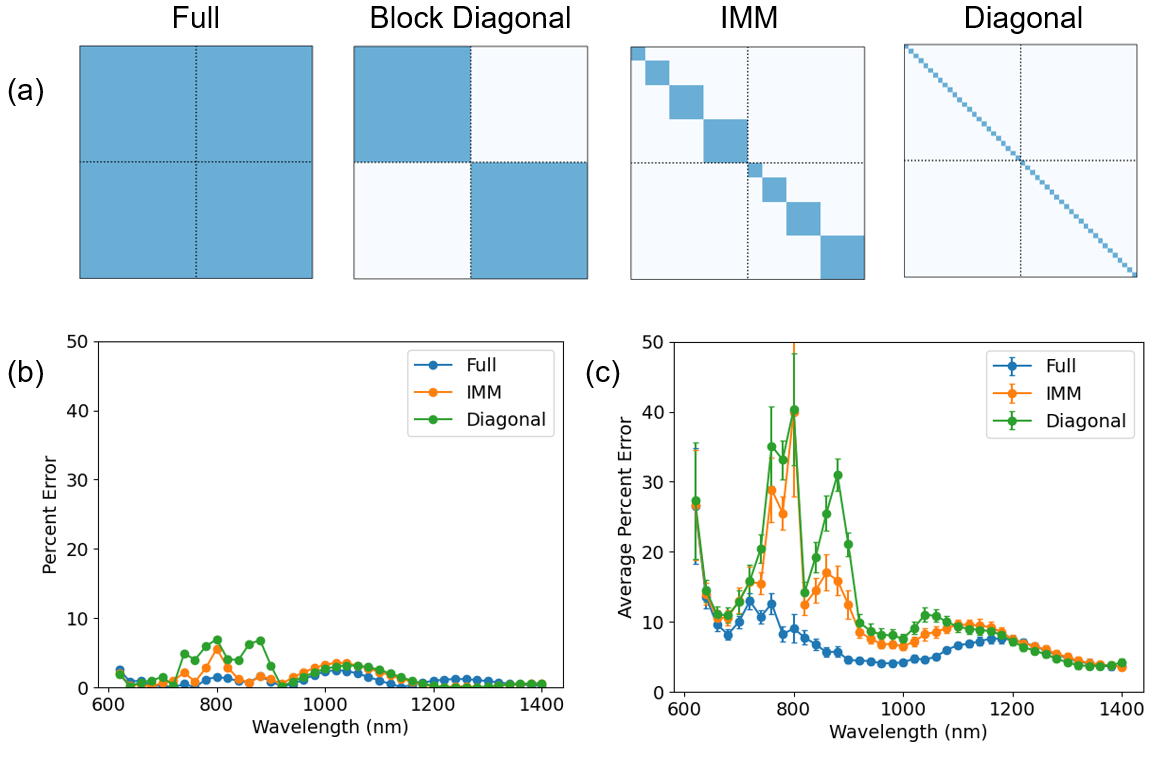}
\caption{Various types of T-Matrices. Non-zero components (used to fit) are highlighted with dark blue and zero components (not fitted) are highlighted with light blue. (a) From left to right: full T-matrix, block diagonal with all the electric and magnetic modes mixed (not further studied in this paper), intra-mode mixing (IMM) T-matrix (where modes with the same order $n$ are mixed), and Diagonal. (b) Error of each T-matrix compared to FDTD results for $C_{scat}$. (c) The average of the errors for 100 different orientations of differential scattering cross section (with error bars as the standard error)}
\label{fig:diagonal}
\end{figure} 

\paragraph{The need to use the full T-matrix.} The data in Figure 3 of the main text represent the full T-matrix fitting. However, it may not always necessarily to fit the full T-matrix and the full fitting may result in over-fitting. To investigate whether the full T-matrix is needed, we explored three different T Matrices; diagonal, intra-mode mixing (IMM, where modes with the same $n$ are allowed to mix), and full. We have excluded a fourth possible T-matrix, a block diagonal one (schematically shown in Figure \ref{fig:diagonal}a), because large differences between this T-matrix and the full T-matrix implies significant electric-magnetic interference, requiring strongly chiral structures \cite{SimonThesis}, which is not the case here. A graphical representation of the nonzero terms for these matrices can be seen in Figure \ref{fig:diagonal}a. A diagonal T-matrix would imply that there is no mode mixing (meaning that all sub-modes that enter contribute only to the scattering of that sub-mode), while an IMM T-matrix implies that all entering modes can only mix among their respective sub-modes.  To evaluate the performance of these fits, we computed the scattering cross section for the structure and plot the error as a function of wavelength for each T-matrix (Figure \ref{fig:diagonal}b). We also computed the differential scattering cross section for an incident plane wave at 100 random orientations and plotted the mean error compared to FDTD results as a function of wavelength (Figure \ref{fig:diagonal}c).

The results indicate that both the diagonal and IMM T-matrices produce significant errors in predicting the higher order magnetic octupole ($H_{oct}$) and hexacecapole ($H_{hex}$) resonances. This error more significantly manifests itself in the accuracy of the differential scattering cross sections. These results mean that there is indeed mode mixing between various order modes and that the full T-matrix includes key details of the scattering behavior of the magnetic  modes of interest, especially at the $H_{oct}$ and $H_{hex}$ resonances, which we will discuss in more details below. This can also be seen in Figure \ref{fig:compare-full-diagonal} that compares the results of the full- and the diagonal- T-matrices for this structure. We observe that the diagonal T-matrix underestimates the contributions of both the $H_{oct}$ and $H_{hex}$ modes at their resonance,  leading to significant errors that were seen in Figure \ref{fig:diagonal}. As such, we opt to use the full T-matrix calculations for the data presented in the manuscript.

\begin{figure}[H]
\includegraphics[width=\linewidth]{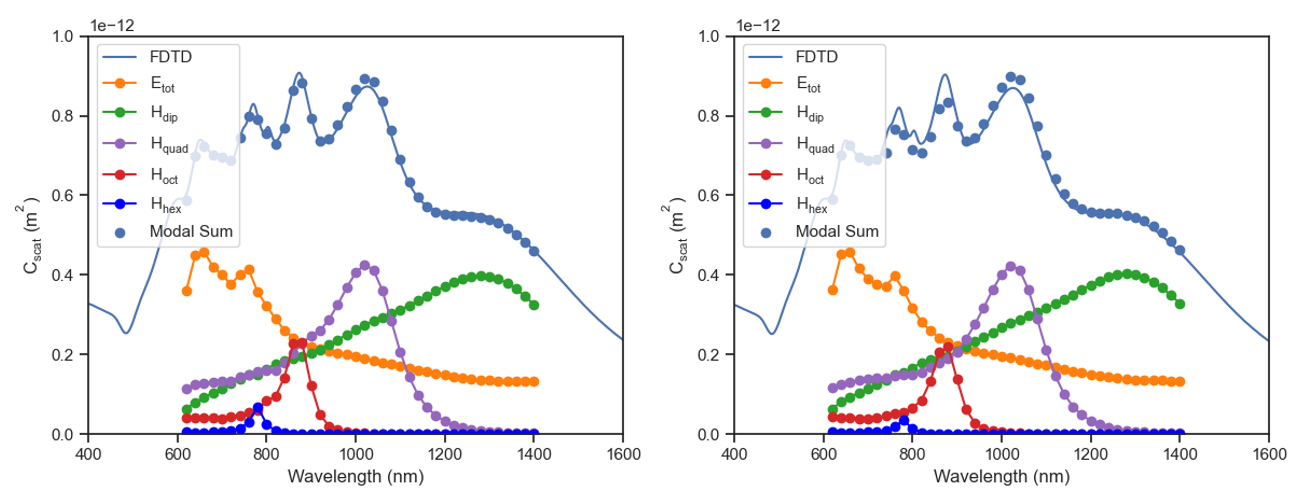}
\caption{The modal scattering cross section obtained from the full T-matrix (left) and diagonal T-matrix (right) as well as the corresponding FDTD results (solid line).}
\label{fig:compare-full-diagonal}
\end{figure}

\paragraph{Block-diagonal T-matrix Approximation}
While the diagonal and IMM T-matrices are inadequate in predicting the properties at the $H_{oct}$ and $H_{hex}$ modes, most of the missing terms are originated in mode-mixing between the magnetic multipole modes themselves. Figure \ref{Fig:hexadecaTMatrix} shows the results of the full T-matrix at the resonant wavelength of the magnetic hexadecapole mode ($\lambda_{hex}=780$~nm). As it can be seen in this figure, the non-zero off-diagonal terms are mostly within the magnetic multipole block. The only exception is mode-mixing between the electric dipole $E_{dip}$ 
modes and the magnetic hexadecapole ($H_{hex}$) modes at the $H_{hex}$ resonance, where there is also significant spectral overlap. as such, for cases where properties at longer wavelengths are of interest, such as manetic multipole resonances with $n\leq3$, one can opt to use the block-diagonal T-matrix and focus mostly on scattering from the magnetic multipole modes and their mode mixing.   

\begin{figure}[H]
\includegraphics[width=\linewidth]{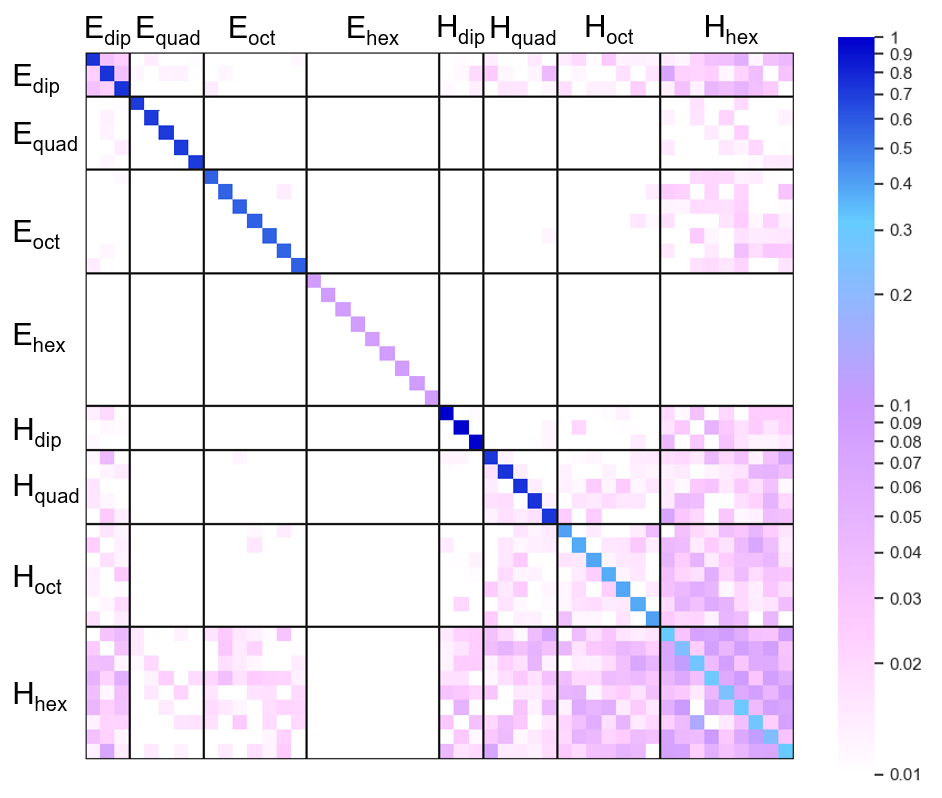}
\caption{The full T-matrix for the $N=126$, $Z=415$~nm, $D=70$~nm structure at the magnetic hexadecapole resonance ($\lambda_{hex} = 780$~nm). The top left quadrant represents the electric T-matrix, the bottom right quadrant represents the magnetic T-matrix, and the off diagonal quadrants represent electric-magnetic mode mixing. As we can see, there is significant mixing between the magnetic hexadecpaole and all other magnetic monopole modes, as well as the electric dipole modes.}
\label{Fig:hexadecaTMatrix}
\end{figure}

\subsection{IV. Effects of Mode Mixing in Single Mode Excitations}

Mode mixing effects are generally not too significant in the total scattering cross section, for the exception of the $H_{hex}$ resonance as discussed above. However, mode mixing can have significant implications for experiments involving specific  polarizations of the incoming or scattered light, where mode mixing can result in constructive or destructive interference patterns, or in directions were a specific mode may have enhanced scattering intensity, attenuating the mode-mixing effect. 

To understand these effects, it is more convenient to simulate single mode excitations, and observe the scattering cross-section from these modes. To do this, we can hand pick a set of incident coefficients and then apply the T-matrix operator to obtain the corresponding scattering cross-section at every point in space, by calculating the scattered electric and magnetic fields. The scattered power is given by $P = \frac{1}{2} \int_{\Omega} Re(\vec{E} \times \vec{H}^*) \cdot d\vec{A}$. We excite our structure with two octupole modes and two hexadecapole modes chosen, one to maximize the amount of mode mixing and one to minimizes the amount of mode mixing. The corresponding data is shown in Figure \ref{fig:mixing}.

\begin{figure}[H]
\includegraphics[width=\linewidth]{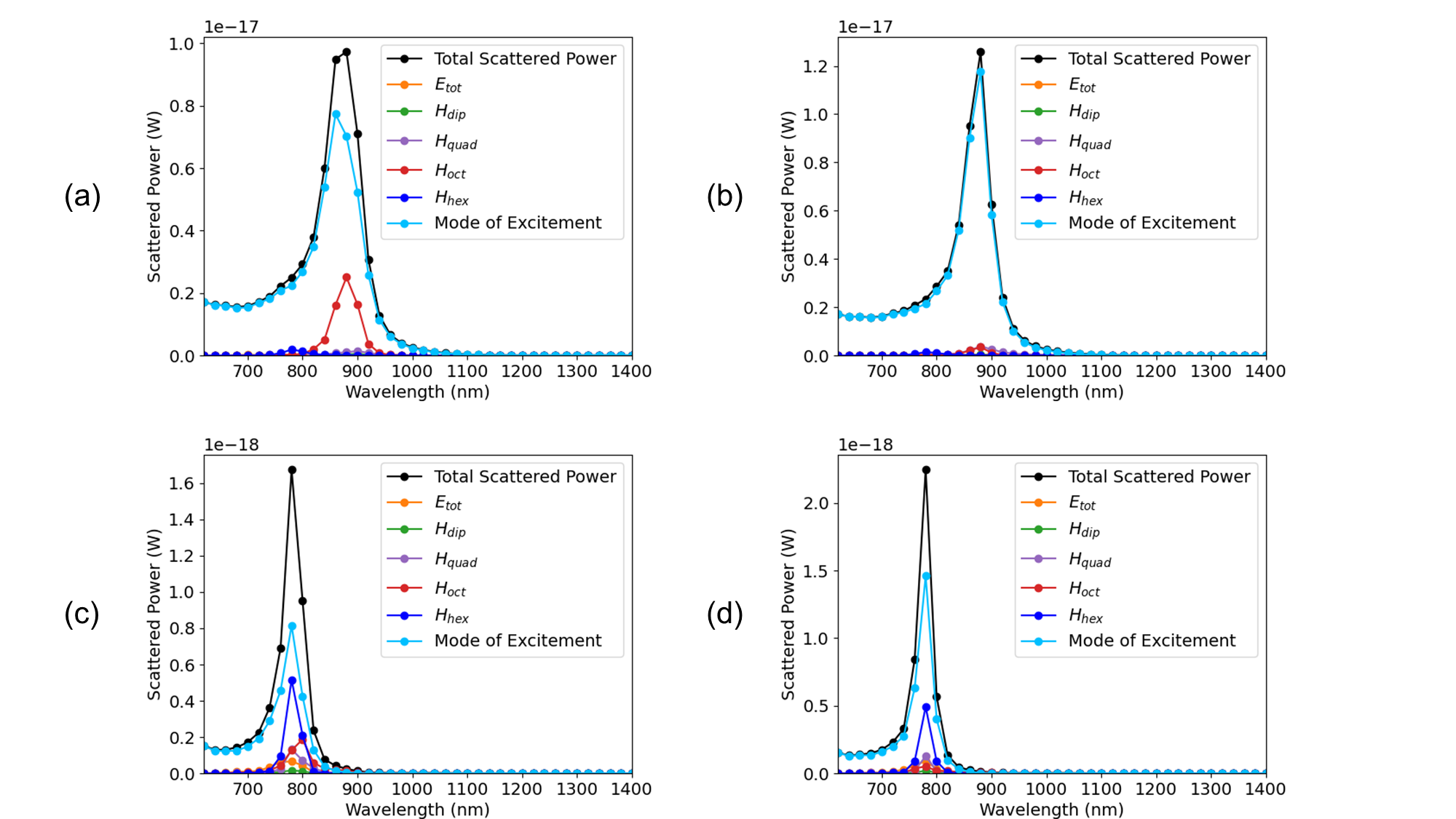}
\caption{The modal scattered power for various modes upon a single-mode excitations. The exciting mode chosen in each case was (a) the incident $H_{oct}$ mode with $(n, m)=(3, 2)$, which maximizes the degree of mode-mixing (b) the incident $H_{oct}$ mode with $(n, m)=(3,0)$, which minimizes mode mixing, (c) the incident $H_{hex}$ mode with $(n, m)=(4,3)$, which maximizes magnetic hexadecapole mode mixing, and (d) the incident $H_{hex}$ mode with $(n, m)=(4, 4)$, which minimizes the scattered power contributed by modes other than the exciting mode.}
\label{fig:mixing}
\end{figure} 

Now if mode mixing effects are not significant then we should expect that exciting a structure with one pure submode will result in only seeing scattered power from that particular submode as calculated from the formula above. This is approximately the case (Figure\ref{fig:mixing}b), for the magnetic octupole submode ($H_{oct}$ of $(n, m)=(3,0)$, which generates a nearly pure scattered mode at the octupole resonance frequency. However, other $H_{oct}$ submodes, as well as nearly all hexadecapole ($H_{hex}$) submodes generate scattered power in  various other modes (Figures \ref{fig:mixing}a,c and d, for example) indicating potent mode mixing effects. Interestingly, the contribution to the electric dipole modes $E_{tot}$, while non-zero for $H_{hex}$ submodes, is not as significant as the contributions generated by other magnetic multipole modes, all of which have broad shoulders towards lower wavelengths, where the magnetic hexadecapole resonance is observed. This is also consistent with the strength of off-diagonal magnetic terms in Figure \ref{Fig:hexadecaTMatrix}, are much stronger than the cross-terms corresponding to the magnetic/electric mode mixing.

\subsection{V. T-Matrix and Current Data for the Magnetic HExadecapole}

\begin{figure}[H]
\includegraphics[scale=0.4]{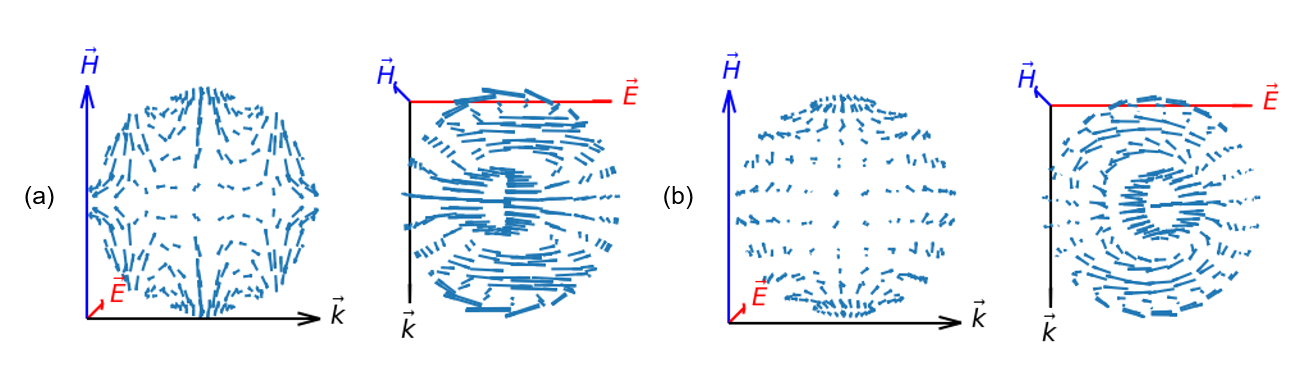}
\caption{The far field displacement current plots of a pure (a) and mixed (b) magnetic hexadecapole at the magnetic hexadecapole resonance ($\lambda_{hex} = 780$~nm). The current maps are taken at the time that maximizes the sum of squared magnitude and are all scaled by the same constant value to the highlight differences.}
\label{Fig:hexadeca}
\end{figure}

Figure \ref{Fig:hexadeca} shows the hexadecapole displacement currents at the hexadecapole resonance ($\lambda_{hex}=780$~nm) as calculated through a diagonal and a full T-matrix (Figure \ref{Fig:hexadecaTMatrix} at the point in time where the sum of square amplitudes is maximized. Note that this is scaled since the hexadecapole is quite weak (as seen in Figure \ref{fig:StructureVaryingDecomp}). 

Mode mixing has a profound effect on the magnetic hexadecapole, $H_{hex}$ as seen in Figure \ref{Fig:hexadeca}b. The field vectors throughout the structure begin to rotate and fall out of phase, rather than resonating all at one point in time, the currents achieve their maximum amplitudes at different points in time resulting in current vectors that are weaker than the diagonal case which demonstrates the degree to which the currents fall out of phase. In addition, as shown in Figure \ref{Fig:hexadecaTMatrix}, $H_{hex}$ has strong mixing with the electric dipole modes at its resonance, which results in non-trivial directional scattering behavior that is not observed in other magnetic multipole modes. There is also significant spectral overlap between $H_{hex}$  and $E_{dip}$ further promoting this mode-mixing behavior.

\subsection{VI. Obtaining Optical Properties from the T-Matrix}

\paragraph{Calculation of the Scattering Cross Section.} Once we have obtained the T-matrix at a given frequency we can use it to determine the scattering cross section $C_{scat}$ for any incident direction, the differential scattering cross section $\frac{d C_{scat}}{d \Omega}$ for any given incident and scattering direction, and the vector scattering amplitude $\vec{X}$ for any incident and scattering direction. We can also obtain the far field scattered electric field $\vec{E}_s(\theta, \phi)$ and far field displacement current $\vec{J}_D(\theta, \phi)$. All these computations hinge on being able to determine the far field scattering amplitude dyad $\overline{\overline{F}}$. In the far field limit the scattering amplitude dyad $\overline{\overline{F}}$ is related to the scattered electric field by the relation:

\noindent
\begin{eqnarray}
\bar{E}_s (\vec{r}) = \frac{e^{ikr}}{r} \overline{\overline{F}} (\theta_s, \phi_s, \theta_i, \phi_i) \cdot \hat{\beta} E_0
\end{eqnarray}

\noindent
where $(\theta_s, \phi_s)$ represents the direction of scatterred light, $(\theta_i, \phi_i)$ represents the direction of incident light, $\hat{\beta}$ is the unit polarization vector, and $E_0$ is the amplitude. We can obtain $\overline{\overline{F}}$ from the T-matrix as follows:
\bigskip

\noindent
\begin{equation}
\begin{split}
\overline{\overline{F}}(\theta_s, \phi_s, \theta_i, \phi_i) = \frac{4 \pi}{k} \sum\limits_{n, m, n', m'} (-1)^{m'} i^{n'-n-1} * \{[T^{EE}_{mnm'n'} i \gamma_{mn} \overline{B}_{mn} (\theta_s, \phi_s) + T^{EM}_{mnm'n'} \gamma_{mn} \overline{C}_{mn} (\theta_s, \phi_s)] \\ \cdot \gamma_{-m'n'} \frac{\overline{B}_{-m'n'} (\theta_i, \phi_i)}{i}  + [T^{MM}_{mnm'n'} \gamma_{mn} \overline{C}_{mn} (\theta_s, \phi_s) + T^{ME}_{mnm'n'} i \gamma_{mn} \overline{B}_{mn} (\theta_s, \phi_s)] \cdot \gamma_{-m'n'} \overline{C}_{-m'n'} (\theta_i, \phi_i) \}     
\end{split}
\end{equation}

In this equation $\overline{B}_{mn}(\theta, \phi)$ and $\overline{C}_{mn} (\theta, \phi)$ are the two vector spherical harmonics that satisfy the vector wave equation. Using $\overline{\overline{F}}$ we can obtain the vector scattering amplitude $\vec{X}(\theta_s, \phi_s) = \overline{\overline{F}}(\theta_s, \phi_s, \theta_i, \phi_i) \cdot \hat{e}_i$. From $\vec{X}$ we can obtain the differential scattering cross section in the $(\theta_s, \phi_s)$ direction $\frac{d C_{scat}}{d \Omega} = |\vec{X}(\theta_s, \phi_s)|^2$ and the scattering cross section $C_{scat} (\theta_i, \phi_i) = \int_{4 \pi} d \Omega [ |\overline{\overline{F}}(\theta, \phi, \theta_i, \phi_i) \cdot \hat{\beta} |^2 ]$  \cite{SimonThesis}. Note that instead of numerical integration we use an exact formula \cite{Tsang,Bohren}:

\noindent
\begin{equation}
\begin{split}
C_{scat}(\theta_i, \phi_i) = \frac{16 \pi^2}{k^2} \sum\limits_{n, m} \{ | \sum\limits_{m', n'} i^{n'} (-1)^{m'} \gamma_{-m'n'} [T^{EE}_{mnm'n'} \frac{\overline{B}_{-m'n'}(\theta_i, \phi_i)}{i} \cdot \hat{\beta} + T^{EM}_{mnm'n'} \overline{C}_{-m'n'}(\theta_i, \phi_i) \cdot \hat{\beta}]|^2 + \\ | \sum\limits_{m', n'} i^{n'} (-1)^{m'} \gamma_{-m'n'} [T^{MM}_{mnm'n'} \overline{C}_{-m'n'}(\theta_i, \phi_i) \cdot \hat{\beta} + T^{ME}_{mnm'n'} \frac{\overline{B}_{-m'n'}(\theta_i, \phi_i)}{i} \cdot \hat{\beta}]|^2 \}
\end{split}
\end{equation}

The formulas for scattering cross section $C_{scat}$ and $\overline{\overline{F}}$ are linear in the scattering mode $n$ and so we can obtain the exact modal contributions to $C_{scat}$, $\vec{E}_s$, and $\vec{J}_D$. 

\paragraph{Calculation of the Displacement Currents.} The T-matrix can be used to obtain the far field displacement current maps which can visualize the currents flowing through the structure as shown in Figure 4. In the far-field $\bar{E}_s (\vec{r}) = \frac{e^{ikr}}{r} \overline{\overline{F}} (\theta_s, \phi_s, \theta_i, \phi_i) \cdot \hat{e_i} E_0$. In a linear medium, $\vec{J}_D = \epsilon_0 \frac{d \vec{E}}{dt} + \frac{d \vec{P}}{dt}$. Assuming a constant $\epsilon_r$ over the frequency range of interest, we have $\vec{J}_D(\vec{r}, t) \propto \frac{d \vec{E}(\vec{r}, t)}{dt}$. Thus, it follows that in the frequency domain we have $\vec{J}_D (\omega, \vec{r}) \propto -i \omega \vec{E}(\omega, \vec{r})$. We can then obtain $\vec{J}_D$ from the T-matrix by computing $\vec{E}_s$. To produce a periodic animation from this relationship we can convert the complex $\vec{J}_D$ vector field to the standard basis and then obtain the amplitude ($J_x, J_y, J_z$) and phase ($\theta_x, \theta_y, \theta_z$) in each direction. We then produce a 3D quiver plot of $<J_x \sin(\omega t - \theta_x), J_y \sin(\omega t - \theta_y), J_z \sin(\omega t - \theta_z)>$ at a fixed value of $r$ in the far-field at various points in time to produce a full animation. 

We can also calculate the contributions of each mode to $\overline{\overline{F}}$, as modes are linear in $n$. Therefore, the cumulative displacement current $\vec{J}_D (\omega, \vec{r})$ is additive over all modes, meaning if we take the amplitudes and phase values from each mode and superimpose all those waves, we recover the total current. Note that this follows from the fact that for two complex numbers $z_1 = a_1 + b_1 i$ and $z_2 = a_2 + b_2 i$ we have that $\sqrt{(a_1 + a_2)^2 + (b_1 + b_2)^2} \sin(\omega t - \atan2 (b_1 + b_2, a_1 + a_2)) = \sqrt{a_1^2 + b_1^2} \sin(\omega t - \atan2 (b_1, a_1)) + \sqrt{a_2^2 + b_2^2} \sin(\omega t - \atan2 (b_2, a_2))$, which can be verified using a symbolic calculator.  

\subsection{VII. Structural Parameters and Full FDTD Data for Model MMs}

\begin{center}
Table S2: Structural parameters for simulated gold MMs
\end{center}

\hspace*{-1cm}\begin{tabularx}{1.1\textwidth} { 
  | >{\centering\arraybackslash}X 
  | >{\centering\arraybackslash}X 
  | >{\centering\arraybackslash}X 
  | >{\centering\arraybackslash}X
  | >{\centering\arraybackslash}X| }
\hline
 Bead Number, $N$ & Core Size, $Z$ (nm) & Bead Size, $D$ (nm) & Mean Nearest Neighbor Gap Distance, $d_{g, avg}$ (nm) & Min Nearest Neighbor Distance, $d_{g, min}$ (nm) \\ 
\hline
126 & 360 & 60 & $6.87 \pm 1.53$ & 3.04  \\ 
\hline
126 & 380 & 60 & $10.05 \pm 1.61$ & 6.04  \\ 
\hline
126 & 400 & 60 & $13.24 \pm 1.68$ & 9.05  \\ 
\hline
126 & 435 & 60 & $18.81 \pm 1.81$ & 14.30  \\ 
\hline
126 & 215 & 35 & $4.80 \pm 0.91$ & 2.52  \\ 
\hline
126 & 300 & 50 & $5.73 \pm 1.28$ & 2.54  \\ 
\hline
126 & 415 & 70 & $7.22 \pm 1.77$ & 2.80  \\ 
\hline
160 & 415 & 60 & $7.38 \pm 1.56$ & 3.67  \\ 
\hline
220 & 415 & 50 & $6.29 \pm 1.43$ & 3.07  \\ 
\hline
310 & 415 & 40 & $6.36 \pm 1.13$ & 3.47  \\ 
\hline
\end{tabularx}
\pagebreak

\noindent
\underline{Series A: Varying Core Size}

\begin{figure}[H]
\includegraphics[width=\linewidth]{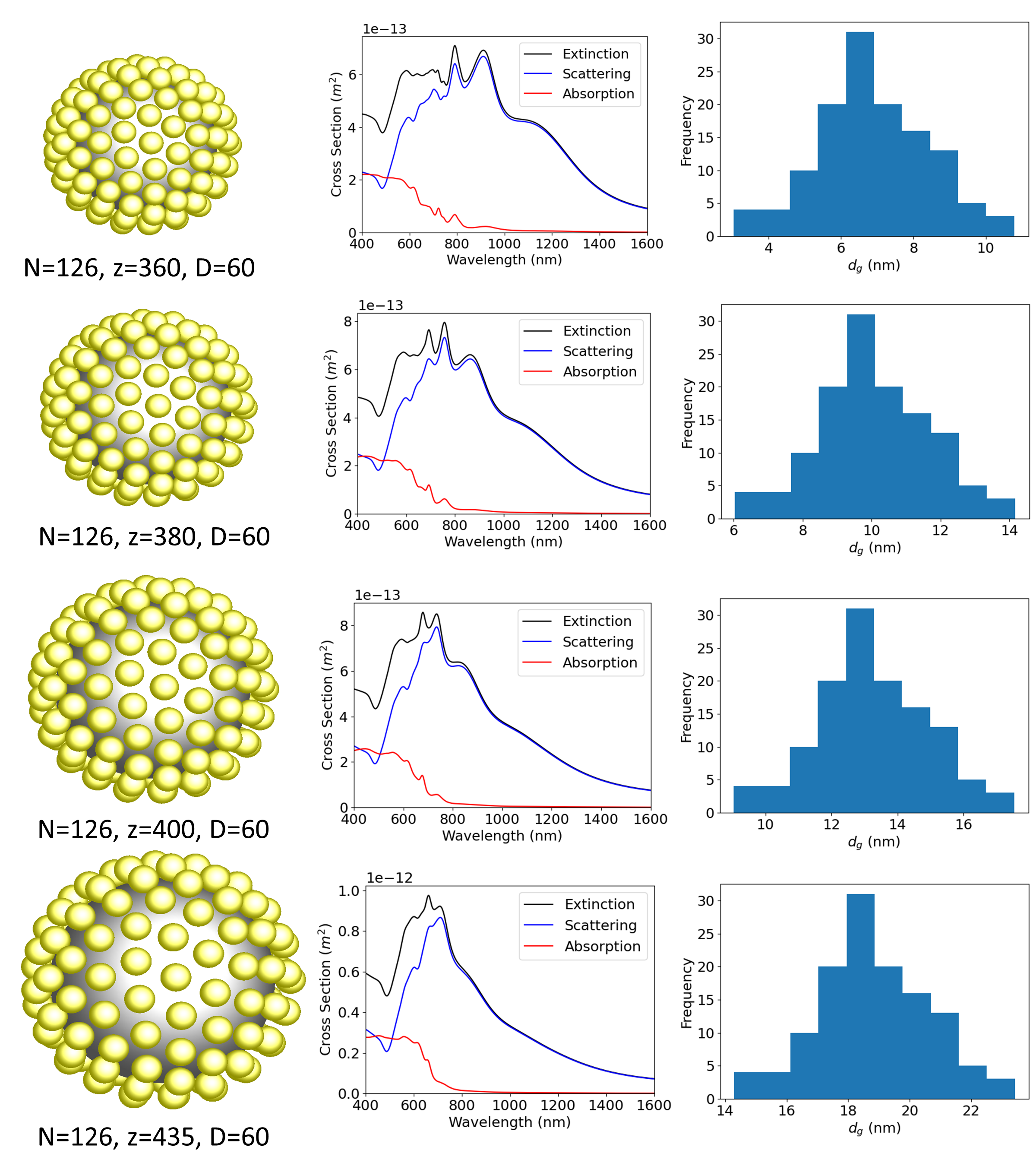} 
\caption{FDTD simulation results and nearest neighbor interparticle gap distance ($d_g$) histograms for Series \textbf{A} (the series where we vary $Z$).}
\label{fig:CoreSizeVaryingFDTD}
\end{figure}
\pagebreak

\noindent
\underline{Series B: Scaling Structure}

\begin{figure}[H]
\includegraphics[width=\linewidth]{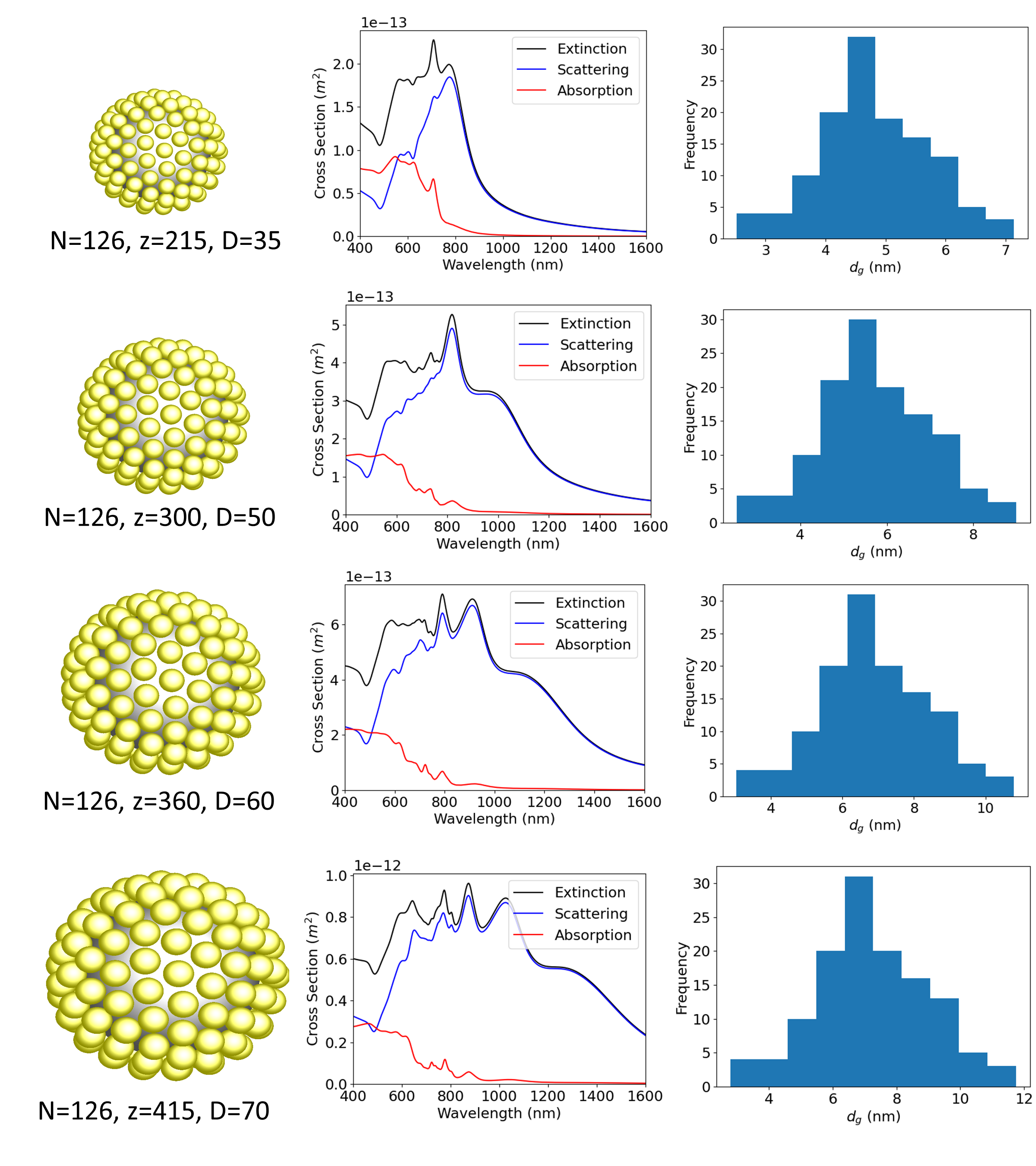} \caption{FDTD simulation results and nearest neighbor interparticle gap distance ($d_g$) histograms for Series \textbf{B} (the series where we keep the packing constant while varying $Z$ and $D$).}
\label{fig:StructureVaryingFDTD}
\end{figure}
\pagebreak

\noindent
\underline{Series C: Varying Bead Size/Number}

\begin{figure}[H]
\includegraphics[width=\linewidth]{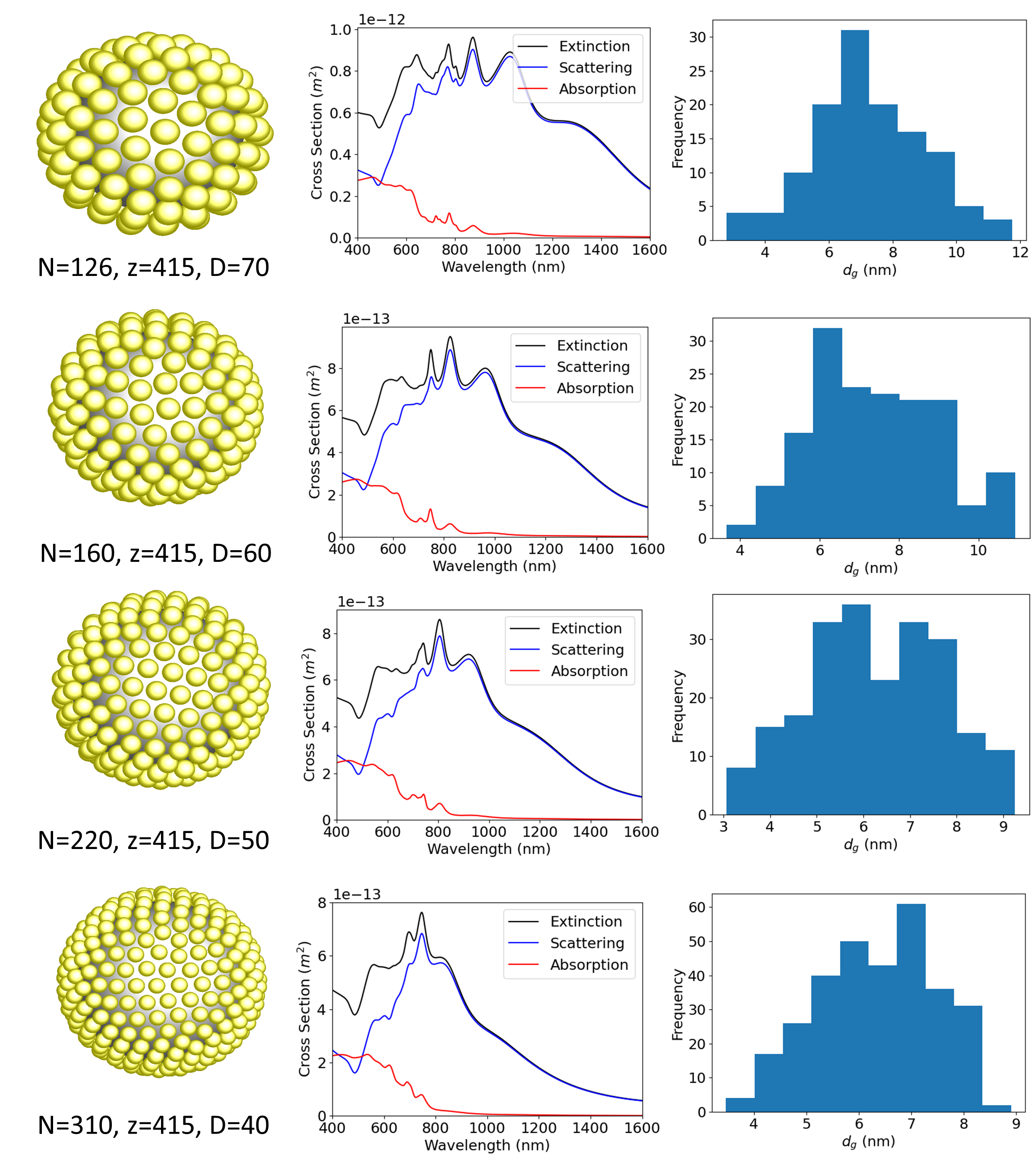} 
\caption{FDTD simulation results and nearest neighbor interparticle gap distance ($d_g$) histograms for Series \textbf{C} (series where we keep $Z$ constant while increasing $N$ and decreasing $D$).}
\label{fig:BeadSizeVaryingFDTD}
\end{figure}

\pagebreak
\subsection{VIII. Full Modal Breakdown for Model MMs}

\underline{Series A: Varying Core Size}
\begin{figure}[H]
\includegraphics[width=\linewidth]{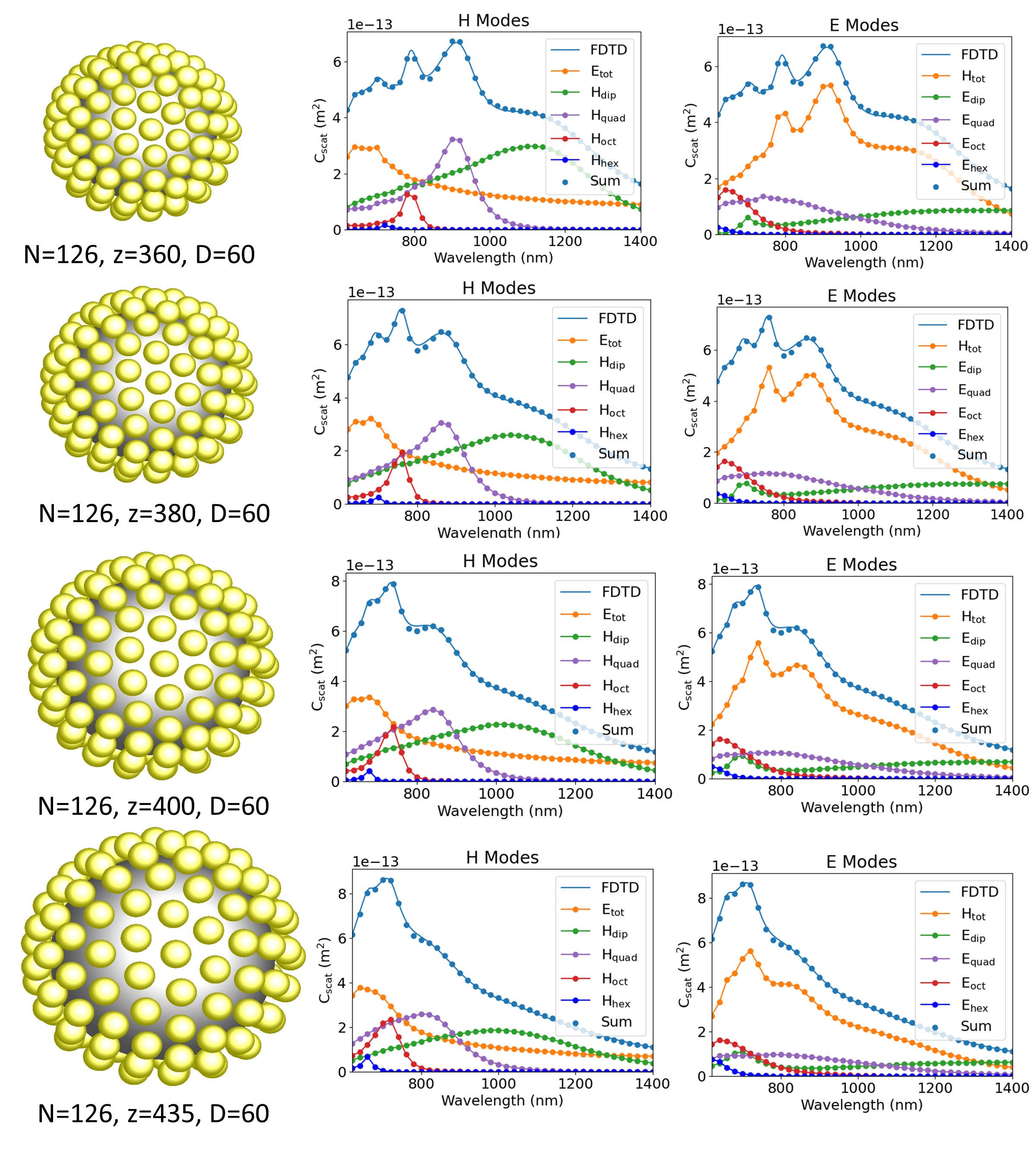} 
\caption{Full modal breakdown (including all electric and magnetic modal scattering cross sections) for Series \textbf{A} (the series where we vary $Z$).}
\label{fig:CoreSizeVaryingDecomp}
\end{figure} 
\pagebreak
\noindent
\underline{Series B: Scaling Structure}

\begin{figure}[H]
\includegraphics[width=\linewidth]{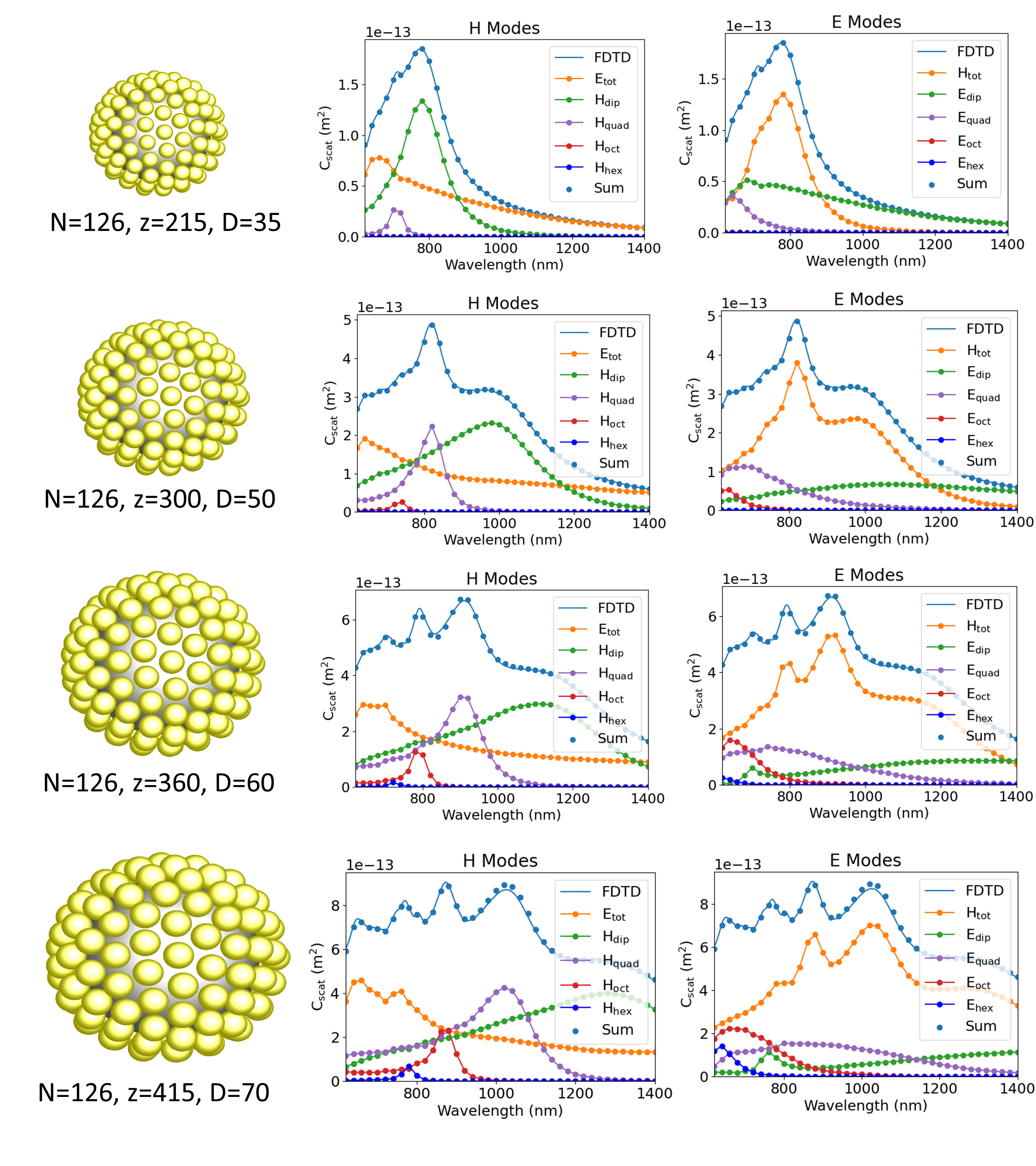} 
\caption{Full modal breakdown (including all electric and magnetic modal scattering cross sections) for Series \textbf{B} (the series where we keep the packing constant while varying $Z$ and $D$).}
\label{fig:StructureVaryingDecomp}
\end{figure}
\pagebreak

\noindent
\underline{Series C: Varying Bead Size/Number}
\begin{figure}[H]
\includegraphics[width=\linewidth]{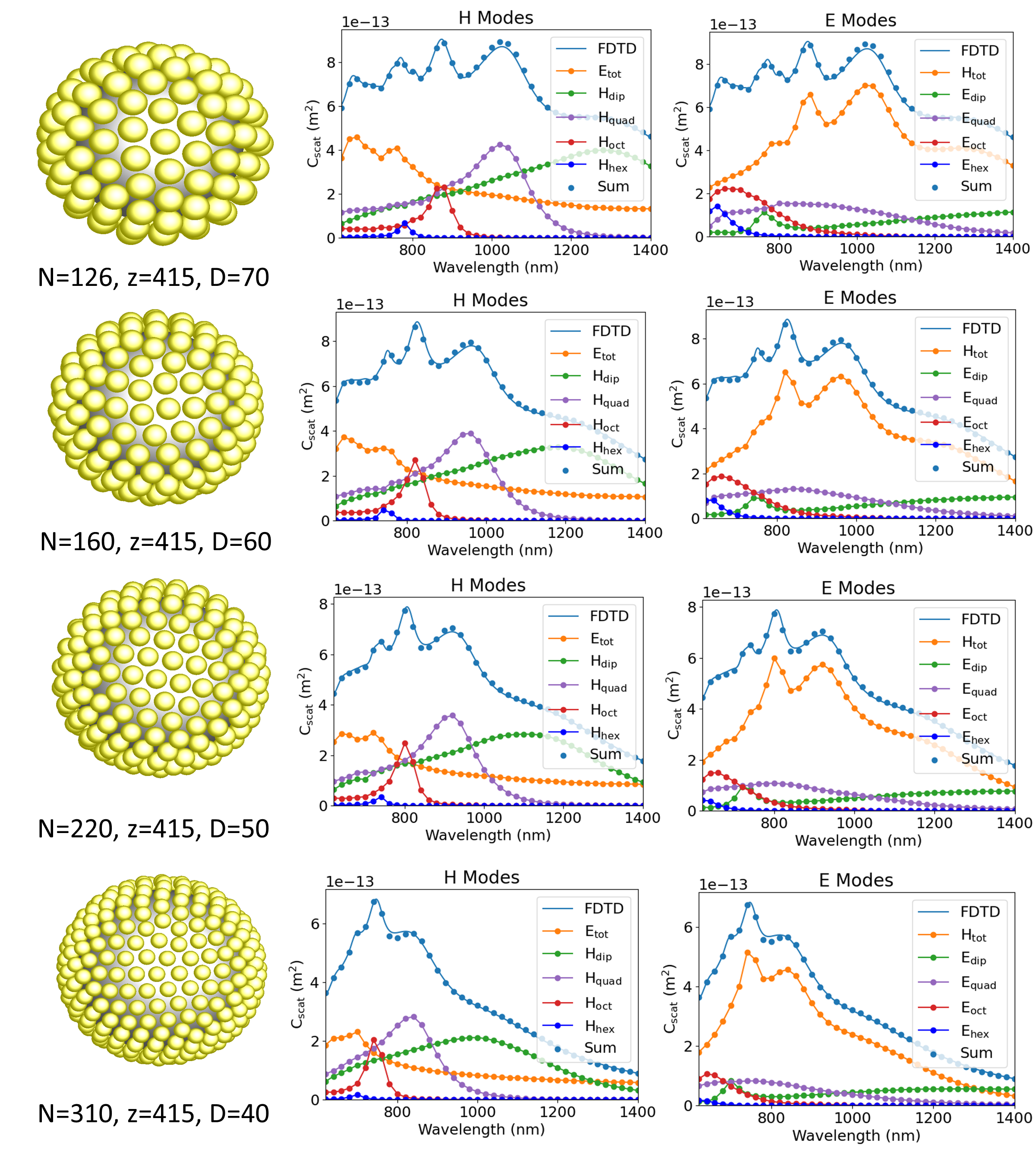} 
\caption{Full modal breakdown (including all electric and magnetic modal scattering cross sections) for Series \textbf{C} (the series where we keep $Z$ constant while increasing $N$ and decreasing $D$).}
\label{fig:BeadSizeVaryingDecomp}
\end{figure}
\pagebreak

\subsection{IX. Mie Based Explanation for Magnetic Multipole Resonance Trends}

As described in the main text, the magnetic component of the incident electric field for a given mode and wavelength peaks at a certain distance from the center of the structure. The resonance narrows and blue shifts as the mode order increases and the resonance for all modes also broadens and red shifts as we move further from the center of the structure. When the beads fall within this resonant zone for a certain mode, the dipoles within the beads can resonate with the incident magnetic mode leading to strong magnetic scattering. 

For a given $n, m$ with a plane wave propagating in the $z$ direction, if we expand the electric field of the incident wave caused by the magnetic field in terms of spherical waves we get \cite{Tsang}:

\begin{equation}
\vec{E}^{inc}_{H}(r, \theta, \phi) = -\sum\limits_{n=1}^{\infty} \sum\limits_{m=-n}^{n} i^n \frac{(2n + 1)}{n (n+ 1)} \hat{e}_i \cdot \vec{C}_{-mn}(0, 0) j_n(kr) \vec{C}_{mn} (\theta, \phi) = -\sum\limits_{n=1}^{\infty} j_n(kr) \sum\limits_{m=-n}^{n} \vec{f}_{mn}(\theta, \phi)
\end{equation}

Thus, we see that for a given mode and sub-mode the radial strength of the incident mode depends only on $j_n(kr)$, a function of wavelength, refractive index of medium, and distance. For a diagonal T-matrix the strength of the scattered mode is directly proportional to that of the incident mode and the proportionality constant is based on wavelength and structural parameters. We plot $j_n(kr)$ as a function of $\lambda$ and $r$ for two different modes (dipole ($n=1$) and octupole $(n=3)$) and two different refractive indices ($N_i=1$ and $N_i=3$) in Figure \ref{fig:propagation}. 

\begin{figure}[H]
\includegraphics[width=0.8\linewidth]{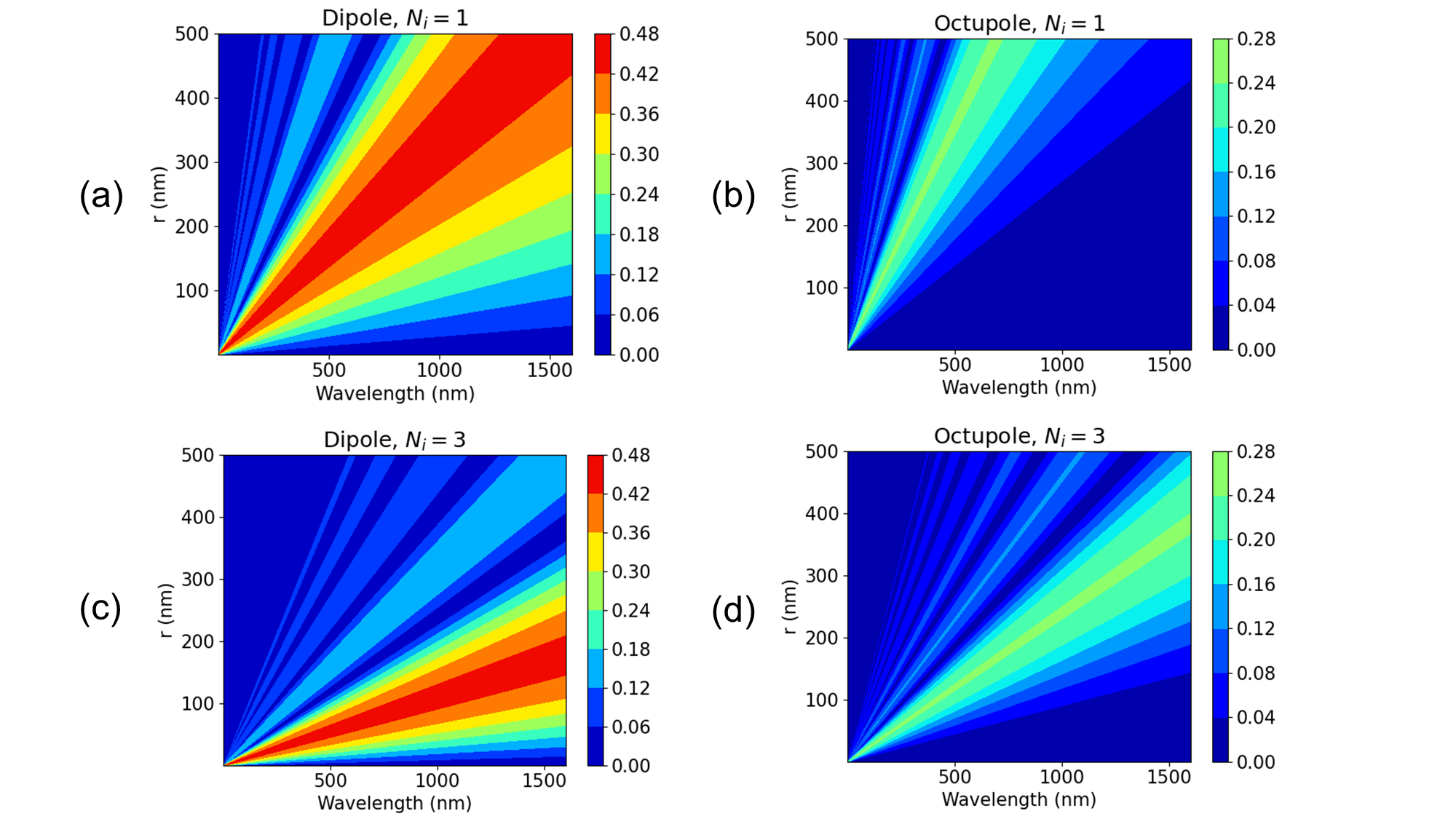}
\caption{A plot of $j_n(kr)$ ($k = \frac{2 \pi N_i}{\lambda}$) for (a) a magnetic dipole with $N_i=1$, (b) a magnetic octupole with $N_i=1$, (c) a magnetic dipole with $N_i=3$, and (d) a magnetic octupole with $N_i=3$. This represents the radial incident field strength of the magnetic component of the electric field of a plane wave.}
\label{fig:propagation}
\end{figure} 

We see that the higher order incident modes do not red-shift as much when $r$ is increased. This is consistent with the observation that higher order magnetic modes do not red-shift as much as lower order modes as the structure size is increased. Moreover, the width of the dipole incident peak at a fixed $R$ is in general significantly larger than that of the octupole peak, which may help explain why higher order modes tend to be sharper resonants. This observation also explains that as $N$ is increased for a fixed $r$, which is analogous to increasing the refractive index of the core, the wavelength of the resonance frequency increases along with its width, resulting in stronger higher order modes with smaller structures, which has been seen in previous studies \cite{Parker}. Thus we can explain many of our observations just by observing the nature of $j_n(kr)$.

\subsection{X. Additional Directional Scattering Information}
\begin{figure}[H]
\includegraphics[width=\linewidth]{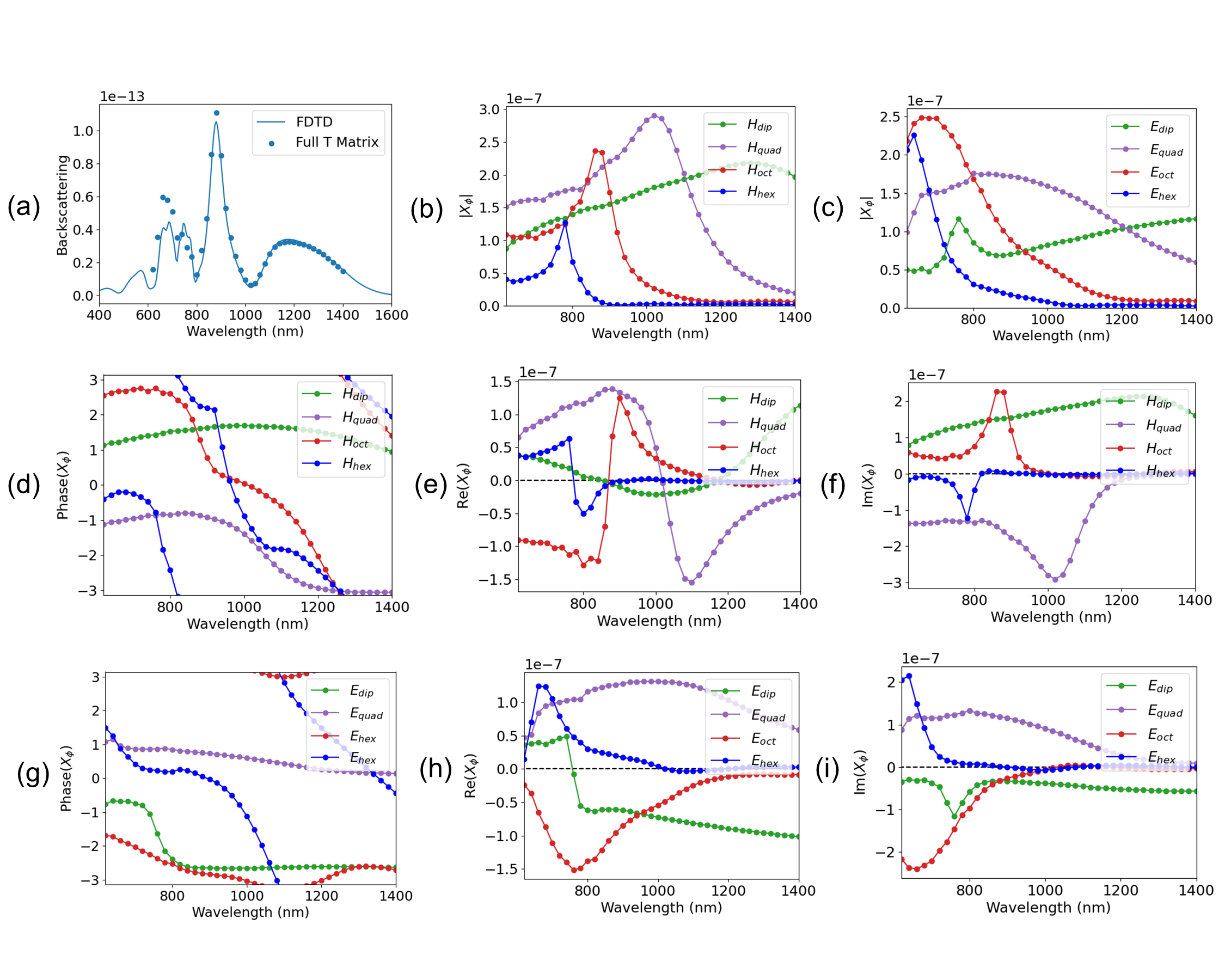}
\caption{(a) Differential backscattering cross section from FDTD (solid lines) and T-matrix (blue symbols) of the 70nm bead MM ($N=126$, $z=415$~nm, $D=70$~nm). The absolute value of the scattering amplitude ($|{X}_{\phi}|$)
of the (b) magnetic and (c) electric modes in the backscattering direction. Phase of the scattering amplitude (${X}_{\phi}$) of the (d) magnetic and (g) electric modes in the backscattering direction. Re(${X}_{\phi}$) of the (e) magnetic and (h) electric modes. Im(${X}_{\phi}$) of the (f) magnetic and (i) electric modes. Note that $\vec{X}$ has a $\theta$ component but is small compared to its $\phi$ component for the backscattering.}
\label{ref:detailed_backscattering}
\end{figure}

\begin{figure}[H]
\includegraphics[width=\linewidth]{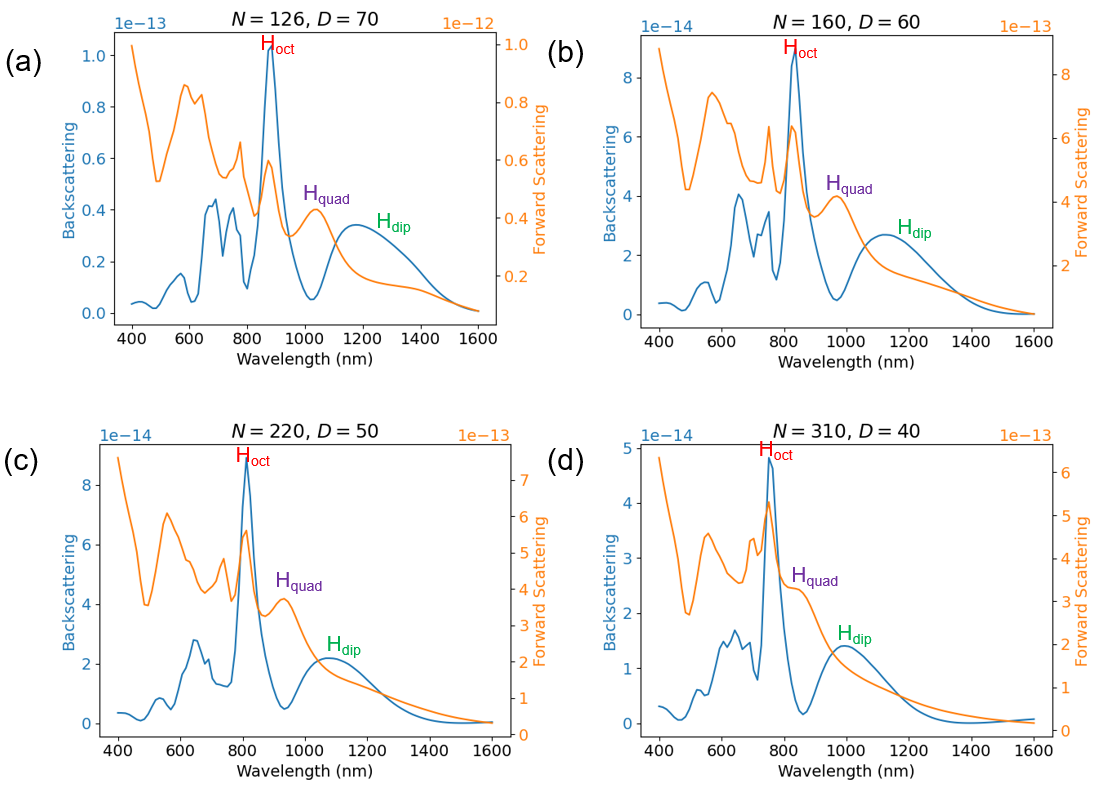}
\caption{FDTD differential forward scattering (orange) and backscattering (blue) cross-sections for MMs with a constant core size of $Z=415$~nm and (a) $N=126$, $D=70$~nm, (b) $N=160$, $D=60$~nm, (c) $N=220$, $D=50$~nm, and (d) $N=310$, $D=40$~nm, corresponding to series {\bf C} in Figure \ref{fig:BeadSizeVaryingDecomp}. As can be seen, the general features of contrast between the forward and backscattering peaks are conserved for a wide range of structural parameters as long as the core is sufficiently large.}
\label{ref:multiple_directional}
\end{figure}

\end{suppinfo}

\end{document}